\documentclass{article}

\usepackage{arxiv}

\usepackage[utf8]{inputenc} 
\usepackage[T1]{fontenc}    
\usepackage{hyperref}       
\usepackage{url}            
\usepackage{booktabs}       
\usepackage{amsfonts}       
\usepackage{dirtytalk}
\usepackage{color}
\usepackage{listings,newtxtt}

\usepackage{hyperref}
\usepackage{nicefrac}       
\usepackage{microtype}      
\usepackage{lipsum}
\usepackage{graphicx}
\usepackage{multicol}
\usepackage{amsmath}
\usepackage{cuted}
\graphicspath{ {./} }
\usepackage{float}

\usepackage[
backend=biber,
sorting=ynt
]{biblatex}

\usepackage{titlesec}
\titlespacing*{\section}{0pt}{1.1\baselineskip}{\baselineskip}
\titlespacing*{\subsection}{0pt}{1.1\baselineskip}{\baselineskip}
\titlespacing*{\subsubsection}{0pt}{1.1\baselineskip}{\baselineskip}

\itemsep=\parsep

\DeclareUnicodeCharacter{2212}{-}
\DeclareUnicodeCharacter{2009}{=}

\title{ \LARGE A Novel Approach to Topological Graph Theory\newline\newline With R-K Diagrams \& Gravitational Wave Analysis\ }

\author{
 Animikh Roy \\
  School of Mathematical \& Physical Sciences\\
  University of Sussex\\
  Brighton BN1 9QH \\
  \texttt{ar505@sussex.ac.uk}
   \And
   Andor Kesselman\\
   Senior Engineer \\
   Pathr.ai \\
  \texttt{andor@henosisknot.com}\\
}

\begin{document}

\lstset{basicstyle=\ttfamily, keywordstyle=\bfseries}
\maketitle

\begin{abstract}
  \textit{\textbf{Graph Theory and Topological Data Analytics} (TDA), while powerful, have many drawbacks related to their sensitivity and consistency with \textbf{TDA \& Graph Network Analytics}. In this paper, we aim to propose a novel approach for encoding vectorized associations between data points for the purpose of enabling smooth transitions between Graph and Topological Data Analytics.\cite{01.1_1stCourse2018algebraicTopo} \cite{01.9_2007MapperPBG} \cite{01_GCarlssonEpstein2011} We conclusively reveal effective ways of converting such vectorized associations to simplicial complexes representing micro-states \cite{01.4_2019SimpCompMicroStat} in a Phase-Space, resulting in filter specific, Homotopic Self-Expressive,event-driven unique topological signatures which we have referred as  \textbf{\say{Roy-Kesselman (R-K) Diagrams}} with Persistent Homology, which emerge from filter-based encodings of \textbf{\say{(R-K) Models}}.  We finally aim to provide an efficient computational framework via the \textbf{\say{(R-K) Pipeline}} which can be used via the \textbf{R-K Toolkit} for filter based ML Driven models for unique topological signature identification and classification problems. This pipeline would be responsible for shape invariant encoding in strong correspondence with topological network divergence as global-macro-state properties of high-dimensional datasets.\cite{01.3_2016TDANewOpportunities} The validity and impact of this approach were tested specifically on  high-dimensional (parameter specific) raw and derived measures from the latest \textbf{LIGO} datasets published by the \textbf{LIGO Open Science Centre} \cite{01.5_LIGOOpenSci} \cite{00_LIGOOpenSciData} along with testing a generalized approach for a non-scientific use-case, which has been demonstrated using the \textbf{Tableau Superstore Sales} dataset.\cite{tableau_community_forums_2021} The results obtained verify the basis of this novel approach in the following ways: \textbf{(1)} Distinct topological structures were created from the data and we have done so both on a store sales data and LIGO data. \textbf{(2)} These topological structures have emergent properties that when evaluated and compared to, have the capacity to provide meaningful insights into the data that standard data analysis techniques would not identify. For example, topological differences were exposed in the analysis that could not be exposed over metric based analysis such as euclidean distance, 'Mahalanobis Distance', and other standard metric based distance measures. \textbf{(3)} The resulting structures provide an extensible representation, which can be applied different methods of analysis, such as classification, identification, segmentation, etc. In the store sales example, results were trained over a non-gradient, combinatorial approach. With respect to store sales data, unique topological structures were derived between distinct purchase events with a loss ( measured in similarity between a set of R-K Diagrams ) in the range of $\lbrack0.78, 0.88\rbrack$. In the case of the classification case study with LIGO data analysis, we recorded a high accuracy of compact binary classifications, and a average similarity measure of BH-BH Mergers falling between the range of  $\lbrack0.9, 0.96\rbrack$ and NS-NS Mergers between a range  $\lbrack0.85, 0.91\rbrack$. Inter-class similarity measures between black holes, neutron stars \& candidate PBH measures were found to be between 0.62 \& 0.71 on with distinctly different R-K Diagrams. Therefore, we believe the findings of our work will lay the foundation for many future scientific and engineering applications of stable, high-dimensional data analysis with the combined effectiveness of \textbf{Topological Graph Theory} transformations.\cite{17.0_2001TGTIntro}\cite{17.1_2012foundationsTGT}}
\end{abstract}

\vskip15mm

\begin{multicols}{2}

 \section{Introduction}

\subsection{Overview}
In this paper, we aim to provide a novel computational framework for efficient and simplified Topological Graph Analysis \cite{01.3_2016TDANewOpportunities} \cite{01.6_GTIntro} \cite{17.1_2012foundationsTGT} \cite{17.3_1996topologicalGT} \cite{17.4_TGTRecentResults}on high-dimensional structured datasets with hierarchical embeddings that are consistent with the mathematical foundations and advancements in both Graph Theory \cite{01.8_ModernGT} and Topology.\cite{01.1_1stCourse2018algebraicTopo} Our objectives and motivation are driven towards the creation of a fundamental \textbf{\textit{"Event-Driven Topological Analysis"}} tool that could automatically generate distinct and unique signatures for the efficient identification and classification of \textbf{\textit{"Events"}} and \textbf{\textit{"Entities"}} based on the choice of \textbf{``Lens''}(As explained in \hyperref[sec:sectionlens]{4.5.5}) with similar or dissimilar attributes derived from scientific or enterprise data with specific ontological properties. The central idea of \textbf{\textit{"Event-Driven Topology"}} is based on the topological clustering \cite{04.0_1975clusteringbook} \cite{06.1_carlsson2008persistentHomo} of all attributes around a central \textbf{\textit{"Event-Node"}} which serves as a static (unchanging) context for all dependent  attribute clusters. Such attribute-clusters have their own \textit{"Directed-Acyclic-Graphs"} \cite{20.0_2013AlgebraOfDAGs} \cite{20.1_2001DAGMechanics} that ensure their uniqueness and interdependencies based on a particular domain Ontology.Thus unique topological signatures are obtained using domain-specific filters or threshold constrains on underlying fundamental structural graphs called \textbf{\say{Roy-Kesselman (R-K) Models}} represented by tensors in n-dimensional space, which are reduced to consistent "Topohedrons" (Polyhedrons that preserve topological invariance under geometric transformations or perturbations of state) with persistent Homology called  \textbf{\say{Roy-Kesselman (R-K) Diagrams}}.  Similar \textit{``Topological Events''} would produce unique \textbf{R-K diagrams} which remain static and unchanging under perturbations caused due to changes in data values at the row level or the influx of data in terms of new/additional rows. On the other hand, different \textit{"Topological  Events"} would produce distinctly different \textbf{R-K diagrams} based on perturbations caused by  the influx of data in terms of new/added rows or additional independent attributes in terms of columns driven by Hierarchical-Feature-Extractions with respect to a particular domain Ontology.

We also aim to present a scalable domain-agnostic pipeline called the \textbf{\say{R-K Pipeline}} with a customisable Toolkit called the \textbf{\say{R-K Toolkit}} for different domain ontologies and for parameter optimizations, which would be responsible for following: (1) Shape invariant encoding (2) Topological Event Identification (3) Topological Classification \& (4) Isometric  Compression of high-dimensional graphs without the loss of essential information or generality.  Each step of our novel approach and methodology have been thoroughly tested on scientific (LIGO Open Science) \cite{01.5_LIGOOpenSci} \cite{00_LIGOOpenSciData} and commercial (Tableau Superstore Sales) datasets, producing efficient and consistent results that could be applied towards future studies on parametrised classification of Event-Signatures and Data-Entities in the field of Topological Graph Theory.\cite{17.1_2012foundationsTGT} This would also allow for the combination and analysis of multiple datasets and data-streams in real-time, such as in the case of Multi-messenger Astronomy.

\subsection{Motivation}

Graph Theory \textbf{(GT)} \cite{01.6_GTIntro} \cite{01.7_GTApplications} \cite{01.8_ModernGT} and Topological Data Analysis \textbf{(TDA) } \cite{01.3_2016TDANewOpportunities} \cite{01_GCarlssonEpstein2011} have recently emerged as independent novel frameworks for extracting hidden meaning and underlying insights from the study of geometric structure, shape and connections of such vast and complex datasets. \cite{02.3_2017introductionTDA} \cite{02.4_TDAResearch} However modern computational tools lack the technology, efficiency and flexibility to consistently carry out Graph Theory Network Analysis with hierarchical connections with localised clustering due to the inherent variability that could encode directed relationships in the affine connexions \cite{23.2_AffineConnection} \cite{23.1_7FundamentalQuants}
of the data-points in phase space, in order to build homotopic manifolds and simplicial complexes\cite{02.6_2009TDAChallenges}.\cite{01.9_2007MapperPBG} \cite{03.1_2009simplicialHomotopy}\cite{01_GCarlssonEpstein2011}  They also lack a consistent framework to mathematically define and classify the global properties of the same network through an effective means to smoothly transit between Graph and Topological structures as established theoretically in Topological Graph Theory (TGT)\cite{17.3_1996topologicalGT} \cite{17.4_TGTRecentResults}. This would prevent the additional burden of having to regenerate the entire data geometry from scratch due to lack of persistent homology between the two models.\cite{02_carlsson2009topology} \cite{03.3_de2007PersistentHomology} \cite{01_GCarlssonEpstein2011}.

\subsection{Objectives}

Being able to showcase TDA and GT capabilities via smooth mathematical transformations on the same data network (consisting of one or more datasets and data streams) without the necessity to recreate its underlying geometric structure encompasses an enormous field of untapped potential taking inspiration from Topological Graph Theory \cite{17.0_2001TGTIntro} \cite{17.1_2012foundationsTGT}  in modern scientific Big-data Analytics.\cite{02.6_2009TDAChallenges}\cite{18.0_2016topologicalBigChem} \cite{18.2_2018TDAonBigData} This academic paper explores the possibility of consistently improving existing GT and TDA technologies with enhanced geometric compatibility while preserving their respective mathematical properties through simple Vectorized Associations in Phase Space.\cite{19.0_2010PhaseSpace} This research also aims to facilitate a smooth transition between these two advanced analytical methodologies by using machine learning to fine tune effective and unique \textit{"Event-Driven"} topological signatures that manifest as self-expressive homotopic\cite{07_bjorner2003Homotopy} \cite{03.1_2009simplicialHomotopy} \say{Roy-Kesselman Diagrams} (R-K Diagrams).

These are shown as a special category of Polyhedrons that maintain Topological Invariance\cite{12.1_2002topologicalInvariaceProjection} \cite{01.0_2010introductionTopoPropertiesInvariance} and Persistent Homology when projected onto a Phase Space. These \textit{"Topohedrons"} or filtered \say{R-K Diagrams} can be generated via filter-based ML driven optimizations on the underlying n-dimensional data set, which are preserved within the underlying Topological Space. This work allows for future research into the preservation of Homotopy of such \textit{"Topohedrons"} under continuous deformations brought about by any changes in Topological Network Entropy due to data perturbations. It also formulates such implications through mathematical and computational models as shown in this paper.\cite{05.1_2007computingTopoEntropy}

The findings of this work have seminal implications on high-dimensional, complex scientific data sets especially in the context of Dark Matter and Gravitational Wave  Analysis\cite{00_LIGOOpenSciData} \cite{00.1_2012GWAnalysisFormalism} with LIGO-Virgo datasets without the necessity of conventional clustering and binning techniques. \cite{00.2_schutz2012GWDataAnalysis}We also justify it's scope and validity on a generalized non-scientific use case on commercial \hyperref[sec:store_sales_section]{Tableau Super Store Data}
. It also replaces the existing Mapper algorithms \cite{01.9_2007MapperPBG} with a holistic analytical framework that goes well beyond partial clustering and addresses \textbf{\textit{\say{Persistent Homology}}} \cite{01.1_1stCourse2018algebraicTopo} \cite{06.4_2005computingPHomology} for Topological shape rendering with built-in  \textbf{\textit{\say{Isometric Data Compression}}} capabilities.\cite{01.0_2010introductionTopoPropertiesInvariance} \cite{21.0_2016TopoCompression} \cite{12.2_compressingTopoNetworkGraphs}

 \section{Mathematical Background}

In this section we shall elaborate on the mathematical formulations and conceptual foundations upon which we aim to build this novel approach towards Topological Graph Theory with some unique data modelling characteristics. This section will also contain all corresponding references to the latest and most relevant research in the respective fields of Topology \& Graph Theory \& Topological Data Analysis. 

\subsection{Eulerian Graph Theory \& the Birth of Topology}

Graph theory is a well-established and versatile branch of mathematics that is primarily concerned with networks of points connected by lines. The subject of graph theory had its beginnings in recreational mathematical problems in the early 17th century, especially those related to the historical drawing of graphs stemming from cartographic representation of geographic regions. \cite{01.10_2001HistoryofGT} However, it has currently grown into a significant area of mathematical research, with applications in Chemistry, Biology, Operations Research, Social Sciences, Computer Science, Natural language processing \cite{22.0_2015_NLPGraphs} and Big-Data Analytics.\cite{22.1_BigDataPredictionGT} \cite{22.2_GTapplicationBigData} The history of graph theory may be specifically traced to 1735, when its founding father, Swiss Mathematician Leonhard Euler solved the Königsberg bridge problem. \cite{01.13_GTBackground}The Königsberg bridge problem was an old puzzle concerning the possibility of finding a path over every one of seven bridges that span a forked river flowing past an island—but without crossing any bridge twice. Euler argued that no such path exists. His proof involved only references to the physical arrangement of the bridges, but essentially he proved the first theorem in graph theory.\cite{01.11_1999historyofTopo}

\begin{figure}[H]
	\centering
	\includegraphics[width=\linewidth]{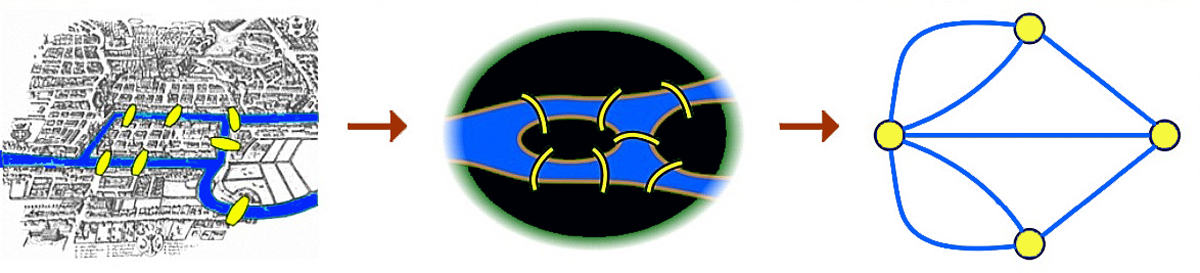}
	\caption{\textit{The above diagram shows the Königsberg bridge problem and its corresponding resolution by Leonhard Euler that lead to the introduction of graph theory and the subsequent birth of the branch of mathematical topology}}
	\label{fig:fig1}
\end{figure}

It is important to note, that in the field of Graph Theory, the term \textit{"graph"} does not refer to data charts, such as the likes of line graphs or bar graphs pertaining to the graphical visualization of data. Instead, it refers to a set of Vertices (V) (i.e., points or nodes) and Edges (E) (or lines) that connect the vertices. When any two vertices are joined by more than one edge, then such a graph is called a \textit{"Multi-graph"}. \cite{01.14_HDMultigraphs}\cite{01.13_GTBackground} A graph without any loops and with a maximum of one edge between any two vertices is called a \textit{simple graph}. When each and every vertex of a graph is connected by an edge to every other vertex, then such a  graph is called a \textit{complete graph}. Moreover, it is important to note in the context of this paper, a direction is assigned to each edge of a graph to produce what is known as a \textit{Directed Graph or Digraph}.\cite{20.0_2013AlgebraOfDAGs} We shall be dealing with such Directed Graphs for the remaining part of this paper.

Another useful concept in Graph Theory that of the \textit{path}. In this field, a path is defined as any route along the edges of a graph.\cite{01.13_GTBackground} A path may follow a single edge directly between two vertices, or it may follow multiple edges through multiple vertices.\cite{01.6_GTIntro} If there is a path linking any two vertices in a graph, then such a graph is called a \textit{connected graph}. A path that begins and ends at the same vertex without traversing any edge more than once is called a circuit, or a closed path. A circuit that follows each edge exactly once while visiting every vertex is known as an Eulerian circuit, and the graph is called an Eulerian graph. An Eulerian graph is connected and, in addition, all its vertices have even degree.\cite{01.16_EulerianGraphs} \cite{01.17_EulerianGraphOrientations} Such Eulerian Graphs shall be taken into consideration in the computational framework of our novel approach.

It is also important to know that the histories of graph theory and topology are closely related, and the two areas share many common problems and techniques even today, \cite{01.12_2013TopoHistHandbook} \cite{01.10_2001HistoryofGT} especially in the domains of mathematics and computer science. Such similarities can be traced back to Euler, who referred to his work on the Königsberg bridge problem as an example of \textit{geometria situs} or the “geometry of position”; while on the other hand, the development of topological ideas during  the 19th century became famously known as \textit{analysis situs}, or the “analysis of position” thereby interlinking these two fundamental pillars of mathematics.\cite{01.13_GTBackground}\cite{01.14_HDMultigraphs} As a further extension to the foundation of GT in 1750, Euler discovered the polyhedral formula $V - E + F = 2$ relating the number of \textit{vertices (V), edges (E), and faces (F)} of a polyhedron (a solid object in 3D geometry with each enclosed surface or face represented by polygons). \cite{01.12_2013TopoHistHandbook}\cite{01.16_EulerianGraphs}

The vertices and edges of all such polyhedrons form graphs on its surface, and this notion led to consideration of graphs on other surfaces such as a torus (the surface of a solid doughnut) and how they divide the surface into disk-like faces. Euler’s formula was soon generalized to surfaces as $V-E+F=2-2g$, where g denotes the genus, or number of “doughnut holes,” which help determine their Euler Characteristic. \cite{01.18_EulerFormula} Having considered a surface divided into polygons by an embedded graph, mathematicians began to study ways of constructing surfaces, and later more general spaces, by pasting polygons together. This was the beginning of the field of Combinatorial Topology, which later, through the work of the French mathematician Henri Poincaré and others, grew into what is known as Algebraic Topology.  \cite{01.1_1stCourse2018algebraicTopo} \cite{01.11_1999historyofTopo} \cite{01.12_2013TopoHistHandbook}Such concepts related to polyhedrons and Euler Characteristics that link the concepts of Graph Theory and Algebraic Topology through Combinatorics, serve as the theoretical basis for the mathematical formulations of our approach.

\subsection{Topological Data Analytics}

Topological Data Analytics or TDA \cite{02.3_2017introductionTDA} \cite{01_GCarlssonEpstein2011} is a mathematical methodology that is often represented as a useful computational framework or tool for high-dimensional data (i.e. data with multiple independent columns, variables or features across different datasets) analysis developed initially by Herbert Edelsbrunner, Afra Zomorodian, Gunnar Carlsson, and his graduate student Gurjeet Singh.\cite{01.9_2007MapperPBG} \cite{02.1_GCarlson2004topoEstimation} \cite{01_GCarlssonEpstein2011}The core idea of TDA  is based on the founding principle of Gunnar Carlson et.al\cite{02.4_TDAResearch} which states that, high-dimensional \& complex data sets (with multiple features and parameters) have intrinsic topological shape that can be represented in n-dimensions using generalized coordinate systems.\cite{01_GCarlssonEpstein2011} However, such shapes can then be assigned extrinsic characteristics and context-based meaning, thus enabling a robust and scalable framework of data labelling, classification and analysis. Such computational frameworks of topological analysis can be further enhanced and automated using Machine Learning and Neural Networks as elaborated in the later sections of this paper.the first data analysis framework using TDA was popularized by Carlsson’s paper in 2009  called \say{Topology $\&$ Data} \cite{02_carlsson2009topology} that later turned TDA into a hot field in applied mathematics, and also found many applications in Data-Science and Big-data Analytics.\cite{02.6_2009TDAChallenges}\cite{18.2_2018TDAonBigData} \cite{17.1_2012foundationsTGT} However, the mathematical foundations that drive TDA had been laid years before by others in the fields of Topology, Group Theory, Linear Algebra and Graph Theory as discussed earlier and referenced in this paper.

As evident today, an important feature of modern science and engineering is that data of various kinds is being produced at an unprecedented rate. \cite{02.6_2009TDAChallenges} \cite{18.2_2018TDAonBigData} This partly because of new experimental methods, and in partly because of the increase in the availability of high powered computing technologies.\cite{18.0_2016topologicalBigChem} \cite{18.1_2017TopoBigdataPipeline} It is also clear that the nature of the data that we obtain today is significantly different. For example, it is now often the case that we are given data in the form of very long vectors, where all but a few of the coordinates turn out to be irrelevant to the questions of interest, and further that we don't necessarily know which coordinates are the most relevant ones for the ideal solution. A related fact is that the data is often very high-dimensional, which severely restricts our ability to visualize it. The data obtained is also often much noisier than in the past and has more missing information (missing data).\cite{02_carlsson2009topology}

This is particularly so in the case of scentific data related to high throughput data \cite{11.0_chazal2017TopoDataScience}from micro-arrays in biology or other sources such as particle accelerators in high-energy physics, or in the case of multi-messenger data-streams in modern-day Astronomy. thus our ability to analyse such data, both in terms of quantity and the nature of the data, is  clearly not keeping pace with the data being produced.\cite{18.2_2018TDAonBigData} In this paper, we will discuss how geometry and topology can be applied to make useful contributions to the analysis of various kinds of data. Our aim is to establish that geometry and topology are very natural tools to apply in this direction, since geometry can be regarded as the study of distance functions, and in any statistical data-analysis model, we often end up applying such distance functions on large finite sets of data. The mathematical formalism which has been developed thus far for incorporating geometric and topological techniques deals with point clouds, i.e. finite sets of points equipped with a distance function. It then adapts tools from the various branches of geometry to the study of point clouds. The point clouds are intended to be thought of as finite samples taken from a geometric object, perhaps with noise.\cite{01_GCarlssonEpstein2011} \cite{01.9_2007MapperPBG} \cite{01.3_2016TDANewOpportunities}

However, our novel approach is different from traditional methods of topological analysis using point cloud data, in that it is primarily \textit{"Event-Driven"} in its technique, which appeals to a limited but widely applicable sub-set of Topological Graph Analysis where in hierarchical relationships can be determined between various dependent and independent column attributes of a dataset and can be then connected to a contextual \textit{"Event" or "Entity"} based on the choice of a \textit{``Lens''}, which will be further explored in our work. In all cases under consideration related to our approach, the  characteristics features of a dataset are associated to a central static context known as an "Event-Node" with independent features and variables associated with such events mapped to it in separate attribute clusters. This gives rise to well-defined hierarchical feature based multi-graphs with nodes and directed edges which represent the \textit{"micro-states"} of an Event, while its \textit{"macro-state"} is determined by the overall topology of the R-K diagrams with distinctive macro features such as holes, voids, loops etc. which can be characterised with the help of  Betti-Numbers which are unique to different classifications determined by domain specific filters and ontological parameters.

Now we shall note some of the key problems that occur when applying traditional statistical \&  geometric methods to data analysis that we aim to address using our analytical framework and pipeline:

$\bullet$ \textbf{\textit{High-dependence on Qualitative Information:}} One important goal of Data Analysis at its core, is to allow the user to obtain knowledge about the data, i.e. to understand how it is organized on a large scale. For example, if we imagine that we are looking at a data set constructed somehow from diabetes patients, it would be important to develop the understanding that there are two types of the disease, namely the juvenile and adult onset forms.\cite{01_GCarlssonEpstein2011}\cite{02.1_GCarlson2004topoEstimation} Once that is established, one of course wants to develop quantitative methods for distinguishing them, but the first insight about the distinct forms of the disease is key which needs to be qualified in non-topological analytical frameworks, where as differences in topological signatures of data pertaining to such patients can enable automated segregation for easier understanding and classification.

$\bullet$\textbf{\textit{Metrics are not Theoretically Justified:}} In physics and Astronomy, the phenomena studied often support clean explanatory theories which tell one exactly what metric to use. In biological problems, on the other hand, this is much less clear. But in certain cases it is far from clear how much significance to attach to the actual data points and distances, particularly at large scales and in developing fields of research such as in the case of Gravitational Wave Astronomy (LIGO, Virgo , etc.). Thus Topological frameworks such as ours would allow for smooth and effective measures of noise reduction in a phase-space independent of any metric dependencies while retaining the vital information isometrically, even in the case of transformations such as projection or compression.

$\bullet$\textbf{\textit{Coordinates are not Natural:}} Although we often receive data in the form of vectors of real numbers. However, it is frequently the case (in physics, astronomy and various branches of modern science) that the coordinates, like the metrics mentioned above, are not natural in any sense.Therefore it is incorrect to restrict ourselves to studying properties of the data which depend on any particular choice of coordinates. Note that the variation of choices of coordinates does not require that the coordinate changes be rigid motions of Euclidean space. It is often a tacit assumption in the study of data that the coordinates carry intrinsic meaning, but this assumption is often unjustified \cite{01_GCarlssonEpstein2011} \cite{02.1_GCarlson2004topoEstimation} \cite{02_carlsson2009topology} and can be effectively addressed in frameworks involving Topological Graph Analysis.

$\bullet$\textbf{\textit{Summaries are more valuable than individual parameter choices:}} In traditional approaches so far, a popular method of clustering a point cloud is the so-called single linkage clustering,  in which a graph is constructed whose vertex set is the set of points in the cloud, and where two such points are connected by an edge if their distance is  $\ge$ $\epsilon$, where $\epsilon$ is a parameter. \cite{02_carlsson2009topology} \cite{02.6_2009TDAChallenges} Some work in clustering theory has been done in trying to determine the optimal choice of $\epsilon$, but it is now well understood that it is much more informative to maintain the entire \textit{"Dendogram"} \cite{20.1_2001DAGMechanics}\cite{20.0_2013AlgebraOfDAGs} of the set, which provides a summary of the
behaviour of clustering under all possible values of the parameter  at once. It is therefore productive to develop other mechanisms in which the behaviour of invariants or construction under a change of parameters can be effectively summarized such as in the case of our novel framework.

\textit{We shall now discuss how Topological Graph Analysis could serve as an important and relevant way for addressing all the above mentioned problems especially to understand how they may be relevant in the case of formulating and justifying this novel approach for specific Analysis on Scientific and Ontology driven datasets that can be used to build useful Knowledge-Graphs and subsequent R-K Diagrams :}

\textbf{(1)} Topology and Graph theory are the best candidates in the branch of Analytics to deal with qualitative geometric information. This includes the study of what the connected components of a space are, but more generally it is the study of connectivity information, which includes the classification of micro-geometric properties such as nodes, edges and directions of DAGs \cite{20.0_2013AlgebraOfDAGs} to signify micro-state differences. While, macro-geometric properties are distinguished based on holes, voids, loops and higher dimensional surfaces within the space which are represented by distinct Euler Characteristics and Betti Numbers \cite{01.16_EulerianGraphs} \cite{01.17_EulerianGraphOrientations} of the specific unique topology. This suggests that extensions of topological \& graph methodologies, such as DAGs, Euler Characteristics, Betti Numbers, Homotopy \& Homology, should be helpful in studying them qualitatively.\cite{03.1_2009simplicialHomotopy} \cite{03_1944Homology}

\textbf{(2)} Topology \& Graph Theory study geometric properties in a way which is much less sensitive to the actual choice of metrics than straightforward geometric methods, which involve sensitive geometric properties such as curvature. \cite{01.1_1stCourse2018algebraicTopo} In fact, topology ignores the quantitative values of the distance functions and replaces them with the notion of infinite nearness of a point to a subset in the underlying space. This insensitivity to the metric is useful in studying situations where one only believes one understands the metric in a coarse way.\cite{02.3_2017introductionTDA}

\textbf{(3)} Topology studies only properties of geometric objects which do not depend on the chosen coordinates, but rather on intrinsic geometric properties of the objects. Therefore it allows for a  coordinate-free analytical approach. \cite{01_GCarlssonEpstein2011} \cite{02_carlsson2009topology}

\textbf{(4)} The idea of constructing summaries over whole domains of parameter values involves understanding the relationship between geometric objects constructed from data using various parameter values. The relationships which are useful involve continuous maps between the different geometric objects, and therefore become a manifestation of the notion of Functoriality, i.e,  the notion that invariants should be related not just to objects being studied, but also to the maps between these objects.\cite{05_mac2013Functoriality} Functoriality is central in algebraic topology in that the functoriality of homological invariants \cite{06.1_carlsson2008persistentHomo}is what permits one to compute them from local information, and that functoriality is at the heart of most of the interesting applications within mathematics. Moreover, it is understood that most of the information about topological spaces can be obtained through diagrams of discrete sets, via a process of simplicial approximation.\cite{03.1_2009simplicialHomotopy} \cite{08.2_1992simplicialAlgebricTopo}

\subsection{Topological Graph Theory}
The connection between graph theory and topology led to a sub-field called Topological Graph Theory.\cite{17.0_2001TGTIntro} \cite{17.1_2012foundationsTGT} An important problem in this area concerns planar graphs. These are graphs that can be drawn as dot-and-line diagrams on a plane (or, equivalently, on a sphere) without any edges crossing except at the vertices where they meet. Complete graphs with four or fewer vertices are planar, but complete graphs with five vertices or more are not. The use of diagrams of dots and lines to represent graphs actually grew out of 19th-century chemistry, where lettered vertices denoted individual atoms and connecting lines denoted chemical bonds (with degree corresponding to valence), in which planarity had important chemical consequences. The first use, in this context, of the word graph is attributed to the 19th-century Englishman James Sylvester, one of several mathematicians interested in counting special types of diagrams representing molecules.\cite{17.2_TGTonlineBritanica}

Topological graph theory deals with ways to represent the geometric realization of graphs. Typically, this involves starting with a graph and depicting
it on various types of drawing boards: 3-space, the plane, surfaces, books,
etc. This field mainly uses topological methods to study graphs. For example, planar graphs
have many special properties. The field also uses graphs to study topology.
For example, the graph theoretic proofs of the Jordan Curve Theorem, or
the theory of voltage graphs depicting branched coverings of surfaces, provide an intuitively appealing and easily checked combinatorial interpretation
of subtle topological concepts \cite{17.3_1996topologicalGT} \cite{17.1_2012foundationsTGT} that serve as the fundamental basis of this novel approach and are establish as a part of our computational framework. 

\subsubsection{Directed Acyclic Graphs}
In our paper, we enable a topological graph theory framework through DAG's at the micro level which can be defined as follows:

A directed acyclic graph is a graph $(N,E) \in G$ where $G$ is a graph $N$ are nodes and $E$ are edges. It is a restricted general graph, in that there are no edge cycles, such that an edge can be described as $E_{N_i,N_j}$ where $i \ne j$ and the path of edges never roll back onto itself. The benefit of a DAG is the composibility of it, and the ability to make complicated and general pipelines from it.

\subsection{Phase Space \& Projections}
In the last century, the development of modern physics has been partially driven by the incorporation of a few key concepts. A phase space is the spatial representation of all possible states of a dynamical system, where each point uniquely identifies a state. A Topological Phase Space can be defined as the n-dimensional spatial representation of the same using a generalized Curvilinear Coordinate system allowing for all possible coordinate transformations and perfect isometric compressions while preserving geometric invariance.
The concept of Phase Space in itself is a simple but powerful idea that emerged in the second half of the 19th century, during the golden era of differential geometry, and it is at the core of modern classical, quantum, and statistical mechanics. The trajectory that a dynamical system describes in the phase space as it evolves with time contains rich information about the system. For instance, by looking at the shape of the trajectories that a pendulum describes in its phase space, we can infer the existence of different dynamical regimes, or the ratio between the length of the pendulum and the acceleration of gravity.

\begin{figure}[H]
	\centering
	\includegraphics[width=\linewidth]{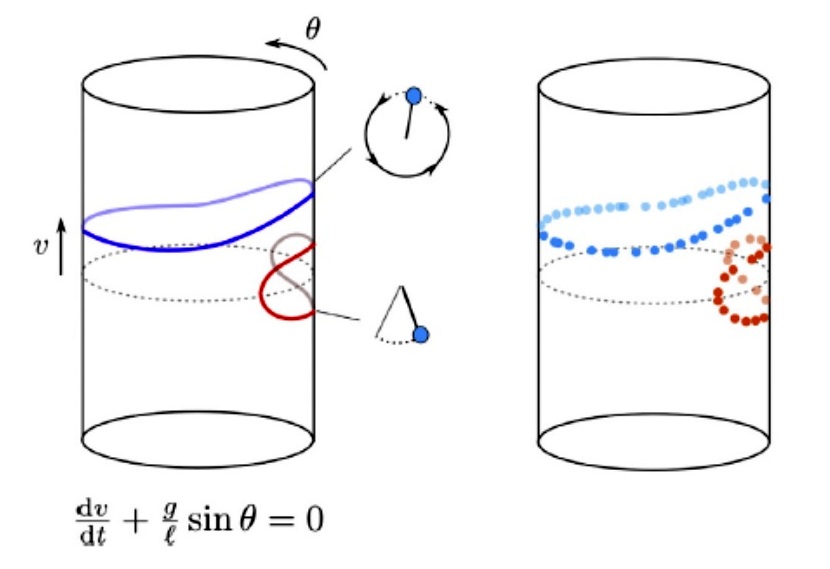}
	\caption{\textit{the above diagram shows that the phase space of a simple pendulum is a two-dimensional cylinder, where the periodic coordinate corresponds to the angle  of the pendulum with respect to the vertical, and the longitudinal coordinate to its angular velocity}}
	\label{fig:fig2}
\end{figure}

Each point in this space specifies a unique combination of the position and velocity and uniquely determines the subsequent evolution. For small angular velocities, the pendulum oscillates back and forth around the equilibrium point. For large velocities, the pendulum describes a circular motion.

\subsection{Homotopy \& Homology}

\textbf{Homotopy} \cite{03.1_2009simplicialHomotopy} \cite{08.1_2003SimplicialHomology} is defined as  a continuous deformation of one continuous function $f(x)$ to another continuous function $g(x)$ without break in topology from shearing and  tearing resulting in discontinuities or the formation of holes or gluing resulting from merger of holes that would also result in discontinuous functions. This gives rise to the notion of “essential-sameness” whereby one geometric shape can be continuously deformed into any other without tearing or shearing just like a doughnut can be converted to a mug with one whole via a smooth transformation with continuous deformations. Conversely, the doughnut cannot be converted into a sphere without gluing in terms of merging the space in between and hence such geometric shapes can be considered “essentially-different” as they cannot be inter-converted into smooth functions via continuous deformations as shown in the figure below.\cite{07_bjorner2003Homotopy}

\begin{figure}[H]
	\centering
	\includegraphics[width=\linewidth]{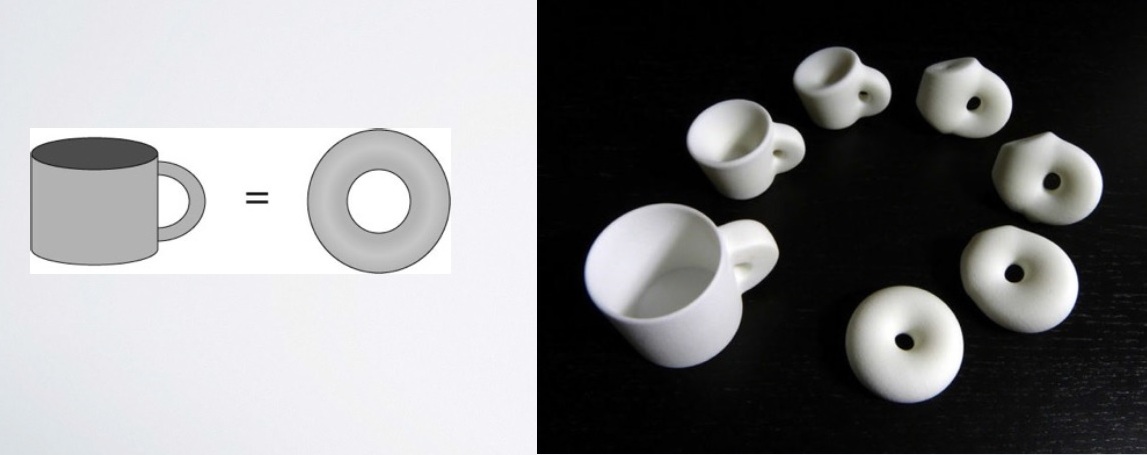}
	\caption{\textit{the above figure shows how the geometric shape of a cup $f(x)$ can be smoothly transformed into that of a doughnut $g(x)$ using continuous deformations without any discontinuities or break in topology resulting from tearing, shearing or gluing as both geometric figures begin and end with 1 hole and thus abide by the notion of “essential-sameness” in Topology. }}
	\label{fig:Homotopy}
\end{figure}

We can describe the formalism in more standard mathematical terms as follows: From the above definition and description, we can say that two continuous maps $f, g : X \rightarrow Y$ are said to be \say{Homotopic} if there is a continuous map  $H :X \times [0, 1] \rightarrow Y$ such that $H(x, 0) = f(x)$ and $H(x, 1) = g(x)$. We also say that a map $f : X \rightarrow Y$ is a homotopy equivalence if there is a map $G : Y \rightarrow X$ so that $f \circ g$ is homotopic to the identity map on Y and $g \circ f$ is homotopic to $f$. Two spaces $X$ and $Y$ for which there exists a homotopy equivalence $f : X \rightarrow Y$ are said to be homotopy equivalent. A space that is homotopy equivalent to the one point space is said to be \textit{contractible}. \cite{07.1_Homotopy} \cite{07_bjorner2003Homotopy}

\textbf{\textit{Definition:}} Thus we can conclude that for any topological space $X$, in Abelian group $A$, with integer $j \ge 0$, there is assigned a group $H_j(X,A)$ which defines the Homotopy.

\textbf{\textit{Functoriality:}} For any A and j as above, and any continuous map $f : X \rightarrow Y$, there is an induced homomorphism $H_j(f,A) : H_j(X,A) \rightarrow H_j(Y,A)$. Therefore, one has $H_j(f \circ g,A) = H_j(f,A) \circ H_j(g,A)$ and $H_j(Id_X;A) = Id_(Hk(X,A))$.These conditions are called collectively \textit{functoriality}.\cite{05_mac2013Functoriality}

\textbf{\textit{Homotopy invariance:}} If $f$ and $g$ are homotopic, then $H_j(f,A)=H_j(g,A)$. It follows that if $X$ and $Y$ are homotopy equivalent, then $H_j(X,A)$ is isomorphic to $H_j(Y,A)$.\cite{01.0_2010introductionTopoPropertiesInvariance} \cite{12.0_alatorre2018TDAinvariant}

\textbf{\textit{Normalization:}}$H_0(\star,A) \approx A$, where $\star$ denotes the one point space.

\textbf{\textit{Betti numbers:}} For any field F, $H_j(X, F)$ will be a vector space over $F$. Its dimension, if finite, will be written as $\beta_j(X, F)$, and will be referred to as the j-th Betti number with coefficients in F. The j-th Betti number corresponds to an informal notion of the number of independent j-dimensional surfaces. If two spaces are homotopy equivalent, then all their Betti numbers are equal with the following properties:

Property 1: For any topological space $X$ with a finite number of path components, $\beta_0(X)$ is the number of path components.

Property 2:  If the first Betti number $\beta_1$ of the letter “B” above is two, and for the letter “A” it is one, for any field $F$, then in this case, it provides a formalization of the count of the number of loops present in the space. \cite{07.2_TopoBettiNumbers} \cite{02_carlsson2009topology}

\textbf{\textit{Homology}}  is defined as a concept in algebraic topology that provides a  powerful tool to formalize and handle the notion of topological features of a topological space or of a simplicial complex in an algebraic way. For any dimension $j$, the $j-dimensional$ “holes” are represented by a vector space $H_j$ whose dimension is intuitively the number of such independent features. For example the 0-dimensional homology group $H_0$ represents the connected components of the complex, the 1-dimensional homology group $H_1$ represents the 1-dimensional loops, the 2-dimensional homology group $H_2$ represents the 2-dimensional cavities and so on.\cite{03.2_2008FindingHomology} \cite{03.3_de2007PersistentHomology} \cite{03.4_2008localizedHomology}

\subsection{Euler Characteristics \& Betti Numbers}

The Euler Characteristic is a useful tool in mathematics that is used to classify topological characteristics of various classes of polyhedrons, based only on a relationship between the numbers of vertices (V), edges (E), and faces (F) of any geometric figure.\cite{07.2_TopoBettiNumbers}

Mathematically, it can be defined as a number $X$, such that: $X = V - E + F$ where $V, E \& F$ represents the number of Vertices, Edges $\&$ Faces of the polyhedron respectively. The Euler Characteristic can also be represented as: $X =2(1-g)$ where  $g= X/2-1$ is known as the genus  of the polyhedron \cite{17.2_TGTonlineBritanica} which can be understood intuitively as the “number of holes” in the polyhedron or geometric figure. The genus theorem states for a polyhedron embedded on an orientable closed surface − or the corresponding combinatorial orientated map −, the Euler characteristic  $X = V − E + F$ is always an even integer, possibly negative, and that, c being the number of connected components of the polyhedron, its genus $g = c − \chi/2$ is a natural number. In fact, g corresponds to the number of holes in the surface of the polyhedron. When g = 0, the surface is without holes, and the polyhedron is said to be planar. When it is also connected $(c = 0)$, it satisfies the Euler formula which is a special case of the Euler Characteristic, given by:  $V−E+F= 2$; for instance, in the case of a cube that has 8 vertices, 12 edges and 6 faces: $V−E + F=8-12+6=2$.\cite{03_1944Homology} \cite{01.18_EulerFormula}

However, our special concern for understanding the genus of a polyhedron is its relation to Betti Numbers and  Euler Characteristics in Topology. Betti numbers are topological objects which were proved to be invariants by Poincaré, and used by him to extend the polyhedral formula to higher dimensional spaces. Informally, the Betti number is the maximum number of cuts that can be made without dividing a surface into two separate pieces. Formally, the $n^{th}$ Betti number is the rank of the $n^{th}$ homology group of a topological space. Betti numbers can be related to Euler Characteristics of Polyhedrons and their genus via the following mathematical relation.\cite{03.7_2018homologicalAlgebra&Data} \cite{07.3_BettiNumber} \cite{07.2_TopoBettiNumbers}

\begin{equation}
  X=\sum_{n \ge 0}(−1)^{n}\beta_{n}(\sum)=2(c - g)
\end{equation}

where the $n^{th}$ dimensional Betti number $\beta_n$ is the dimension of the $n^{th}$ homology group $H_n(\sum)$ of the $SC \sum$. These are important metrics that would characterize the topology of the data and would be significant in various steps of our computational pipeline in the following sections.\cite{01.16_EulerianGraphs} \cite{01.18_EulerFormula} \cite{17.2_TGTonlineBritanica} \cite{03.2_2008FindingHomology} \cite{08.1_2003SimplicialHomology}

\subsection{Simplicial Complexes}
The concept of Simplicial Complexes \cite{03.1_2009simplicialHomotopy} \cite{08_1971simplicialComplex} are quite essential to Homology that applies to all topological spaces (singular homology) relies on the linear algebra of infinitely generated modules over the ring $Z$ in defining homology groups, and for this reason it is not useful from a computational point of view. Computations can be carried out by hand using a variety of techniques (long exact sequence of a pair, long exact Mayer-Vietoris sequence, excision theorem, spectral sequences), but direct computation from the definition is not feasible for general spaces.\cite{03.2_2008FindingHomology}
However, when one is given a space equipped with particular structures, there are often finite linear algebra problems which produce correct answers, i.e. answers which agree with the singular technique. A particularly nice example of this applies when the space in question is described as a “simplicial complex”.\cite{08.2_1992simplicialAlgebricTopo} \cite{08.3_2018simplicialComplexes}

\subsection{Persistent Homology}
Persistent homology \cite{03.3_de2007PersistentHomology} \cite{03_1944Homology} \cite{03.3_de2007PersistentHomology} can be explained as a method for computing stable topological features of a dataset through evaluating transformations over varying parameters, such as spatial resolutions against point cloud data. It gives a multi-scale view of the topology of a space by capturing the evolution of topology with increasing (or decreasing) resolutions. It  is a powerful tool to compute, study and encode efficiently multi-scale topological features of nested families of simplicial complexes and topological spaces. It not only provides efficient algorithms to compute the Betti numbers of each complex in the considered families, as required for homology inference in the previous section, but also encodes the evolution of the homology groups of the nested complexes across the scales.\cite{03.7_2018homologicalAlgebra&Data}
The input data for the computation of persistent homology is often represented as a point cloud. The output is a set of real number pairs (the birth and death times) documenting the spatial resolutions where each topological feature first appears (birth) and when it disappears (death). The pairs are usually plotted either as a set of lines, called bar-codes, as a set of points in a 2D plane, called a persistence diagram, or as a persistent landscape. Most implementations of the persistent homology are done over point cloud data via the Mapper algorithm \cite{01.9_2007MapperPBG}, however the core concept of persistent homology extends past it's traditional implementation on point cloud data in ways such as data related to knowledge graphs.\cite{01.1_1stCourse2018algebraicTopo}\cite{03.2_2008FindingHomology}

\textbf{\textit{Example:}}  We shall now consider a typical example of how Persistent  Homology has been implemented thus far in the domain of Topological Data Analytics. So, in order to demonstrate the concept of  persistent homology, let us imagine we have collected a bunch of data points that we refer to as a data cloud. For the next step, let us imagine that there is a control parameter called the proximity parameter $\epsilon$, which defines the radius of an imaginary ball centered at each of these data points.\cite{01_GCarlssonEpstein2011} \cite{02.1_GCarlson2004topoEstimation}\cite{03.6_2015persistentHomo}

Now, when we gradually increase $\epsilon$, the balls will grow outwards and eventually touch other balls. The overlapping of these balls form a unique topological characteristic that is unique to this dataset, and hence we can use this unique topological characteristic to differentiate nuances in the topologies of different point clouds. This filtration process can be demonstrated and visualized in the figure below:\cite{06.2_carlsson2009Multipersistence}

\begin{figure}[H]
	\centering
        \includegraphics[width=1.0\linewidth]{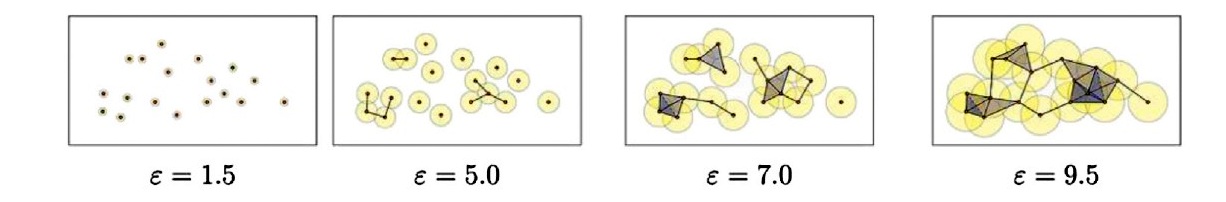}
	\caption{\textit{The figure above shows a schematic diagram representing a data cloud, and how the filtration process results in outcomes of various overlapping of balls from different proximity parameters $\epsilon$ (shown in the upper column). The bottom column is barcodes scanning through a full-range of proximity parameter $\epsilon$ values. $\beta_0$ and  $\beta_1$ denote the 0-dimensional and 1-dimensional Betti numbers, which can be deduced from the subfigures to be roughly $18\rightarrow11\rightarrow4\rightarrow1$, and $0\rightarrow0\rightarrow1\rightarrow2$, respectively.}}
	\label{fig:example_pipeline_fig}
\end{figure}

Thus standard practices in TDA use a similar  encoding and filtration process, as shown above to convert a point cloud that is made from brain functional signals, or a correlation matrix from examples such as time series data, to filtration diagrams. From these filtration diagrams, we can calculate barcodes, persistence diagrams, and other TDA metrics for further applications of Persistent Homology.\cite{03.4_2008localizedHomology}\cite{03.3_de2007PersistentHomology}

Thus, Persistent homology can be explained  as a method for computing topological features of a data set (point cloud) at different spatial resolutions. It gives a multi-scale view of the topology of a space by capturing the evolution of topology with increasing (or decreasing) resolutions. It  is a powerful tool to compute, study and encode efficiently multi-scale topological features of nested families of simplicial complexes and topological spaces. It not only provides efficient algorithms to compute the Betti numbers of each complex in the considered families, as required for homology inference in the previous section, but also encodes the evolution of the homology groups of the nested complexes across the scales.\cite{03.2_2008FindingHomology}\cite{03.3_de2007PersistentHomology}\cite{09.2_chazal2008CompGeometry} \cite{11.0_chazal2017TopoDataScience}

\subsection{Topological Projection \& Isometric Compression of Graphs}

The mathematics of projection\cite{12.1_2002topologicalInvariaceProjection} is defined in terms of a mapping of a set or other mathematical structures such as operations, metrics or topology into a subset (or sub-structure), which is equal to its square for mapping Function  Composition  which can be mathematically defined as a function $h(x)$ such that $h(x) = g(f(x))$ That is, the functions $f \: X \rightarrow Y$ and $g \: Y \rightarrow Z$ are composed to yield a function that maps x in X to $g(f(x))$ in Z.\cite{12.2_ProjectionMath}  Projections can be intuitively explained as casting a shadow of  geometric points or objects  on a plane or a sheet of paper. For example the shadow projection of a point on a plane or a sheet of paper is the point itself; whereas, the shadow projection of a 3-D sphere is a closed circle which is equal to the square of its mapping function and can be used to obtain the original topology of the 3-D sphere without the loss of essential information.\cite{12.0_alatorre2018TDAinvariant}

The mathematical appeal and utility of projection is that it's fundamentally elementary and geometric with a finite number of degrees of freedom; once the acceptance domain and the dimensions of the spaces used in the construction are chosen which allows for easier manipulation of higher dimensional data and applying essential filters and transformation in the domain of TDA. This also helps us to obtain all essential information from high-dimensional data sets while carrying out efficient noise reduction. It also preserves and amplifies the essential nature of such datasets for analysis and obtaining useful insights from its underlying attributes and relationships.\cite{01.0_2010introductionTopoPropertiesInvariance}\cite{02.3_2017introductionTDA} \cite{02.4_TDAResearch}

In the field of Topology, we mostly work with the special class of point sets obtained by cut and projection from the integer lattice $Z^{N}$ which is generated by an orthonormal base of $R^{N}$. In short we can call a projection method pattern $T$ on $E = R^{d}$ a pattern of points (or a finite orientation of it) given by the orthogonal projection onto E of points in a strip $(K × E) \cap Z^{N} \subset R^{N}$, where E is a subspace of $R^{N}$ and $K × E$ is the so-called acceptance strip, a flattening of $E$ in $R^{N}$ defined by some suitably chosen region K in the orthogonal complement $E \perp$ of $K$ in $R^{N}$. The pattern $T$ thus depends on the dimension N, while also depending on the positioning of E in $R^{N}$ and the shape of the acceptance domain K. When this construction was first made, the domain $K$ was taken to be the projected image onto $E \perp$ of the unit cube in $R^{N}$ and this choice gives rise to the so-called canonical projection method patterns in Topology and TDA that are useful concepts which are essential in establishing a feasible and seamless way to carry out near-lossless topological projections and isometric compression of data-structures represented by DAGs and multi-graphs in this novel approach.\cite{12.0_alatorre2018TDAinvariant} \cite{12.1_2002topologicalInvariaceProjection}

The application of isometric compression of graphs \cite{03.5_kramar2013persistenceComputing} \cite{12.2_compressingTopoNetworkGraphs} is further useful as the mathematics of Topology allows for unique possibilities that provide methods for creating compressed representations of data sets that retain all features of the original data set via efficient techniques such as Isometric Compression. These represent the various relationships among points in the data set. The representation is in the form of a topological network or combinatorial graph, which is a very simple and intuitive object to work with using graph layout algorithms and allows for the comparative analysis of all essential attributes of the original high-dimensional dataset without loss of essential information. This provides a unique and necessary advantage over standard graphical visualizations and data plots which only address and compare 2 or more limited subset of variables, dimensions or columns of the high-dimensional dataset.\cite{21.0_2016TopoCompression}

 \section{Proposing a Novel Approach}

In this section we propose a Novel Approach in Topological Graph theory with Roy-Kesselman Diagrams (R-K Diagrams) as a useful extension to standardised methods of Topological Analysis as described in the previous sections and in consistency with the mathematical formulations and theorems of Topology and Graph Theory which serve as the foundation of this novel approach.

\subsection{Motivation}

As mentioned earlier, the motivation for this approach takes root in the appeal of \textit{“Event-Driven Topology”} which is based on the topological clustering of all attributes around a central ``Event-Node'' that serves as a static (unchanging) context for all dependent attribute clusters. The choice of event nodes are determined by a choice of ``Lens'', for example, in the case of the \hyperref[sec:store_sales_section]{Tableau Super Store Data}, a lens could be classified as a purchase event related to an order. Alternatively, a lens could also be changed to represent an entity such as a customer with all associated attribute clusters representing customer data. Such attribute clusters can dynamically evolve in time due to perturbations caused by the influx of data in terms of new/added rows or additional independent attributes in terms of columns driven by Hierarchical-Feature-Extractions with respect to a particular domain Ontology.

\subsection{Objectives}

The objectives of our Novel Approach can be summarised as follows:

  \textbf{(1)}To build a computational framework based on the mathematical foundations of Topological Graph Theory that allows for the flexibility to switch between graph theory analysis and TDA on all datasets pertaining to physical systems as defined in  \hyperref[sec:PhysicalSystems]{section 4.2.1}  and a subset of non-physical systems with inherent ontology and domain specific knowledge graphs as defined in  \hyperref[sec:NonPhysical]{section 4.2.2}.
  
  \textbf{(2)} To provide a novel framework that allows \textit{"Event-Driven" Topological Signatures} based on the choice of  lenses  as defined in \hyperref[sec:sectionlens]{section 4.5.5} along with domain specific node masks and filter functions as defined in  \hyperref[sec:Filters]{section 4.5.2}.
  
  \textbf{(3)} To extract hierarchical features from a given dataset as discussed in \hyperref[sec:HEF]{section 4.4}, while linking them to a central "Event-Node" and to generate a hierarchy that is centred around an event based on the choice of a lens that allow for effective generation of unique \textit{"Event-Driven" Topological Signatures}.
  
  \textbf{(4)} Creating a framework that would allow for the analysis of differences in micro-geometric properties (such as nodes, edges and DAG's) via graph analysis while allowing for the study in differences of macro-geometric property (such as holes, voids \& loops) with the help of \textit{Euler characteristic} and \textit{Betti numbers} as discussed in \hyperref[sec:BettiNumber]{section 4.5.7}.
  
  \textbf{(5)} By generating unique signatures across a set, the RK-Diagrams could be utilized to form a basis for traditional Machine Learning models such as identification, clustering, classification, and segmentation as discussed in sections \hyperref[sec:TopoIdentifictaion]{identification} and \hyperref[sec:Classify]{classifications}. The possibilities of applications would only be constrained by the input layer being an R-K diagram (or graph).

 \section{Computational Pipeline}

\subsection{R-K Toolkit}

 To build the RK-Diagrams and RK-Models, we implemented a generalizable package called RK-Toolkit \cite{rktoolkit}. The R-K toolkit is a framework to build a component called an RK-Pipeline (described in section 4.4), and can be found open sourced, at \url{https://github.com/andorsk/rk_toolkit}. The RK-Toolkit was built in python, and converts the core mathematical concepts described in this paper into computational representations that can be used on real data. Similar to how other libraries such as Sci-Kit \cite{a2021_scikitlearn} provide base components like the Pipeline object \cite{a2021_61}, to be extended later by a user or by native extensions, RK-Toolkit operates similarly by providing the base building blocks such as a R-K Pipeline and R-K Diagram, to be extended by the user or by native representations. As an example, if one were to want to classify a Hodgkins vs. Non-Hodgkins disease along with their 4 stages of metastasis as distinct topological signatures that are divergent from healthy patients, one could construct their own R-K Pipeline out of the toolkit, to potentially generate unique topological structures and R-K Diagrams. The R-K Toolkit's design was done explicitly with the intent to be extended later, with the expectations that new and more robust methods will be included into framework over time.Any computation carried out using the R-K Toolkit leading to a set of actions to build and generate R-K diagrams can be defined as the \textbf{\textit{R-K Workflow}}.

The R-K Toolkit provides many advantages over starting from scratch. We submit that there are 9 main advantages to using the R-K Toolkit for building future models.

\begin{enumerate}
    \item{The R-K Toolkit provides in build validation for components, such as adherence to a DAG and valid transforms}
    \item{A pipeline framework is provided to chain steps together with accordance to the computational steps to be described below}
    \item{Visualization's are provided to assist with rendering of RK-Models and converting them into RK-Diagrams.}
    \item{Pre-built transforms provide baseline transforms for testing }
    \item{Various Optimizations and Machine Learning implementations will be present in the RK-Toolkit for various use cases}
    \item{Built in mechanisms to provide inverse operations}
    \item{Our toolkit allows for plug and play of well defined domain ontologies pertaining to knowledge graph or custom ontologies pertaining to a structured dataset}
    \item{This framework can be extended to include many datasets as long as it conforms to the data requirements}
    \item{The R-K Pipeline is built in such a way that it could in theory support streams for real time classification and identification use cases, however future work would be required to update this feature in further details.}
\end{enumerate}

To apply the toolkit to a dataset, one would import the relevant components into their code (similar to how you would Sci-Kit), define a RK-Pipeline, and then run the pipeline. It is entirely in the control of the user \textit{how} they desire to build their pipeline as most pipelines will vary across domains and data. The intent with this paper, and with the core concepts of the R-K Pipelines, are that despite the divergence of implementations, the foundational building blocks of an R-K Pipeline will always remain the same.

We acknowledge that this is the first implementation of the R-K toolkit, but it has shown good promise with useful results on the below use cases, but future work is required to make it more robust for it's application across different domains.

\subsubsection{R-K WorkBench}

To help users get started quicker, we provide a docker image called RK-Workbench \cite{andorsk_2021_andorskrkworkbench}, which wraps the ML-Workspace \cite{mltooling_2021_mltoolingmlworkspace} with packages relevant to building RK-Diagrams built into the core image. The README file details how to use it for the purpose of independent use by researchers and programmers. 

\subsection{Data Qualification Criterion}

Let us now address the data qualification criteria and its associated constraints for the R-K pipeline that would allow for the accurate execution of our analytical framework pertaining to "Event-Driven" Topological Graph Analysis. In this approach, we shall not be addressing the study of unstructured point-cloud data as addressed in the traditional TDA frameworks of the past \cite{01.9_2007MapperPBG} \cite{02_carlsson2009topology} \cite{02.3_2017introductionTDA} but focus on a subset of data with well-defined hierarchical features that can be extracted automatically as a part of this pipeline and linked to the corresponding interdependencies of columns and rows in relation to the chosen context for the purpose of segregation (ref:4.5.5) and classification of meaningful events and entities within such datasets.

We start by defining $D_i (\forall i \in N)$ as the input dataset to the R-K Toolkit; where $M_j (\forall j \in N)$ is a measure within the dataset. For example, Mass and Spin are measures on the LIGO dataset.\cite{00_LIGOOpenSciData}Then for any such dataset $D_i$ containing $M_j$ measures to be accepted into the R-K Model Framework, it must adhere to the following constraints:

\begin{itemize}
	\item It must be able to derive a hierarchy of relationships between measures such that the Dendogram of hierarchy $H$ describes how one measure relates to another.
	\item It must be able to segregate independent variables and their corresponding dependencies in separate attribute clusters of the "Event-Node".
	\item It must have at least 3 mutually-independent variables pertaining to the well-defined attributes of "Physical Systems" \&  "Non-Physical Systems" as detailed below.
\end{itemize}

\subsubsection{Physical Systems}
\label{sec:PhysicalSystems}

The Datasets pertaining to “Physical Systems” must contain at least \textit{3 mutually independent variables} related to the 7 Fundamental Quantities of Physics \cite{23.0_DimensionalAnalysis} to meet the the minimum criteria for generating an  R-K Diagram in accordance with consistent Topological Properties that can be distinguished using appropriate filtering \& divergence criteria.

In physics, a physical system is a portion of the physical universe chosen for analysis. Everything outside the system is known as the environment. The environment is ignored except for its effects on the system. Any system defined in the realm of science and engineering can be referred to as a physical system with independent variables that have the following fundamental dimensions: \textit{mass, length, time, temperature, electricity, luminous intensity \& count/amount of substance/matter}. \cite{23.1_7FundamentalQuants} An example of hierarchical clustering using the well defined relationships of dependent and independent variables pertaining to physical systems is shown below with respect to a fundamental structural graph in the R-K model framework, which can be filtered specifically to obtain unique event-driven R-K Diagrams. 

\begin{figure}[H]
	\centering
	\includegraphics[width=0.5\textwidth]{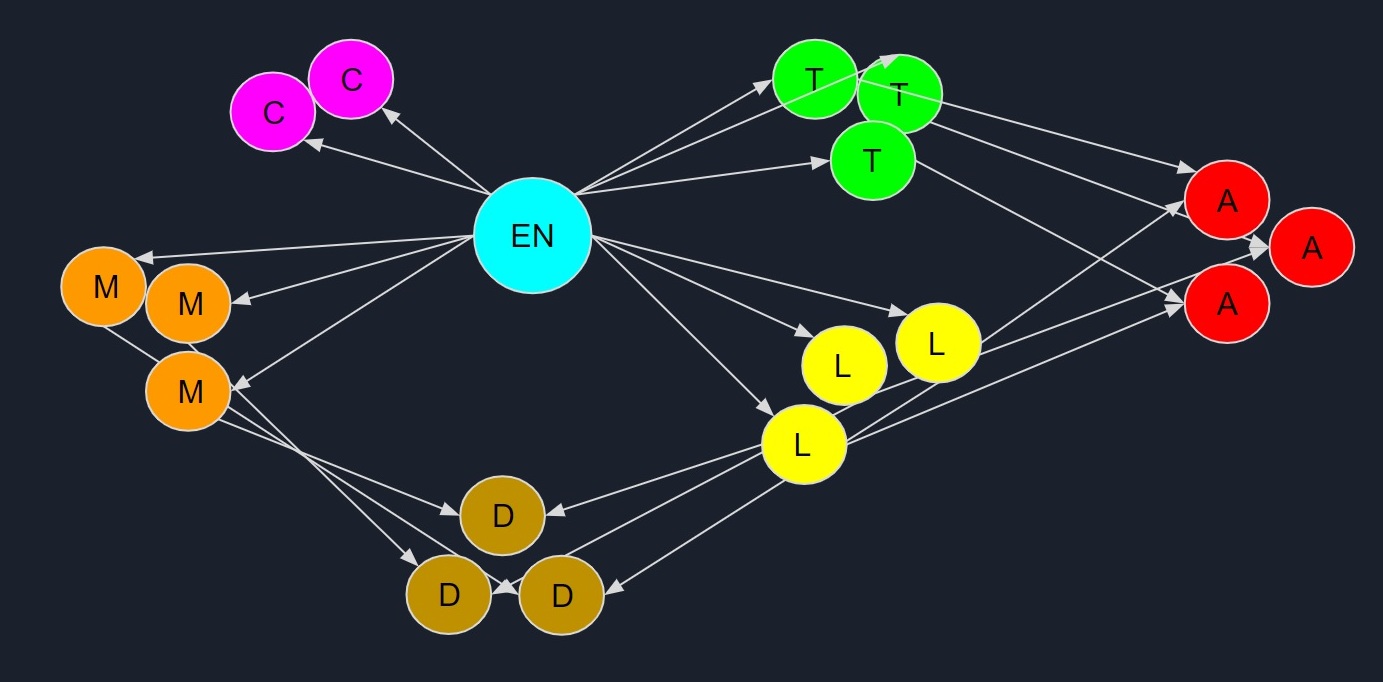}
	\caption{\textit{The above structural graph represents and unfiltered R-K model, where EN stands for the central Event Node based on the choice of an arbitrary  physical phenomenon or entity. T, L, M \& C Represent the Mutually Independent Clusters  of Time, Length, Mass and Charge. A \& D represent the derived/dependent clusters of Acceleration and Density respectively.}}
	\label{fig:PhysicalSystems}
\end{figure}

\subsubsection{Non Physical Systems with Knowledge-Graphs}
\label{sec:NonPhysical}

The Datasets pertaining to “Non-Physical Systems” must contain a well-define Knowledge Graph related to a specific domain Ontology that contains parent nodes with 3 or more mutually independent attributes to meet the the minimum criteria of  generating an  R-K Diagram in accordance with consistent Topological Properties that can be distinguished using distinct Divergence criteria.

Such independent attributes should allow for the classification and separation of variables based on direct or derived relationships with fundamental physical dimensions and associated independent variables such as:

E.g. Age of a person can be separated as a time variable while the weight of the person is a derived from the independent mass variable and the height of that person can be separated as a length variable. All other attributes have a secondary/ derived dependence on the fundamental independent variables. However, the R-K Pipeline would allow for the addition/classification of any  additional domain specific independent variables based on a specific domain ontology relevant in a non-scientific commercial use-case such as in the case of the Tableau Super Store Dataset.\cite{TableauSuperStore}

\subsubsection{Derived Relationships}
Derived relationships are relationships that are determined by the data itself, and are learned during analysis. For example, it may be possible to derive an ontology from physical system consisting existing fudemental measures such as mass, time, and length in order to obtain derived measures such as force, momentum, and acceleration.

\subsubsection{Data Assumptions}

A directed relationship exists between metrics $M$ such the formula $Edge_{M_{i} \to M_{j}}$ would describe a directed relationship between $M_{1}$ and $M_{2}$. For the approach to be effective, there exists some combinations of filter and linkage functions, such that optimized they can segregate topological structures.

Furthermore, it is assumed that there exists a method to compare two variables (i.e over distances). Practically, this means all data needs to have some numeric based encoding to be compared.

\subsubsection{Limitations}
Such as described by the Section 4.2 in Data Criteria, this imposes the following limitations on data fed into the pipeline:

Low dimensional data, with less than 3 independent dimensions will not be compliant with the data criteria and are not suitable for this approach. Three mutually independent dimensional attributes are required to make simplicial complexes and add structure ot the data and so lacking rank would impact the structure. Additionally, data with unclear relationships between measures may be less effective than data with clear relationships between measures because a lot of the power of the R-K Models come from a hierarchical embedding, which if unclear, limits the ability for the models to perform as they should.

As an example, in the Iris dataset \cite{iris_dataset}, all measures are built off the fundamental quantity: length. In spite have 4 columns (sepal width, sepal length, petal width, petal length), all columns are inter-dependent and are not mutually independent and therefore cannot be clustered into separate independent attributes of the central iris event node, other than length, thereby resulting in non-unique Topological signatures and the failure to generate  R-K Diagrams.

\subsubsection{Ontology}

In the figure below, we describe a dendrogram with a specific ontology. An ontology in the context of an R-K Model, describes the relationship between two or more measures in the form of a hierarchy. Ontological frameworks have numerous advantages, such that it makes domain assumptions on data explicit and/or to share a common understanding of structural information across metrics. \cite{ben_mahria_chaker_zahi_2021}. The ontologies can either be learned such as with the Mahria et. al paper titled \textit{A novel approach for learning ontology from relational database: from the construction to the evaluation} or formulaized by domain experts \cite{ben_mahria_chaker_zahi_2021}. There are numerous ontological databases available for describing relationships between known domains. The Web Ontology Language (OWL) is a specific set of web standards and language devised to standardize ontologies through \textit{OWL Documents}. \cite{owl_semantic_web_standards_2012}.

In the figure below, we define a strict ontology in the store sales data. This provides the basis on which all R-K Models are formed. Any R-K Model most have some strict ontological backbone, such as the below, and compliant with data assumptions in 4.2.4 to be valid.

\begin{figure}[H]
	\centering
        \includegraphics[width=0.5\textwidth]{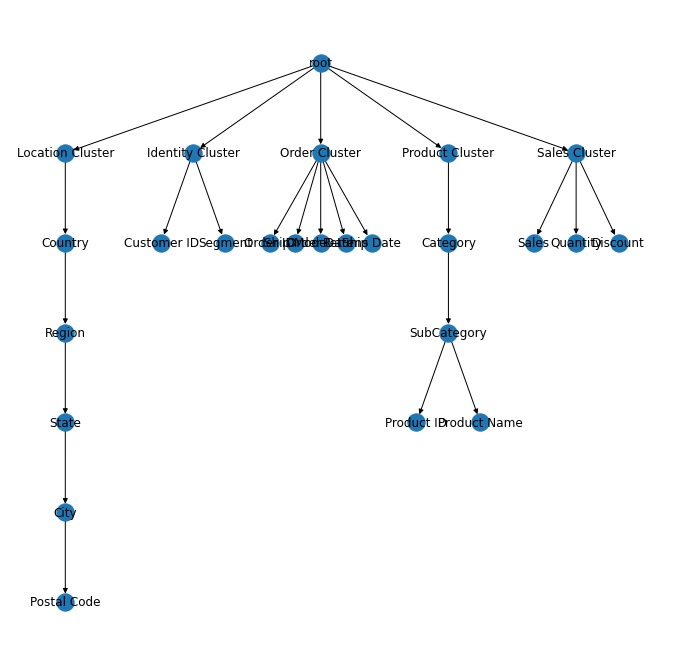}
	\caption{\textit{}}
	\label{fig:nodemask}
\end{figure}

\subsection{R-K Pipeline}

The foundational component of any R-K Diagram is an R-K Pipeline. At a base level, the R-K Pipeline can be understood as a Directed Acyclic Graph (DAG) as defined in section 2.3.1, which provides transformational components that result in an composite model we call an R-K Model. Transforms in an R-K Pipeline can be chained against each other, as long as egress from one component complies with the ingress specifications from another component. We can mathematically represent this with the following representation: $\lbrace C_{0}, C_{1}, C_{2}..., C_{n} \rbrace \in RK_{i} $ where $C_{i}$ represents a pipeline component and the egress of $C_{i}$ is compliant with a set of constraints imposed by $C_{i+1}$'s ingress.
\begin{figure}[H]
	\centering
        \includegraphics[width=0.5\textwidth]{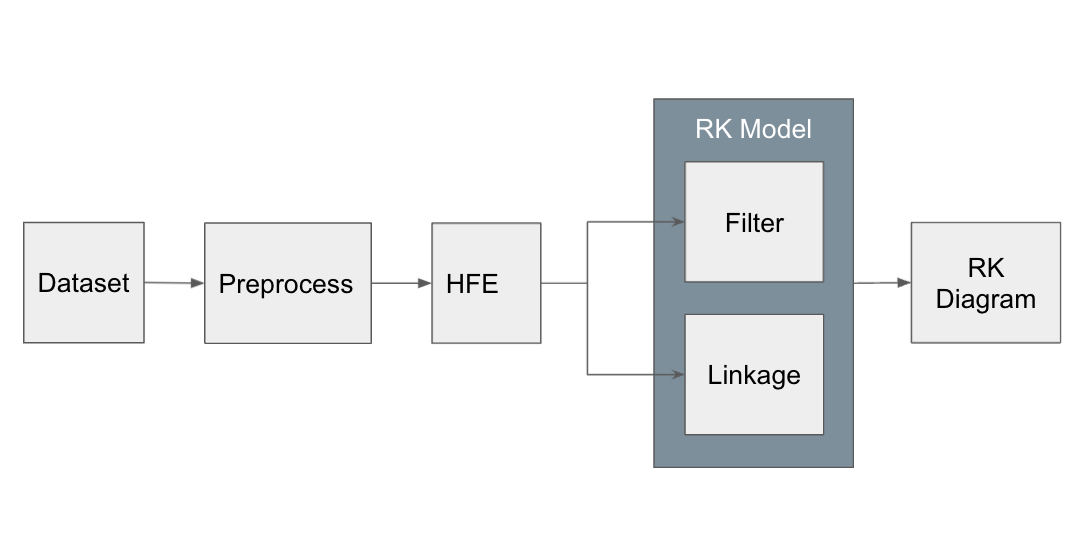}
	\caption{\textit{The above represents a simplest R-K Pipeline. In this pipeline, data first gets preprocessed, then moved to a Hierarchical embedding function. From the hierarchy, the standard filters and links provide an R-K model. It is then forked between a compression node and a visualization node. Thus the end result is two R-K Diagrams, one compressed version and one raw version}}
	\label{fig:example_pipeline_fig}
\end{figure}

\subsection{Hierarchical Embedding Function}
\label{sec:HEF}

As discussed in Section 4.2.6, there must exist an explicit ontology related to the measures of the dataset.  Let us describe $D$ as the input dataset to the RL Toolkit. Let us describe $m_{i}$ as a measure within the set of measures $M$ such that $m_{i} \in M$. To be compliant with this proposed framework, there must exist an ontology $O$ that describes the relationship of the measures w.r.t each-other. For example, w.r.t location data, an ontology could present that a \textbf{$m_{city}$} is a child of  \textbf{$m_{state}$} which is a child of \textbf{$m_{country}$}. In the datasets that were analysed, we often provided an ontology which merged similar measures into the same ``cluster'', for example $mass_{1}$ and $mass_{2}$ were ontologically defined as part of the $Mass$ cluster in our LIGO analysis. In the Tableau Super Store Sales Dataset, we defined a custom ontology that we used for our representations. Using the R-K toolkit, one can either use pre-defined ontologies or provide custom ontologies for their data.

An Hierarchical Embedding Function ($f_{H}$) generates a graph $G$ from a set of measures $M$ such that the resulting graph $G$, is a DAG which represents a bijective mapping between $m_{i} \longleftrightarrow G_{i}$ such that all $V_{i} \in G$ can be traced back to the original value in the dataset, where $V$ is a vertex. More succinctly, $f_{H}(M) \rightarrow G$ such that $f_{H}(G_{v_{i}}) \rightarrow m_{i}$ and $f^{-1}_{H}(G_{v_{1}}) \rightarrow m_{i}$. The advantage of this requirement is that any structural graph can be inverted back into original measures used to generate the graph.

Across the pipeline, the graph generated by $H_{f}$ is referenced as a \textit{Structural Graph} ($S$). The structural graph provides the baseline ontological structure that forms the basis for all other transformations in the pipeline.

Computationally, to maintain the bijective mapping, we maintain internally a history of steps applied during the transformation such that we can easily provide the inverse function by returning the steps in reverse.

\subsection{R-K Model}

The R-K Model is the basis for any R-K Diagram. It represents the composite object that can be used to render an R-K Diagram. In that sense, an R-K Diagram the rendering of an R-K Model, and the R-K Model underlying datastructure for that render.

All R-K Models contain the following 3 components: Structural Graph, Node Masks, and Derived Links
\begin{figure}[h]
	\centering
        \includegraphics[width=0.5\textwidth]{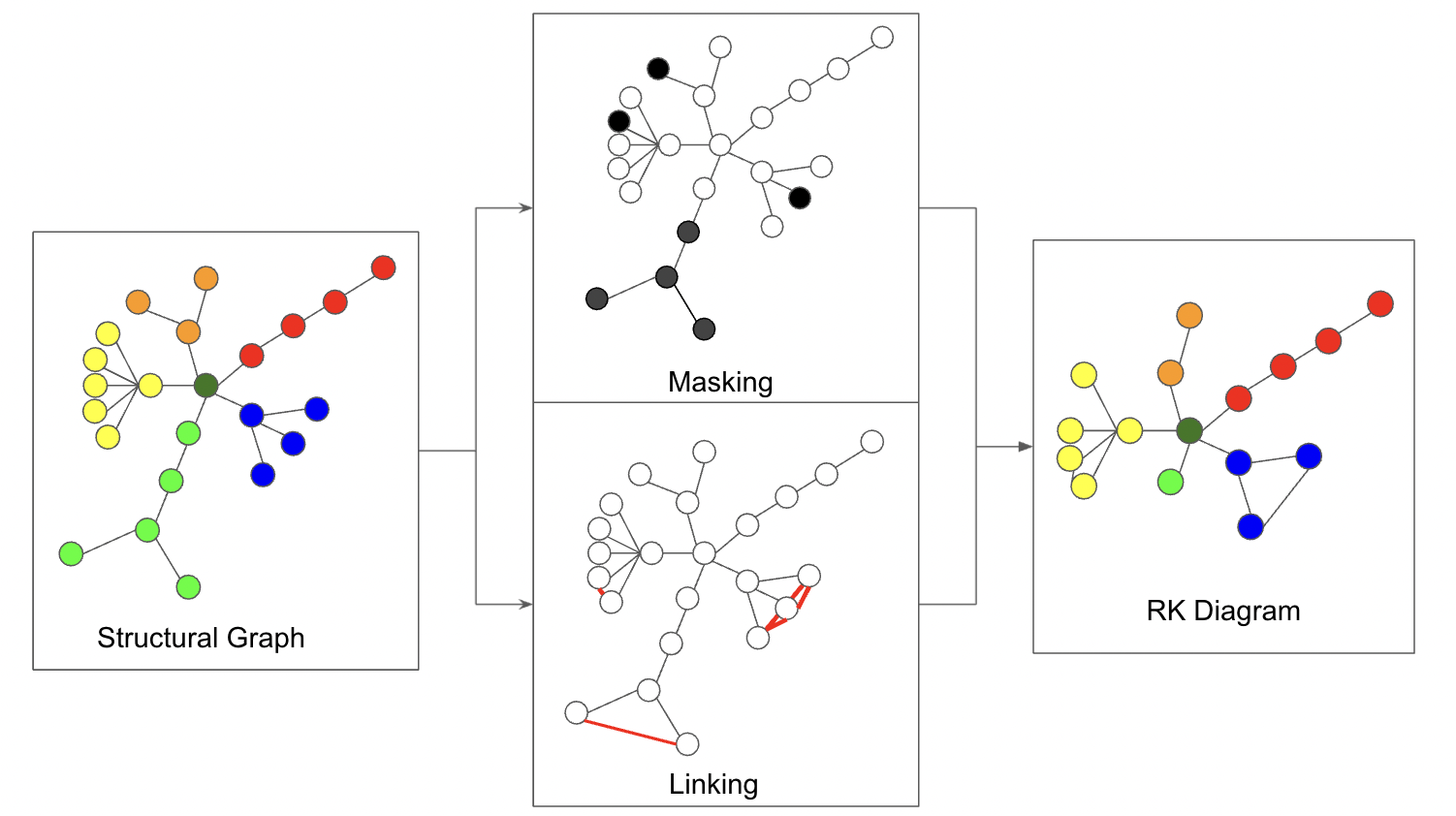}
	\caption{\textit{The above image represents the building blocks of an R-K Model. To the left, a structural graph, which is defined in 4.7.1. In the middle, filters and linkage functions are applied resulting in the final R-K Model. The R-K Diagram is simply a rendering of the model}}
	\label{fig:fig1}
\end{figure}

Each component is described in detail below.

\subsubsection{Structural Graph}

The structural graph ($S$) is the base graph derived through the Hierarchical Embedding Function, also known as the structural graph. The structural graph provides the baseline ontological structure that forms the basis for all other transformations in the pipeline.

Because node masks are reductive operations, the number of nodes in the structural graph represents the \textit{maximal number of nodes in the R-K Diagram} such that $|nodes| \in S >= |nodes| \in R-KDiagram)$. The structural graph however does not represent that maximal number of edges. The number of possible edges in the R-K Diagram is bounded by the number of combinations of nodes in the structural graph.

The R-K Filters (described below), can bound the possible combinations of edges to less than the number in the original combinatorial graph.

In the case below, we demonstrate two forms of structural graphs. The first is the initial structural graphs that were derived off the Store Sales data outlined in the next section. The second structural graph is built off of similar data, however some nodes were expanded out. Both are viable candidates for structural graphs.

\begin{figure}[H]
	\centering
        \includegraphics[width=0.5\textwidth]{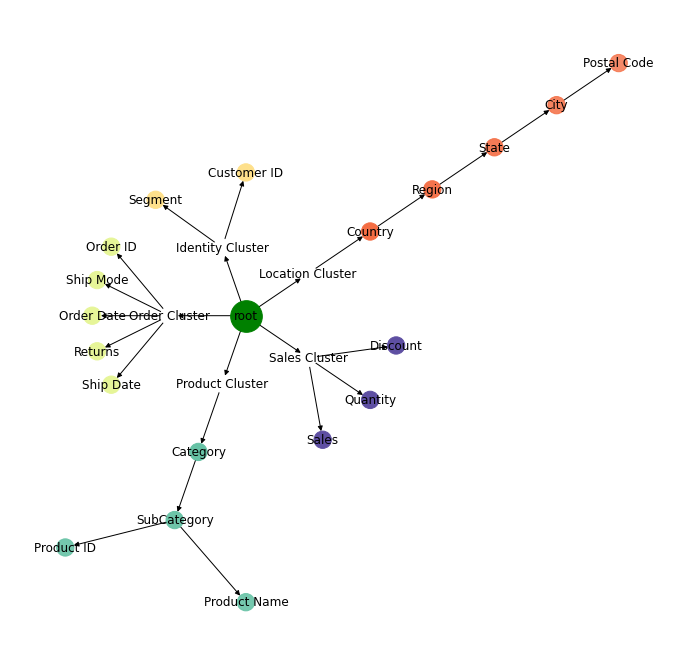}
	\caption{\textit{The base structural graph of the Store Sales Data}}
	\label{fig:fig1}
\end{figure}

\begin{figure}[H]
	\centering
        \includegraphics[width=0.5\textwidth]{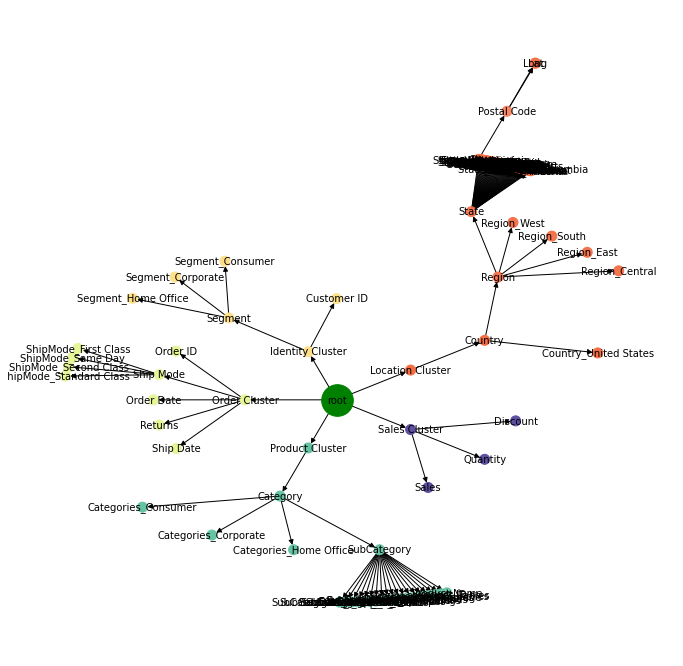}
	\caption{\textit{An expanded structural graph}}
	\label{fig:fig1}
\end{figure}

\subsubsection{Node Masks and Functions}
\label{sec:Filters}
A node mask represents a masking structure that when applied to a structural graph $S$, reduces the number of nodes into a subgraph $S^{'}$. The node masks are binary operators, which when set to \textit{true}, filter a node and it's direct children.

To derive the node masks, we produce a set of filters $F_{n}(G)$, which takes in a graph and returns a mask. We define Filter Functions  w.r.t. Simplicial Complexes formed by a combination of nodes and edges, such that filtration of a simplicial complex K is a nested family of subcomplexes $(Kr)r \in T$, where $T \subseteq R$, such that for any,r; $r \prime \in T$  if $r \le r \prime$ then $Kr \subseteq Kr\prime$,  and  $K=Ur \in TKr$. The subset T may be either finite or infinite. More generally, a filtration of a topological space M is a nested family of subspaces $(Mr)r \in T$,  where $ T  \subseteq R $, such that for anyr; $r\prime \in T$, if $r \le r\prime$ then $Mr \subseteq Mr\prime$ and, $M=Ur_{\in}TMr$. For example, if $f : M \rightarrow R$ is a function, then the family $Mr = f-1 ((-\infty , r \rbrack), r \in R$ 

In the pipeline, one can combine multiple filter functions together. The union of the filters provide the final node mask, such that $ F_{n}(G) \cup N_{m}$, where $N_{m}$ is the node mask.

\begin{figure}[H]
	\centering
        \includegraphics[width=0.5\textwidth]{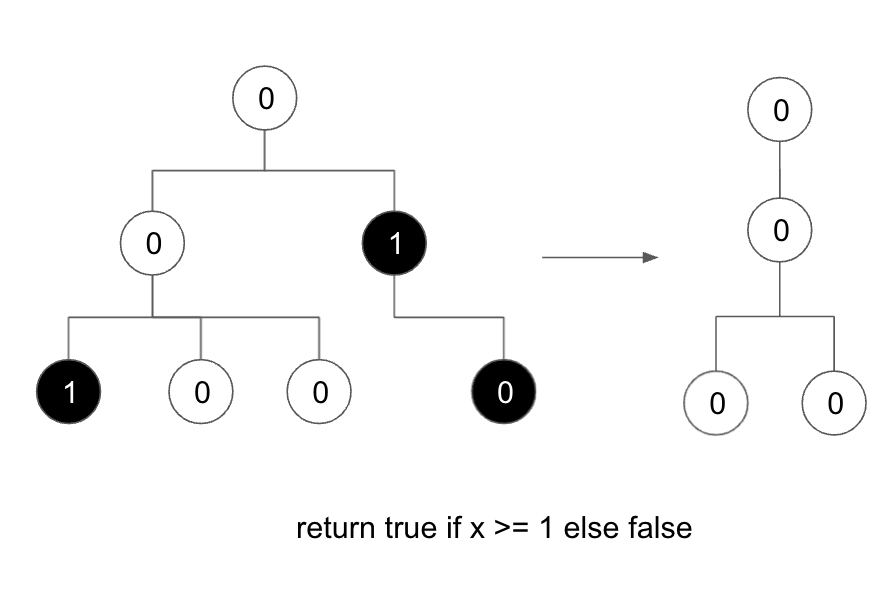}
	\caption{\textit{An example node masks with a filter of >= 1. All nodes and children of the nodes are removed that have a value >= 1 and applied at the root level. Different filters can be applied to different subgraphs.}}
	\label{fig:nodemask}
\end{figure}

\subsubsection{Linkers}

In order to define linkages we must begin with the following mathematical consideration. Let $G = (V, E)$ be an undirected graph without multiple edges or loops. Let $n =|V|$ and $e= |E|$. The linkage of G is defined to be the maximum min-degree of any of the subgraphs of G (the min-degree of a subgraph is the least degree of any of its vertices; the degree of a vertex is taken relative to the subgraph). The width of G is defined to be the minimum, over all linear orderings of the vertices of G, of the maximum, with respect to any vertex v, of the number of vertices connected with v and preceding it in the linear ordering. It has also been mathematically proven in Topology that the width of a graph is equal to its linkage.

Furthermore, according to the French Mathematician Erdös, it has been proven that every graph G has a subgraph with min-degree at least equal to the density $[e/n]$ of G. Therefore, the density is a lower bound for the linkage (equivalently, width) of graphs with given numbers of edges and vertices. Given a positive integer j, if in the definition of width we consider not the number of vertices preceding and connected with v but rather the least number of vertices preceding and connected with any cluster of at most j consecutive vertices extending to the right up to v, we get a graph parameter known as j-width.

\begin{figure}[H]
	\centering
        \includegraphics[width=0.5\textwidth]{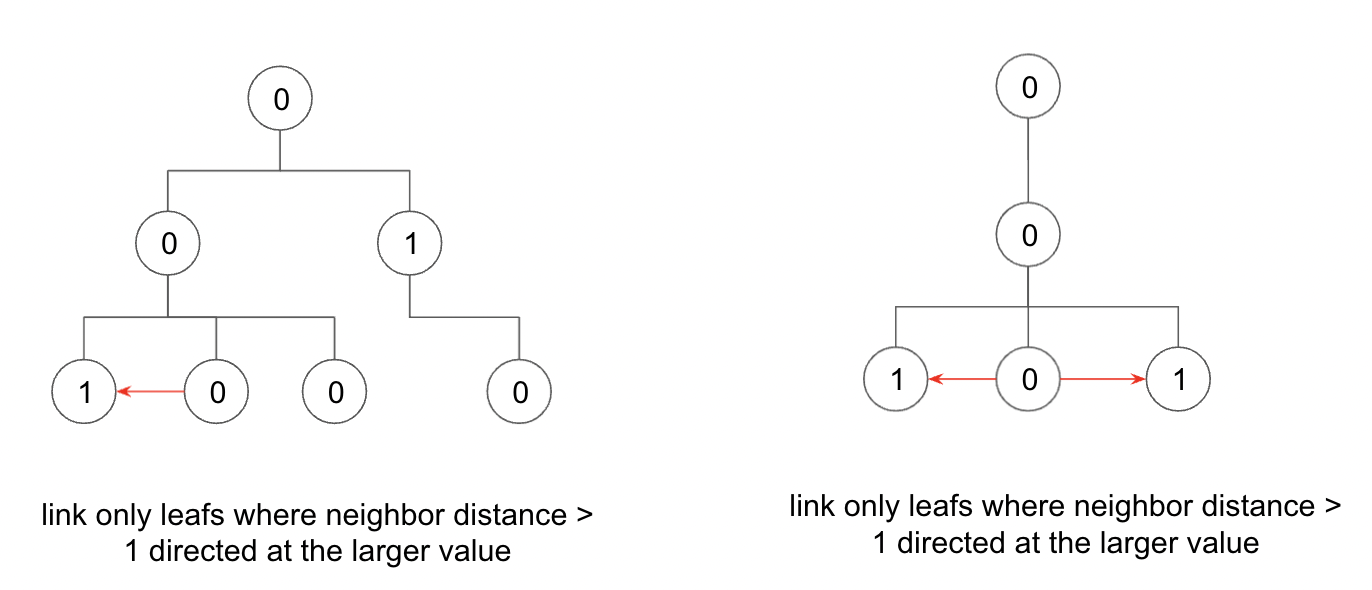}
	\caption{\textit{Example of a linker function. The same linker function are applied to two different graphs, providing directed edges across leafs.}}
	\label{fig:linker}
\end{figure}

\subsubsection{Linkage Math}

Let be a layout of a graph $G (V, E)$, i.e., a linear ordering $v_{1} ,..., v_{n}$ of its vertices. The width with respect to of a set S ${v_{i}^{-k+1},...,v_{i}}$ of k consecutive vertices (notationally widthl(S)) is defined to be the number of vertices $\lbrace v_1,...,v_{i-k} \rbrace$ in the set adjacent to vertices in S $(1 \le i \le k \le n)$. Informally, the width of S with respect to is the number of vertices preceding S and adjacent to elements of S.

Informally, the width of S with respect to is the number of vertices preceding S and adjacent to elements of S.Now, let j be an integer such that $1 \le j \le n$. The j-width of a vertex v with respect to is the least width of any set of at most min (i,j) consecutive vertices extending to the right up to vi, i.e.,

$j-widthl(vi) = min \lbrace widthl(vi-k+1,...,vi):k =1,..., min(i,j)\rbrace$

The j-width of G with respect to l is defined to be the maximum of j-widthl(vi) over all vertices v of G.

The j-min-degree of a subgraph H of G is the minimum ext-degreeH(S) over all sets S of vertices of H with $1 \le |S| \le j$. Obviously, the 1-min-degree of a subgraph is the least degree of its vertices.The j-linkage of G is the maximum j-min-degree of any subgraph of G. Obviously, the l-linkage of G is the maximum min-degree of any of its subgraphs. The l-linkage of G is simply called the linkage of G.Furthermore, it has been proven mathematically that for any graph $G(V, E)$,, the j-width of G is equal to its j-linkage.

\subsubsection{Choice of Lens}
\label{sec:sectionlens}

The choice of lens is critical for the basis of ontology and the R-K Model. Is is the lens at which an ``event'' is determined from, such that ``Lens'' defines the root node such that all branching nodes and the hierarchy are birthed from the choice of lens. As an example, in the store sales data, there may exist many lenses, such as a Customer lens, or a Transaction lens. Each lens would provide unique events associated with such events, with different structural graphs as the foundation for each R-K Diagram.

\begin{figure}[H]
	\centering
        \includegraphics[width=0.5\textwidth]{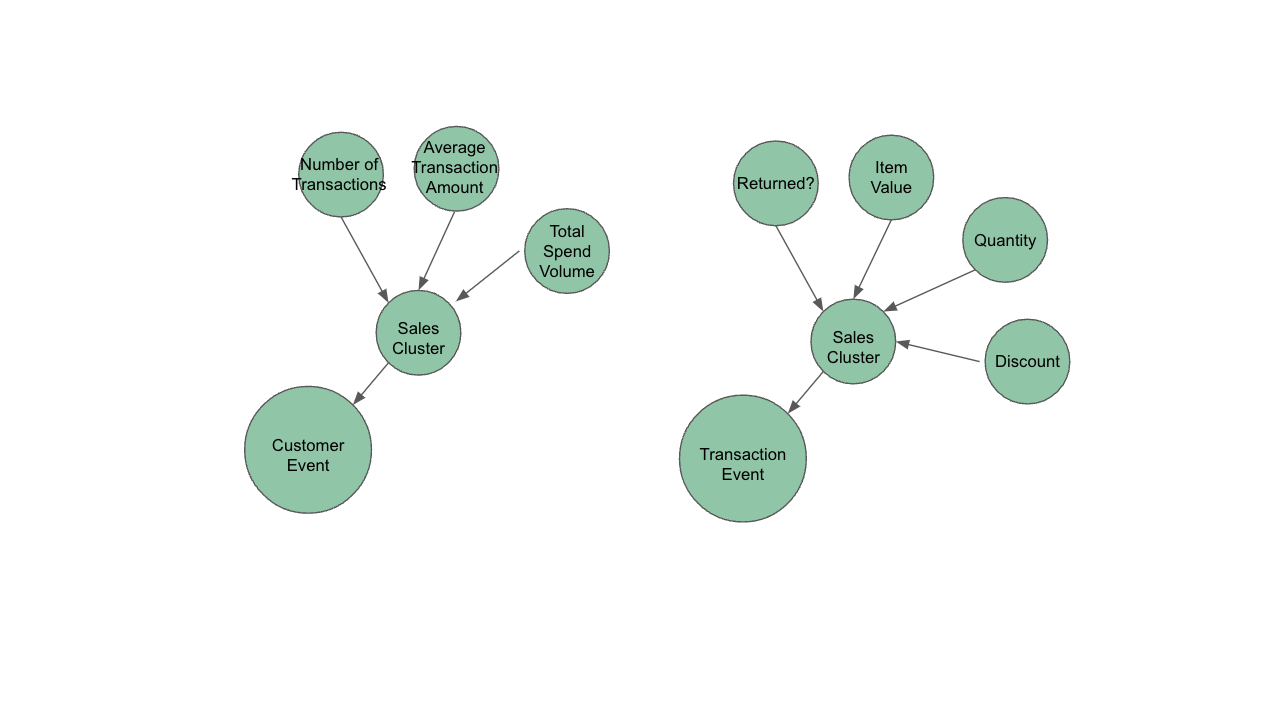}
	\caption{\textit{The diagrams above show the effect of the lens on the structural graph. To the left, a lens is chosen from a specific ``Customer'' event. To the right, a lens is chosen from a ``transactions'' perspective. The values and metrics associated with each lens are related to the respective lens. }}
	\label{fig:linker}
\end{figure}

\subsubsection{R-K Diagram}

An RK-Diagram is the render of an RK-Model. Given an RK-Model, an R-K Diagram renders the R-K Model in 2 or 3d space. As an R-K Model is multi-dimensional representation of data, an R-K diagram can display many dimensions in 2D, without data loss that a typical projection model would have.

We tend to use a radial layout for our demonstrations, but any graph layout can be used, with a preference toward deterministic layouts. We prefer deterministic layouts, because it allows easier qualitative comparisons of R-K Diagrams and their differences.

\begin{figure}[H]
	\centering
        \includegraphics[width=0.5\textwidth]{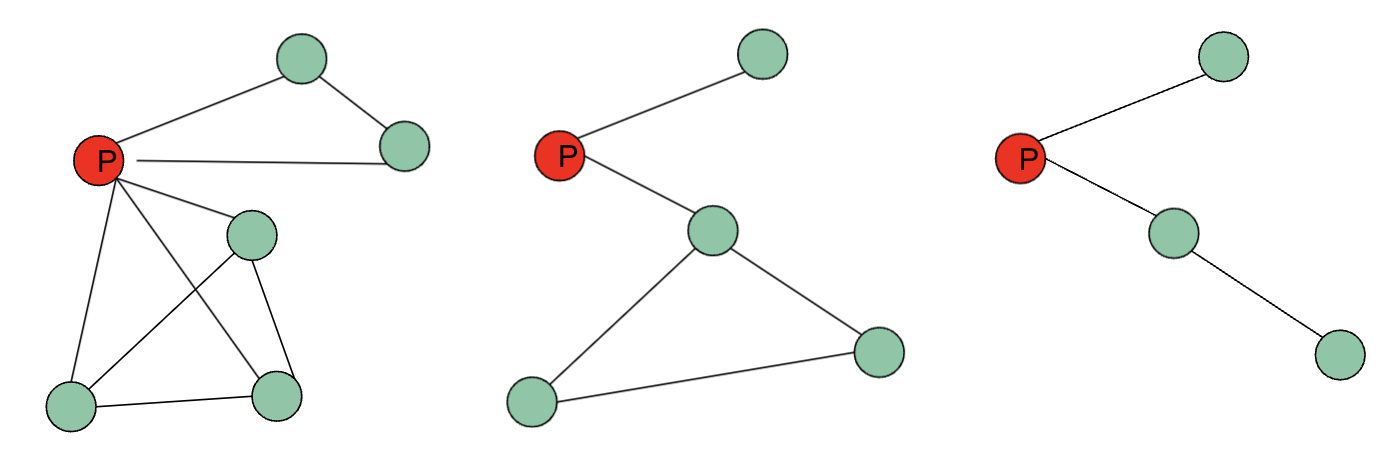}
	\caption{\textit{In the R-K Diagrams above, we use a deterministic layout which makes it easier to spot the differences between R-K Diagrams based on visual features}}
	\label{fig:linker}
\end{figure}

\subsubsection{Betti Number Analysis for Macroscopic Topological Classification}
\label{sec:BettiNumber}
In this paper, we have only explored topological similarity and divergence measures that are calculated with respect to a combined study of nodes, directed edges, properties of DAGs and corresponding Topological distance measures between any 2 R-K diagrams. However, the R-K pipeline and R-K models are built in a way to allow for the smooth establishment of homotopy equivalence between any 2 well defined spaces unlike point cloud data where such equivalence is not established by default. Hence, this allows for effective ways to compute the number of independent j-dimensional surfaces for higher dimensional embeddings on R-K diagrams in Phase-Space. 

This would in tern also allow for smooth ways to drive and obtain the $j-th$ Betti Numbers which would lead to a formalised method to count the number of loops present in unique event-based topological signatures. Thus new metrics can be established to compute and classify the global macroscopic properties of R-K diagrams, thereby leading to the quantification of additional parameters for machine learning templates which bring about further enhancements to the R-K pipeline for the classification and segregation of different R-K diagrams with more distinct divergence measures.
 
\subsection{Measuring R-K Distance}
\label{sec:rk_distance}

Critical to the tuning and understanding of R-K Diagrams is the ability to quantitatively measure distance (and in the dual, similarity) across R-K Diagrams. There is a plethora of existing research that has been done on understanding topological and graph distances in Phase Space. For example, in Hilaga et. al, they analyzed a Reeb graph, and using TDA, computed the geodesic normalized distance, where the geodesic distance of a surface from the formula $\mu(v) = \int_{p \in S} g(v,p)dS$ where $g(v,p)$ returns geodesic distance of points u and v on a mesh. \cite{hilaga_shinagawa_kohmura_kunii}. This approach was not well suited for our analysis, because we do not create a mesh in an R-K Diagram, however demonstrates one of many potential ways to compare topological structures against eachother.

Another approach by Máté et. al use multiple approaches in \textit{A topological similarity measure for proteins}, in which they use a geometric approach using the Jaccard distance and Hausdorff distance based on the barcodes used in persistent homology. \cite{máté_hofmann_wenzel_heermann_2014}. Various other appropriate metrics exist that we will not cover in this section but refer to in the citations. \cite{mazandu_mulder_2012} \cite{reuter_2009}

In our implementation, we used a weighted distance function $d(G_{1}, G_{2})\times{w}$ by applying composite distance function based on geometric and value distance, where the geometric distance was implemented over a Jaccard distance based upon nodes and edges, and the value distance was computed using the Mahalanobis distance similar to Salleh et. al\cite{Salleh1_et_al}. The final distance equation can be supplied through the following:

\begin{equation}
D(G_{i},G_{j}) = f(T,V)\times{w}
\end{equation}

where $f(T,V)$ is a composite function based on topological distance $T$ and value distance $V$ weighted by vector $w$.

Our approach was inspired by Máté et. al's geometric measures and after concieving of such a method, later found Salleh et. al's using Mahalanobis distance with Jaccard's distance in the paper that validated a simliar approach in \textit{Combining Mahalanobis and Jaccard Distance to Overcome Similarity Measurement Constriction on Geometrical Shapes} \cite{Salleh1_et_al}, in which they combined Mahalanobis and Jaccard distance in a weighted average to provide similarity measures across metrics.

\subsubsection{Topological Distance}
\label{subsec:topological_distance}

As mentioned in \hyperref[sec:rk_distance]{Measuring R-K Distance} section, we used Jaccard distance to provide the geometric distance across R-K Diagrams. The Jaccard distance is one of many possible distance functions that can be applied toward graph distances. It is simple but effective in many machine learning algorithms and is a widely applied algorithm across many domains. \cite{roughgarden_valiant_2021}. The formula can be described as:

\begin{equation}
J(A,B) = \frac{| A \bigcap B |}{| A \bigcup B |}
\end{equation}

where A and B are a tuple that represents an edge $V_{i} \to V_{j}$ that $V_{i}$ is the source node and $V_{j}$ is the sink.

To be appropriate for the R-K Diagrams, it is critical to evaluate the Jaccard distances against the \textit{directed edges}, such that the distance measure is sensitive to topological differences due to direction. If a Jaccard distance is applied only at the vertex level, key information about the directed edges and the linked vertexes would be lost. This would be ineffective in the R-K Diagram approach hence we utilized Edges for Jaccard measurements, such that critical features in the distance measurement are preserved.

\subsubsection{Value Distance}
\label{subsec:valuedistance}

The value distance is intended to amplify the effects of the distance measure when topology isn't sufficient to demonstrate differences. For example, in the case of store sales, a purchase could be very similar topologically, but very different in terms of magnitude as the actual sales value differ radically across R-K Diagrams. By comparing the magnitude of the nodes as well as the topology, it provides a clear distinction when topological differences are not sufficient.

There are various methods to compute the value distance that can be effective. Salleh et. al use the Mahalanobis distance as a method to measure the ``Value Distance'' of a graph.  We applied the Mahalanobis distance to provide the distance mesaure. Mahalanobis can be computed by the formula: \cite{Salleh1_et_al}

\begin{equation}
d_{MH}(G_{1}, G_{2}) = \sqrt{(\bar{x}-\bar{y})^{T}\Sigma^{-1}(\bar{x}-\bar{y})}
\end{equation}

where: $\bar{x}$ and $\bar{y}$  are the means of the data, $(\bar{x}-\bar{y})^{T}$ is the transposed of the differences of the mean and $\sum^{-1}$ is the inverse covariance matrix.

We compute the Mahalanobis distance across the entire dataset, and then normalize the values to between 0 and 1 such that we are bound between $[0,1]$ pre-weighting.

\begin{figure*}[t]
	\centering
        \includegraphics[width=1\textwidth]{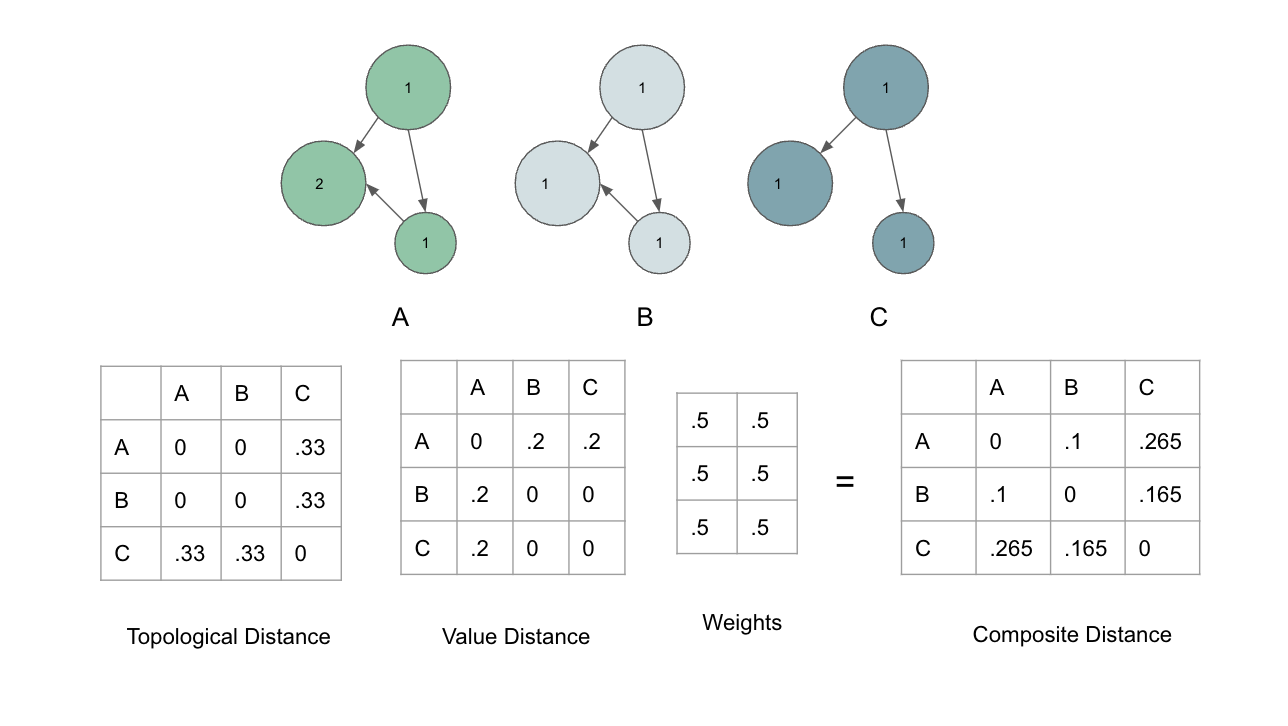}
	\caption{\textit{In the top row, all three diagrams have the same topologies but very different values. The value function, encodes this as an important piece of information when comparing the distance of the diagrams between each-other. On the bottom, the bottom left is more topologically similar to the middle such that the cost topologically to the center is 0 where the cost of the bottom right would be greater than 0 because there is a missing edge. }}
	\label{fig:sample store sales dataset}
\end{figure*}

\subsubsection{Combining the values into a composite distance score}
\label{subsec:distance_score_composite}

We finally combine the Value Distance and the Topological Distance into a composite distance score by the function:

\begin{equation}
D(G_{i},G_{j}) = f(T,V)\times{w}
\end{equation}

where $T$ and $V$ are bound by the range $[0,1]$ and $\sum{w} = 1$. This provides us a final distance score in the range $[0,1]$. This can be expressed by the following:

\begin{equation}
D(G_{i}, G_{j}) = \frac{\sum_{i=1}^{n}x_{i}w_{i}}{\sum_{i=1}^{n}w_{i}}
\end{equation} \cite{Salleh1_et_al}

Where
\begin{equation}
\sum_{i=0}^{n}{w_{i}} = 1
\end{equation}
\subsubsection{Similarity}
\label{subsec:similarity}

As our results are normalized between 0 and 1, we can compute the similarity as the dual of the distance, such that $Sim(G_{i}, G_{j}) = 1-D(G_{i}, G_{j}) = 1-f(T,V)\times{w}$.

\subsection{Future work on distances}
\label{subsec:future_distance_work}
We acknowledge that there is still much research that can be done to improve our distance models and provide more robust measures for R-K Diagram comparisons. The proposal in the above sections outline a starting point for future research.

\subsubsection{Homotopy \& Persistent Homology of R-K Diagrams}
The  Hierarchical Embedding Function ($f_{H}$) of R-K diagrams always ensures that it generates a graph $G$ from a set of measures $M$ such that the resulting graph $G$, is a DAG which represents a bijective mapping between $m_{i} \longleftrightarrow G_{i}$ such that all $V_{i} \in G$ can be traced back to the original value in the dataset. This is also consistent with respect to ontological relationships between dependent and independent variables which form the basic simplex clusters of our fundamental structural graphs - $G$. The advantage of this requirement is that any structural graph can be inverted back into original measures used to generate the graph. Now, since the structural graph provides the baseline ontological structure that forms the basis for all other transformations in the R-K pipeline.

Hence, it ensures that all R-K models consistently render R-K diagrams that are Homotopic with respect to their topological properties and would allow for the smooth computation of persistence diagrams between topologically similar R-K diagrams such as the topological signatures of  BH-BH mergers or NS-NS merger signals from LIGO data. These signatures would preserve the same global topological properties of invariance and persistence under continuous deformations of shape, due to the well defined nature of simplicial complexes consisting of interconnected DAGs which eventually give rise to the fundamental structural graphs of all R-K diagrams.Thus it could be quantified both mathematically and computationally and explored as a part of future research on the study of Homotopy \&  Persistent Homology of different R-K Diagrams.

\subsubsection{Isometric Compressions, Inverse Function, and Decompression}

With R-K Diagrams, we employ isometrically compressible techniques which maintain state change history to provide compression and decompression functions. The compression techniques must be lossless. We can define a compression function $C(G)$, that given a graph returns a compressed version of the graph $G\prime$ where the number of nodes and edges is equal to or less than $G$. We restrict compressible techniques to require an inverse function, such that $G\prime + \nabla{G} = G$. This is critical for the ability to transition between levels of compression.

Because an R-K Diagram can represents a projection that can display many dimensions in 2D space, compression of the R-K Diagram also isometrically projects the diagram into a lower dimensional space, however the visualization can still be visualized at the same space.

Graph compression models have been well evaluated in academia. For example, Gelbert et. al, \cite{gilbert_levchenko} explore various compression techniques focused on semantic compression, based on: degree, a parameter Beta (which weighs neighbourhood of a cluster), paths, an algorithm called KeepOne and KeepAll, and redundant vertex elimination. \cite{gilbert_levchenko}

In the first version of the R-K Toolkit, we have shown a limited implementation of graph compression via a 1 degree leaf compressible technique, which compresses all 1 degree leafs into a single leaf branching from the parent node. The steps are maintained internally so that the $G$ can be reconstructed from $G\prime$. We intend to extend the compressible techniques to others, including those implemented in the Gilbert et al. paper, but did not in the interest of executing the first completed version of the pipeline.

\begin{figure}[H]
	\centering
        \includegraphics[width=0.5\textwidth]{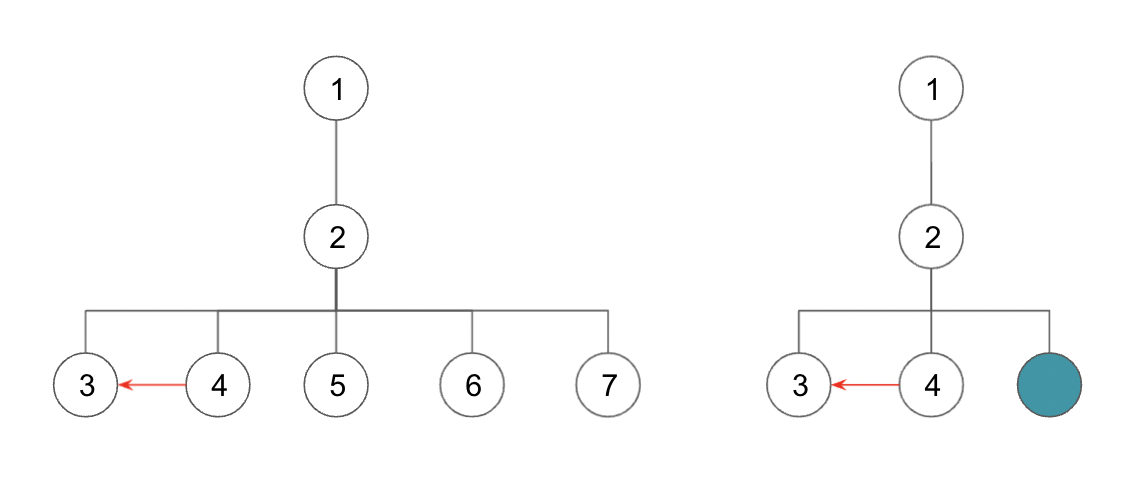}
	\caption{\textit{We employed a simple compression model. In this example, nodes 5,6, and 7 are compressed into a single node.}}
	\label{fig:linker}
\end{figure}

\subsection{R-K Extensibility and Flexibility}

The R-K Toolkit and concepts explored in the implementation are a general framework and way of thinking about the data, not a specific implementation. However, each component is intended to be implemented in a way that can be optimized for a particular domain and use case.

 \section{Case Study: Super-Store Sales Data Analysis}
\label{sec:store_sales_section}

In order to first demonstrate a generalised application of the \textit{R-K pipeline}, \textit{R-K toolkit} and \textit{R-K WorkBench}, we chose to apply our computational framework on a popular non-physical commercial dataset, that serves as a benchmark for standardization of various analytical tools and softwares in the domains of Data-Science \& Big-Data Analytics. Therefore we shall show the implementation \& applications of our generalised R-K Pipeline to a standard super-store sales dataset released by Tableau.\cite{TableauSuperStore} This dataset was chosen for \textbf{3 primary reasons:}

\begin{enumerate}
        \item{Since the initial analysis was done on LIGO data, we decided to choose an unrelated and orthogonal dataset to prove that the core concepts of the R-K workflow can traverse across domains. The R-K Diagrams are meant to be generalizable across many domains and we wanted to prove it was possible.}
        \item{Store sales require less domain knowledge for demonstration purposes. This makes it an easier dataset to teach the core concepts.}
        \item{It is a widely applied in research and is extremely accessible}
\end{enumerate}

\begin{figure*}[t]
	\centering
        \includegraphics[width=1\textwidth]{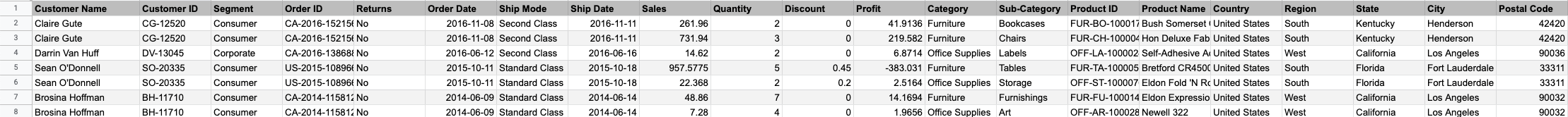}
	\caption{\textit{Head of the Store Sales Dataset}}
	\label{fig:sample store sales dataset}
\end{figure*}

The corresponding code along with our analytical framework and its applications on this dataset can be found on \href{https://github.com/andorsk/store_sales}{Github}. Our results show that we can generate distinct topological signatures for each transaction event, and that the encoding provides descriptive properties that standard distance evaluation methods would not capture.

\subsection{Data Description}

The data for store sales consisted of the following information: \cite{tableau_community_forums_2021}

\begin{itemize}
        \item Location Data for where the transaction occured, in the form of country, region, state, city, and postal code.
        \item Identity information such as customer id and segment id.
        \item Information on the order such as the shipment type, returns, shipment date, and shipment mode.
        \item Product information for the line item, specifically the category, sub category, product id, and product name
        \item The information of the sale of the line item: sales, quantity, and discount
\end{itemize}

The data is synthetic data. All data was merged together into a single dataframe within python. The head of the contents can be found below, post merge. The dataset has 9994 entries. A basic description of the numeric entries is available at figure 7. The non-numeric data is preprocessed and encoded into numeric data in the preprocessing step. W

\begin{figure}[H]
	\centering
        \includegraphics[width=1.0\linewidth]{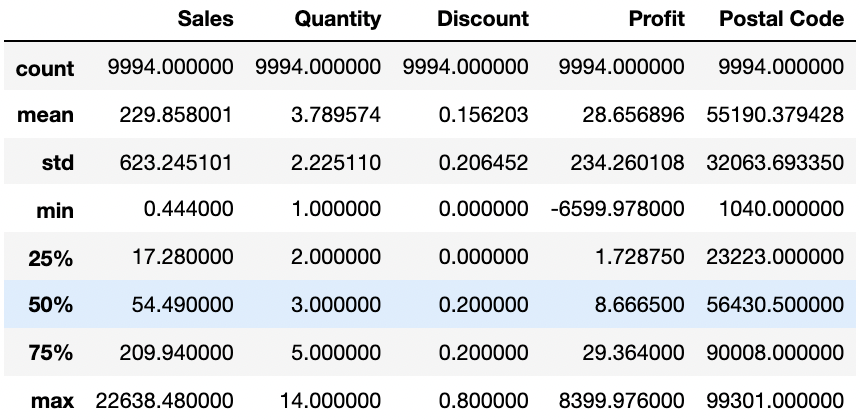}
	\caption{\textit{Summary of the Store Sales Data}}
	\label{fig:fig5}
\end{figure}

\subsection{Objective of Analysis}

The objective of the analysis was to localize and generate R-K Diagrams that provide unique and descriptive topological signature for the data in the form of an R-K Diagram. Through the R-K Pipeline, these structure would preserve a unique topological representations that allows attributal comparisons across events. We expect that the encoding provided through the R-K Pipeline would capture information that normal techniques will fail to capture, and allow for better cross-event comparisions based upon both value and topological distances.

After deriving the unique topological structures, through machine learning, the goal of this analysis is to finally embed the structures in $\mathbb{R}^{2}$ space such that topological differences maximize the inter-R-K Diagram distance and topological similarities minimize the inter-R-K Diagram distance in Phase Space as elaborated in the subsection \hyperref[subsec:phase_space]{Phase Space \& Projections}.

\subsection{Lens Implementation}
Since R-K-Diagrams represent event based topological signatures, therefore the choice of lens is critical to how the R-K-Diagrams are formed. In this dataset, \textbf{each transaction} was represented as an event. We chose ``transactions'' as an event lens because of the clear use case which represents a discrete temporal purchase event in a sales store. We could have chosen alternative lenses such as a customer, however coercing a customer data source into an event that can be used in the R-K models, involve extracting features from all transaction events of a customer into an ``Customer Observation Event''. Such a use case, while possible, is non-temporal and requires aggregation before applying the lens.

\begin{figure}[H]
    \centering
            \includegraphics[width=0.5\textwidth]{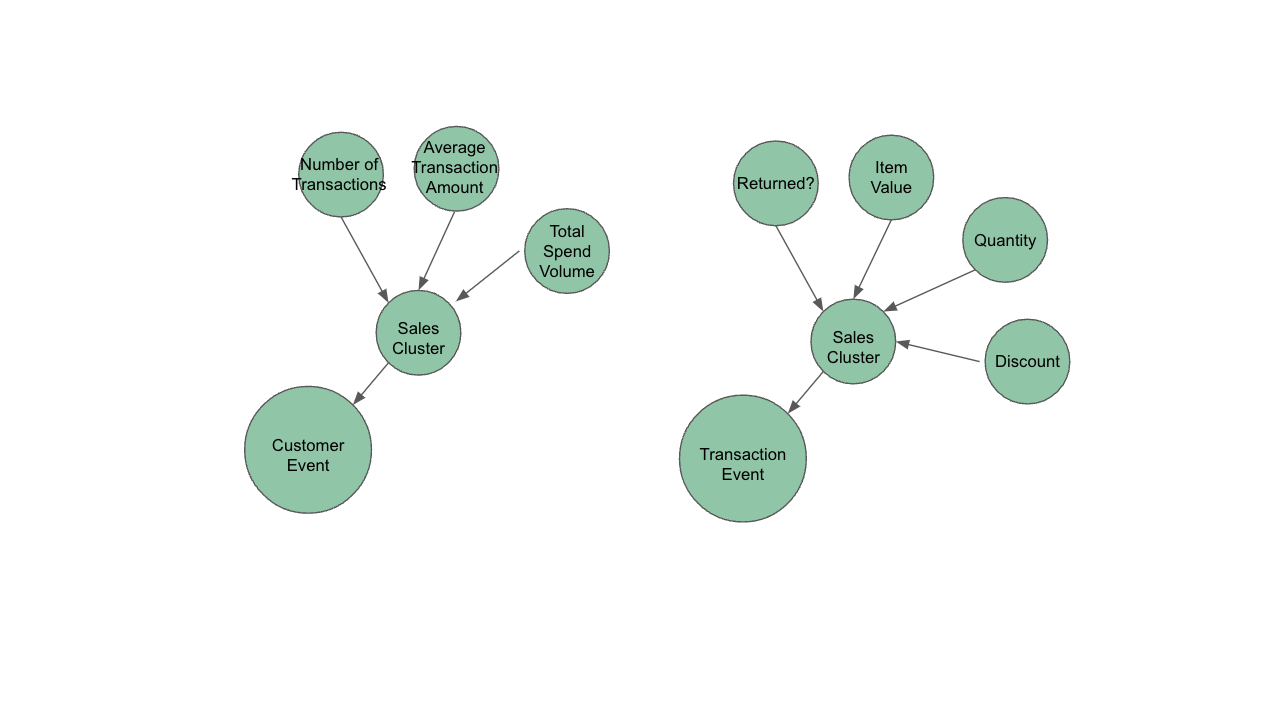}
    \caption{\textit{The lens choice used for this analysis. We chose to review the data from a transaction level, but various other ``lenses'' can be applied where appropriate.}}
    \label{fig:lens_choise}
\end{figure}

\subsection{Preprocessing}

One hot encoding was applied to categorical data. Dates were converted to numbers. Identifiers unique to the event, or with very large cardinality, such as the Customer Id, Product Id, Segment Id, and Product Name, were dropped. Location data was converted into latitudal and longitudinal positions using Google Map's API's and applying a zip code lookup. Failure to search the latitude and/or longitude of a particular zip resulted in removal from the dataset.

The resulting dataset after transformations became 9918 rows and 90 columns. The data was then normalized over a MinMaxScaler so that training the models were more effective.

\subsection{Deriving the Hierarchical Function}

We generated an ontology based on our domain understanding of store sales data. The graph was documented as a JSON file and provides the \textit{Structure Graph} of the R-K-Diagrams. The ontology was coerced into a transformation node that take in a Pandas dataframe and returns a graph object, with attributes encoded acocording to the event that was sent to the transform.

The hierarchy was derived by utilizing the domain specific ontological graph. An initial ontology was generated based upon domain knowledge and based on the application of financial domain knowledge on Super Store Sales data. The ontology graph can be found in the following link: \url{https://github.com/andorsk/store_sales/blob/ask-store-sales-work/data/heirarchy.json}{hierarchy.json}

\begin{figure}[H]
	\centering
        \includegraphics[width=0.5\textwidth]{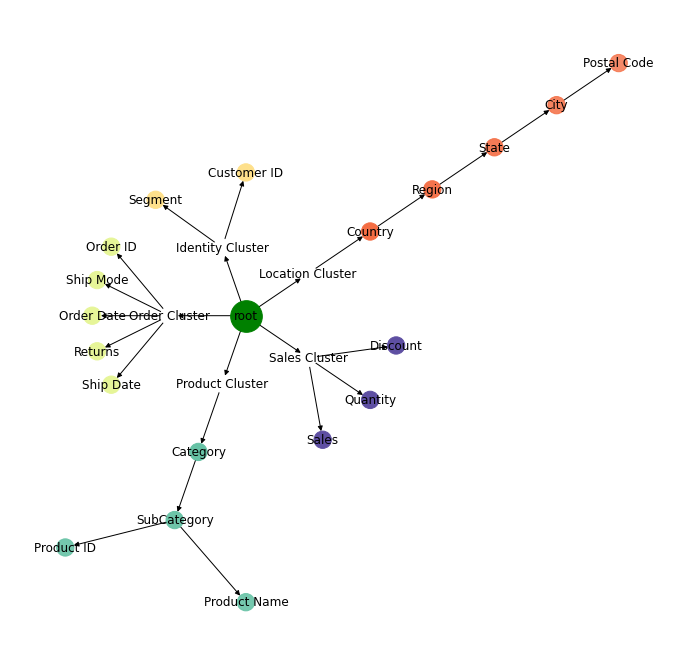}
	\caption{\textit{The base heirarchy provided for the store sales data}}
	\label{fig:nodemask}
\end{figure}

After generating the structural graph, it can be visualized using the R-K Toolkit as shown below:

\begin{figure}[H]
	\centering
        \includegraphics[width=1.0\linewidth]{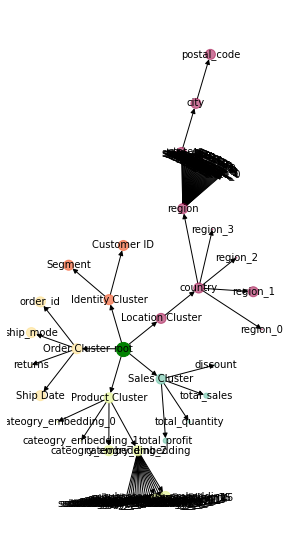}
	\caption{\textit{At the top, you have the core structural graph. Below, are 10 corresponding events without transformation, using the strucdtural graph. You can see the Mahalanobis Distance above the graph. You can see the distance pre-filters and linkages is <metric> when a distance measure bound between 0 and 1. The distances are purely derviced from a value distance, and because the filters and linkage functions provide topological deviations, and have not been applied yet.}}
	\label{fig:fig5}
\end{figure}

As you can see in the figure above, we have expanded out the original hierarchy due to the categorical data. This can be compressed in later steps back to the original form.

\subsection{Pipeline Construction with Filter and Linker Designs}

We used a basic filter (The RangeFilter) and linker called (SimpleChildLinker) design as explained below.

\subsubsection{RangeFilter}

A range filter is one of the simpliest filters that is provided in version 1 of the R-K Toolkit. We assign the filters to each level of the hierarchy that contains numeric data.

\[
\begin{cases}
   true  & \text{if } v \notin \lbrace m, M \rbrack\\
   false & \text{else}
\end{cases}
\]

where:

\begin{itemize}
        \item{$v$ is the value of the node}
        \item{$m$ is the minimum boundary condition for the filter}
        \item{$M$ is the maximum boundary condition for the filter}
\end{itemize}

\subsubsection{LeafLinker}

We also used the simple leaf linker, which is provided in the R-K Toolkit. The simple leaf linker compares only the leafs of a tree, and provides an edge $E$ in the case the euclidean distance is less than the specified threshold:

\begin{figure}[H]
	\centering
        \includegraphics[width=0.5\textwidth]{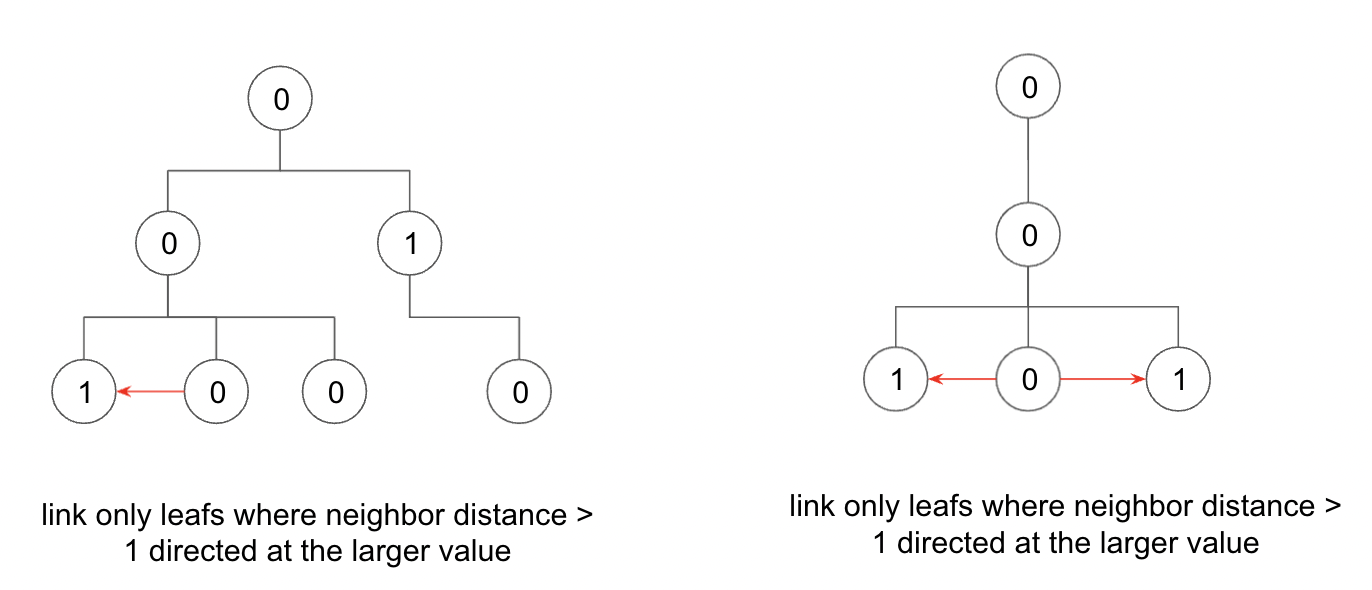}
	\caption{\textit{Leaf Linker compares the distances between values with $\epsilon$ < $\theta$, and links the leafs of the final clusters based upon the threshold.}}
	\label{fig:linker}
\end{figure}

Mathematically, this linkage is represented below:
\[
\begin{cases}
     E & \text{if } \epsilon \le \theta\\
     None & \text{else}
\end{cases}
\]

\paragraph{where}
\begin{itemize}
  \item{$n_{i}$ and $n_{j}$ are two leaf nodes }
  \item{$\epsilon = || \sqrt{n_{i}-n_{j}^{2}} ||$}
  \item{$\theta$ is a defined tolerance for linkage}
\end{itemize}

The LeafLinker's O notation would be $O(n^{2})$, however the number of total computations more concretely is $\sum_{i=0}^{i=|C|}\frac{N!}{2(N-2)!}$ where $|C|$ is the number of clusters and $|N|$ is the number of leafs within the cluster.

\subsection{Visualizing an Untrained Network}

The following image below demonstrates the first 10 rows of the store sales data with untrained thresholds.

\begin{figure}[H]
	\centering
        \includegraphics[width=0.5\textwidth]{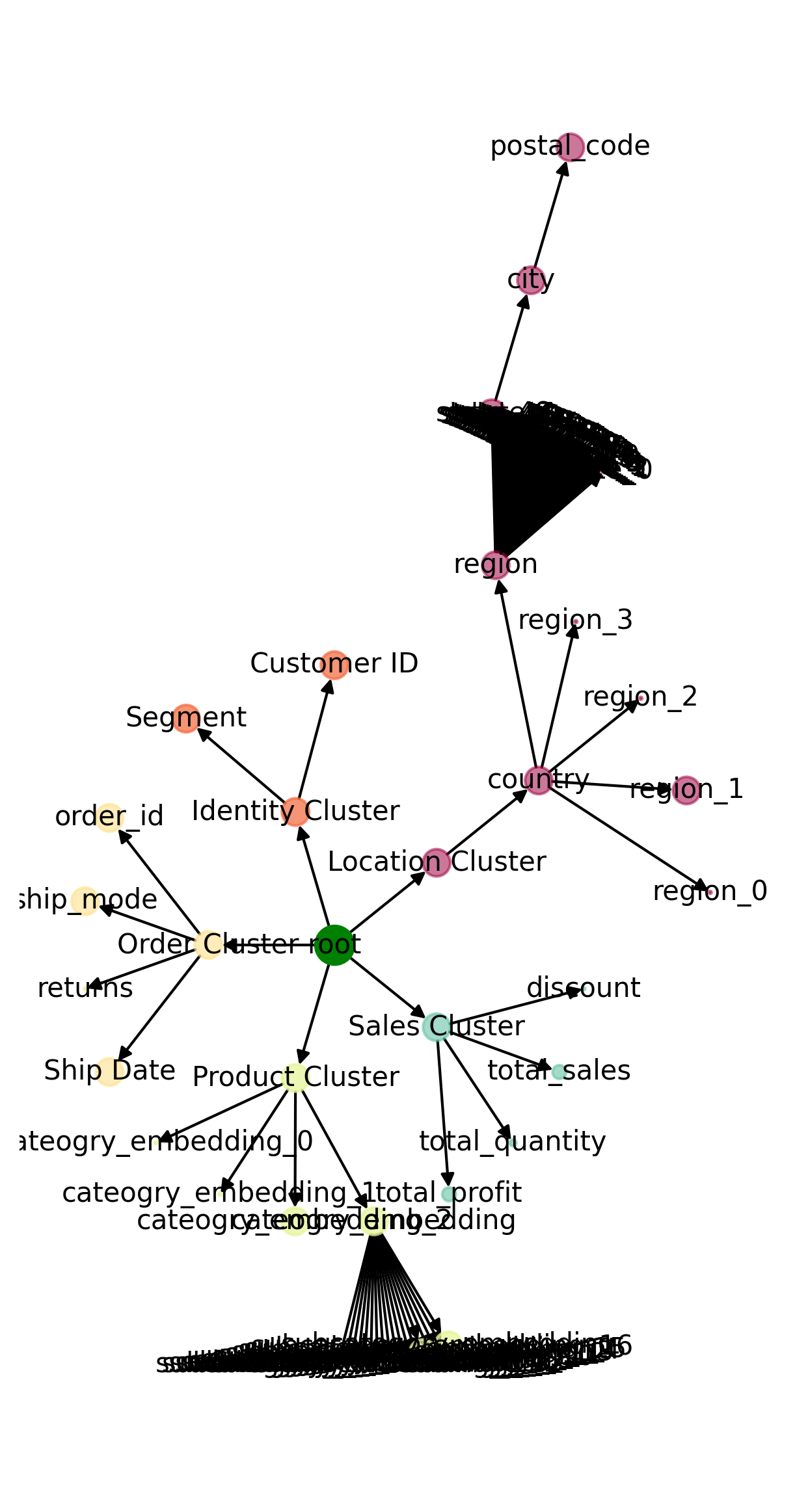}
	\caption{\textit{The base structural graph of the store sales data.}}
	\label{fig:fig5}
\end{figure}

\begin{figure}[H]
	\centering
        \includegraphics[width=0.5\textwidth]{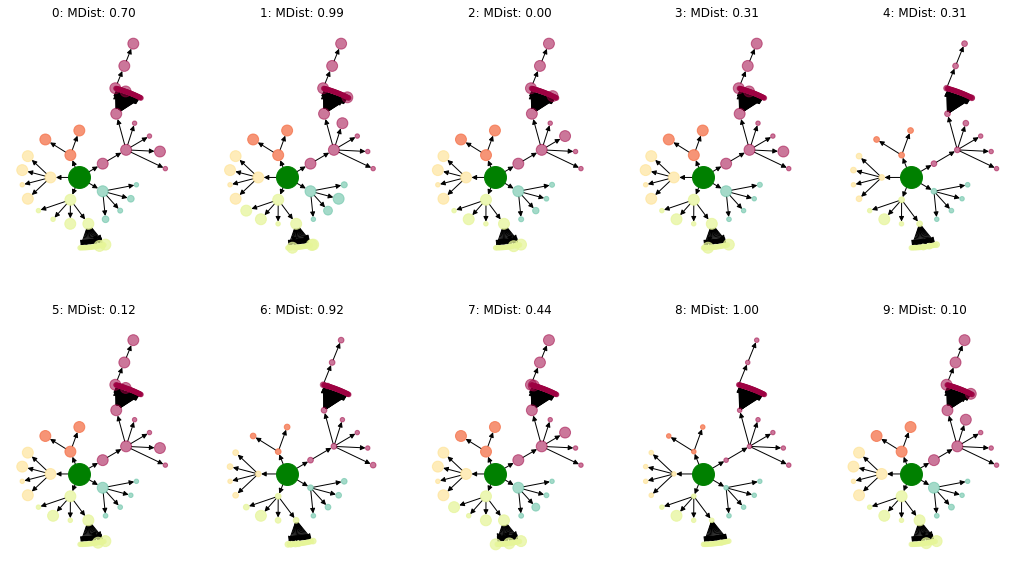}
	\caption{\textit{The first 10 events on an untuned network. M Dist is the normalized ( MinMaxScaled) Mahalanobis distance of the data in data space. Prior to training, the total average distance across graphs is 0.88. Topologically, there are no differences because no filters and/or linkages have been applied yet. The Topological Similarity measure in the above R-K Diagrams are all equal to 1. All deltas are derived from the distances of the values.}}
	\label{fig:fig5}
\end{figure}

After filters and linkages have been applied, the untuned loss goes down from \textbf{0.88} to \textbf{0.78} with a prelimianry Simple Child Linker and Range Filter applied to each node with the bounds [0,1].

\begin{figure}[H]
	\centering
        \includegraphics[width=0.5\textwidth]{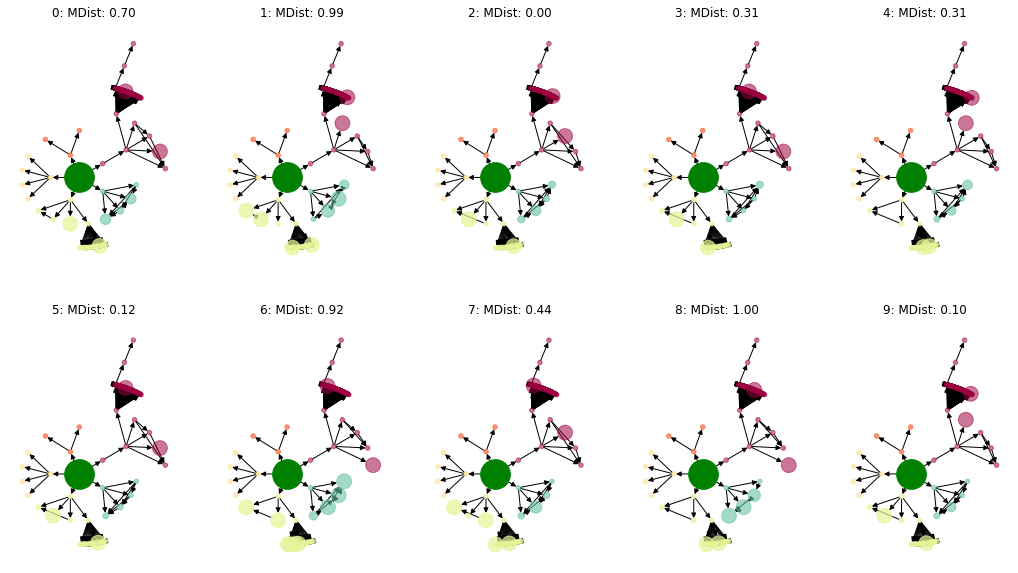}
	\caption{\textit{The first 10 events on an untuned network after filters and linkages applied. The initial guess was a RangeFilter of [0,1] applied to all nodes and a SimpleChildFilter with 0.5 as the initial starting point}}
	\label{fig:untuned_10}
\end{figure}

With filters and linkages, we get an improvement of \textbf{.1} from a pure value distances by introducing topological divergence through filters and linkers.

From a qualitative review, we can look event 6 and event 4, which according to Mahalanobis distance are quite different (.9 and .3) however topologically are quite similar. Conversely, 3 and 4 are very similar in Mahalanobis distance but very different topologically. This divergence would not normally be visable in a traditional distance metric.

We can look at the individual transaction with more detail to get further understanding:

\begin{itemize}
\item Event 6 was a large purchase of $697.074$ dollars made in Houston. It was a highly discounted transaction and a multi-category purchase with 11 items.
\item Event 4 was a medium purchase of $21.376$ dollars made in Glendale. It had a $40$\% discount, was 5 items in a single category. 6 was very different to 4 in terms of Mahalanobis distance but very similar topologically. These are similar topologically because of the high discount rate, and the multi-category transaction type.
\item Event 3 was a smaller purchase of $3.928$ dollars made in New York. A $20$\% discount was applied and it was a single category purchase. Thus, Event 3 was very similar to event 4 in terms of Mahalanobis distance but very different topologically. This was very different topologically to Event 4, because unlike event 4; event 3 was a single item purchase.
\end{itemize}

In conclusion, using our loss function defined in \hyperref[sec:rk_distance]{Measuring R-K Distance}, we find that the average distance between R-K Models is \textbf{0.22}, and inversely, our similarity is \textbf{0.78} after applying filters and linkers. Before filters and linkers, topological distances across the models are 0. After filters and linkages, topological distances are 0.2 unweighted. Emergent properties from the data begin to form at this stage, however, to improve the results, we applied optimization and tuning of hyper-parameters via Machine Learning algorithms such that the appropriate weights to apply to the filters and maximize diveragnce across R-K Diagrams are learned programatically through the data.

\subsection{Training The Pipeline}

In order to maximize divergence across R-K Diagrams, we trained our models using Facebook's Nevergrad \cite{a2020_nevergrad} optimizers. Thus, as demonstrated in the below sections, by providing an objective function and iterating over the various R-K models in a stochastic batch, we were able to reduce total loss of the set from \textbf{0.78} to \textbf{0.7}.

This gets us closer to providing a unique topological signatures across R-K Diagrams post training. This post-trained R-K Pipeline, could in theory be utilized across live stream data, to provide unique topological structures to incoming data from a similar dataset. This analysis is merely the beginning of optimization techniques that are possible with the R-K toolkit. Our goal for the following section is to show the potential of R-K Diagrams and Machine Learning, with the expectations that the methodology and implementation will improve with future research.

The loss history can be seen below and the description of how this loss function was obtained can be seen in the next sections.

\begin{figure}[H]
	\centering
        \includegraphics[width=0.5\textwidth]{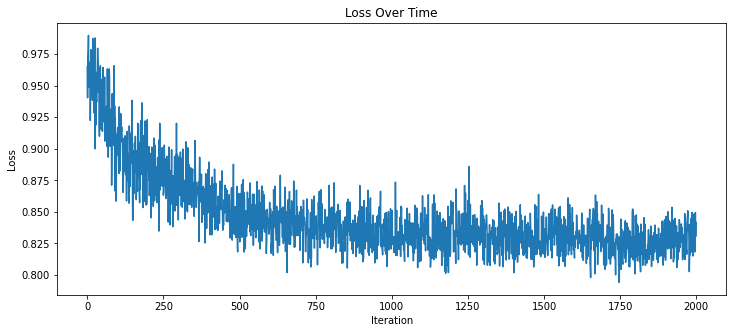}
	\caption{\textit{Loss function over 1000 iterations}}
	\label{fig:loss_func}
\end{figure}

\subsubsection{Objective Function}
\label{sec:ObjectiveFunction}

We define an objective function as the following:

\begin{equation}
\frac{\sum^{n}_{i=0}\sum^{n}_{j=0}d(g_{i}, g_{j})}{ \frac{n!}{2(n-2)!}}
\end{equation}

The goal of the objective function, as defined above, is to maximize divergence across R-K Diagrams by minimizing the similarity across diagrams. This is determined through a distance function defined in the \hyperref[sec:rk_distance]{Measuring R-K Distance} subsection, which takes into account topological and magnitudal similarities across R-K Diagrams using a weighted distance function. We chose an even distribution of $[0.5, 0.5]$ for $w$ as a prior, as there is no reason to bias the weights apriori.

Over iteration of $\theta$, we will attempt to minimize the overall loss. Assuming an infinite number of iterations, we would hope that we maximize divergence across R-K Diagrams such that no R-K Diagram is exactly the same except for the same data, which would deterministically produce the same R-K Diagram.

The objective function provided above has large scope for future improvements and research, as ultimately the goal embedding the data into an R-K Diagram would be to maximize topological differences across differences in the shape of the data, and minimize differences of R-K Diagrams across similarity in attributes.

\subsubsection{Optimizer}
\label{sec:Optimizer}

Because topological distance functions do not exhibit continuous gradients we employed a gradient free optimization using Nevergrad. \cite{a2020_nevergrad}. NGOpts is an optimizer built by Facebook and the default suggested optimizer by Facebook. We ran with 5 max workers (parallelization) and a budget (number of allowed evalautions) of 1000.

\subsubsection{Post Training Results (Pre-Compression)}

The final results post training resulted in a loss of \textbf{0.825}. The diagrams post training are shown below:

\begin{figure}[H]
	\centering
        \includegraphics[width=0.5\textwidth]{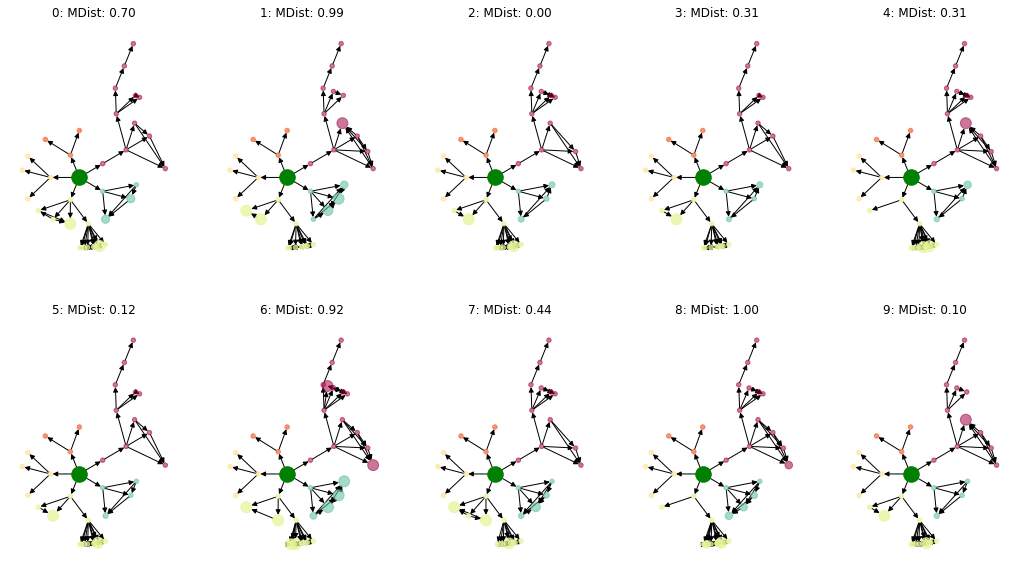}
	\caption{\textit{Tuned Networks (Pre-Compression)}}
	\label{fig:fig5}
\end{figure}

As you can see in the above, by maximizing distances across R-K Diagrams, we can see distinct topological signatures across events. This allows us to perform a variety of different methods of analysis later on, such as classification (given an R-K diagram, assign a label), estimation (given an R-K Diagram, estimate some value based), or other ML related techniques. There are many possibilities, if one considers the R-K Model as an input layer to a model.

In the case above, we can embed the R-K Diagrams in a 2D space and cluster them based upon attributes of the R-K Diagram.

\subsubsection{Isometric Compression}

Unlike standard dimensionality reduction techniques, compression techniques applied to the R-K Diagrams do not lose resolution in the data and can also maintain a consistent number of dimensions (2 or 3) of diagrams regardless of the number of model dimensions. This makes it far easier to evalaute models in higher dimensional space qualitatively without data loss and is one of the primary advantages to the pipeline.

We used the steps outlined in \hyperref[subsec:compression]{Isometric Compressions, Inverse Function, and Decompression} for graph compression. We used a limited compression technique via a 1 degree leaf compression technique, which compressed all 1 degree or ``unlinked'' leafs into a single compressed leaf branching from the same parent. The steps are maintained internally so that the original structural graph can be reconstructed from compressed version, thereby providing the inverse transformation. With future research, we may employ various other compressible techniques (outlined in 4.5.9) to the store sales data to acheive better results.

\begin{figure}[H]
	\centering
        \includegraphics[width=0.5\textwidth]{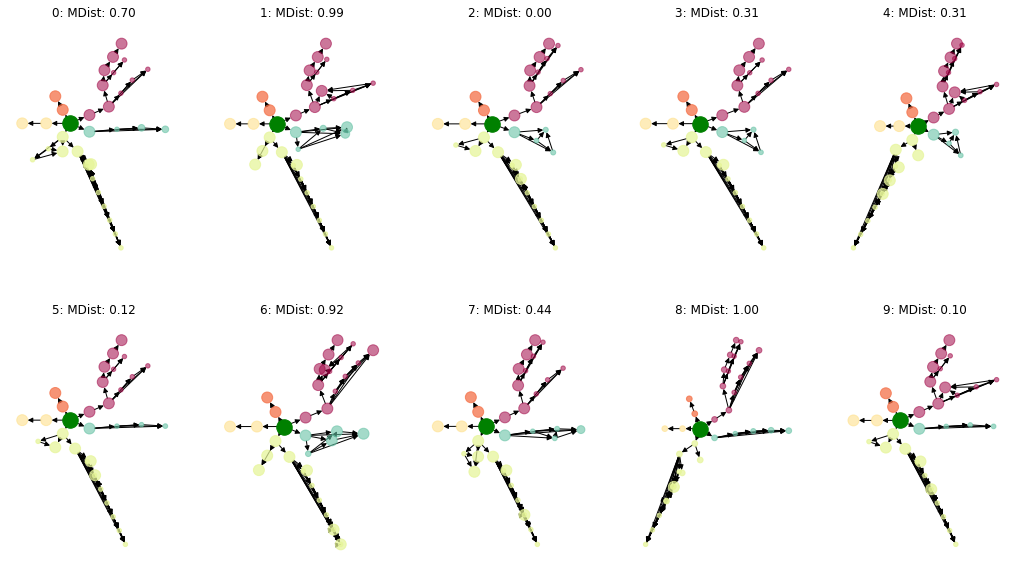}
	\caption{\textit{Post Compression Diagrams}}
	\label{fig:compresseion_example_fig}
\end{figure}

\subsubsection{ML over R-K Diagrams for Variant Use Cases}
\label{sec:classification}

By generating distinct R-K Diagrams, it is possible to employ many standard Machine Learning algorithms for numerous use cases such as classification, clustering, segmentation, and identification. For example, we could use cluster the R-K Diagrams into segments. Such segments could then be used to optimize store operations or sales across the data.

Other use cases such as classification also become possible. The classification use case is used in LIGO (described below), which used R-K Diagrams to describe Binary Merger Event Classifications. Since a pipeline is event level, stream data can be fed into the pipeline to generate new R-K diagram, with a single event generating a unique topological signature. As more data comes in, these models can be refined to provide greater sensitivity to topological differences.

\subsubsection{Conclusion and Future Work}

In conclusion, the store sales dataset has demonstrated a novel approach using event driven Topological Graph Theory Analysis on a canonical and standard, non-scientific dataset. It proves our first case of generalizability outside the initial models made for the LIGO analysis.

Future work can be done to improve the R-K pipeline and algorithms such that more distance and accurate signatures are derived from the store sales data. Further more, applying the signatures to various use cases such as classification or identification would be a valuable exercise and is in scope for future papers.

     \section{Case Study: LIGO Data Analysis}

    The previous case study demonstrated an example of "Event-driven" topological signature identification on Tableau Super-Store Sales Data \cite{TableauSuperStore} with the help of unique filter-specific R-K Diagrams generated using the R-K pipeline. In this second case study, we aim to extend the identification of unique topological signatures with a classification use-case to address an important scientific problem in the field of Gravitational-Wave Astronomy, that has been exploding with research and interest in recent years. \cite{00.3_GravitationalWaveResearch}\cite{00.4_GWRevolution} Hence, we shall now demonstrate a more specialized and scientific application of the R-K pipeline on a  Physical Dataset (as elaborated in \hyperref[sec:PhysicalSystems]{section 4.2.1} with certain domain specific modifications intended for  Gravitational-Wave Analysis\cite{00.1_2012GWAnalysisFormalism} \cite{00.2_schutz2012GWDataAnalysis} on LIGO data, which would aid in the identification and classification of  Compact Binary Merger Events \cite{24.0_BinaryMergerIdentification} \cite{24.1_BinaryMergerClassify} such as Black Holes, Neutron Stars \& Candidate Primordial Black Holes. We have thus applied the R-K pipeline to data-strains and datasets released by the LIGO Open Science Centre \cite{01.5_LIGOOpenSci} \cite{00_LIGOOpenSciData} for the the successful demonstration of our computational framework.This dataset was specifically chosen over other scientific data for the following reasons:

    \begin{enumerate}
        \item {To extend the scope of our pipeline and demonstrate a classification use-case on a field of scientific research that would be both relevant and useful in modern day Astronomy}
        \item{To choose a cutting edge area of modern scientific research which involved very large volumes of data with high-dimensional features and properties spanning across multiple data-strains and datasets to prove that the core components of the R-K Work-flow can traverse and analyse such scientific case-studies with distinct advantages over traditional approaches}
        \item {To showcase the flexibility of the R-K pipeline with modifications necessary for specialised scientific study}
        \item{To demonstrate the ease, flexibility \& scalability of the R-K pipeline to extend its scope of applications on Multi-messenger Astronomy }
        \item{It is open-source and widely accessible to the scientific community for analysis and applications }
    \end{enumerate}

    \subsection{Methodology}
    The methodology of this implementation has been focused to address the following domain specific challenges specifically with respect to the classification of Compact Binary Event Merger signals pertaining to \textbf{\textit{Black Holes, Neutron Stars and Primordial Black-Hole (Dark Matter):}}

    \begin{enumerate}
        \item {Very long wave-forms}
        \item {Computational complexity}
        \item {Difficult to compress signature strain data and eliminate noise}
        \item {Difficult to new apply traditional techniques of machine learning and neural-networks over a wide variety of parameters and constraints or implement any theoretical or observational constraints to the same analytical model}
        \item {Standard Neural Nets \& ML Models cannot be used for classification due to lack of training data, templates and learning models}
        \item {Immense manual effort needed over moths of research and verification for Compact Binary Event Merger signal identification with no computational framework for the automated classification of Binary-Mergers}
    \end{enumerate}

    In order to obtain a holistic methodology for addressing the above challenges and constraints with respect to an efficient computational pipeline to ingest enormous amount of raw-data, segregate confident merger signals and then classify them into a particular category of compact objects (i.e. Black Holes, Neutron Stars, Primordial Black Holes etc.); it is important for us to segregate and specify different data sources for the purpose of our analytical framework. Therefore, the data under consideration has been divided into 3 categories along with its corresponding analytical procedures:
    \begin{enumerate}
        \item Primary Analysis which is carried out on Raw-Strain Data
        \item Secondary Analysis pertaining to the intrinsic physical properties of the compact binary coalescences
        \item Tertiary Data
    \end{enumerate}

    \subsection{Data Description}
    As mentioned in the previous section, the LIGO data was divided into 2 parts for primary and secondary analysis.

    \subsubsection{Primary Data Analysis}

    We can define Primary Analysis on LIGO data as follows:
    The Primary dataset consisted of strain data with the following information

    \begin{itemize}
        \item the interferometer photodiode output of each detector produces gravitational-wave strain data as a time series sampled at 16384 Hz for LIGO data.
        \item For the Advanced LIGO detectors, the calibration is valid above 10 Hz and below 5 kHz.
        \item The detectors also record hundreds of thousands of auxiliary channels, time series recorded in addition to the strain signal, that monitor the behaviour of the detectors and their environment.
        \item Strain data is down-sampled to 4kHz from the original 16 kHz to reduce the size of LOSC files while (arguably) losing no science content, since the higher frequencies are dominated by
        optical noise.
        \item Thus, a 4,096-second file contains 16,777,216 GW-strain samples from a single
        detector, represented as floating-point "doubles" (eight bytes each).
        \item Invalid data due to detector malfunction, calibration error, or data acquisition problems are
    tagged so that they can be removed from analyses.

    \end{itemize}

    Any descriptions pertaining to the above dataset will be referred to as \textbf{\textit{"Primary Data"}} in this paper.

    \subsubsection{Secondary Data Analysis}
    We can define Secondary Analysis on LIGO data as follows:
    The Secondary analysis was carried out on tabular data  on the intrinsic physical parameters of compact binary mergers using Bayesian Methods of Parameter Estimation. \cite{00.7_LIGOBayesianAnalysis} \cite{00.6_LIGOAnalysisPipeline} \cite{24.4_CompactBinaryParameterEstimates} \cite{24.5_GWParameterEsitmation} \cite{24.6_LIGOParameterEstimates} consisting of the following information:

    \begin{itemize}
        \item The most important information w.r.t. secondary analysis are the merger event labels on the Gravitational Wave Transient Catalogue as they would serve as the parent node in correspondence with their GPS merger times. \cite{00_LIGOOpenSciData} \cite{00.2_schutz2012GWDataAnalysis} e.g. Event Names : GW150914 \& GW 170817 corresponding to GPS times of 1126259462.4 \&  1187008882.4 respectively.
        \item The tabular data consisted of Mass (1) $(M_1)$, Mass 2 $(M_2)$, Network SNR, Distance (Mpc), $\chi_eff $ , Inspiral, Final Spin, Total Mass $(M_\odot)$, Chirp Mass $(M_\odot)$, Detector Frame Chirp Mass $(M_\odot)$, Final Mass $(M_\odot)$, Redshift $(Z)$, False Alarm Rate and (yr-1) Pastro  measures.
        \item However, due to the current sensitivity issues and LIGO VIRGO constraints \cite{24.8_PBHdetectionparameters} our analysis of compact binary classifications was primarily done between black holes, neurton stars and primordial black holes based on mass, spin, q-value ratios and red-shift measures with electromagnetic counterparts being added from Multi-messenger sources for further distinction of Neutron Stars w.r.t. Kilonova events that was left out of scope due to our focus on LIGO data in the initial version of this computational framework.
        \item Hence based on the ontological clustering of dependent and independent clusters the R-K models pertaining to secondary analysis of LIGO Data consisted of 4 independent clusters of Mass, Spin, Q-value ratios (derived from $(M_1)$ \& $(M_2)$)  and Red-shift.
        \item The following 4 events: GW170729, GW170817, GW190521 \& GW190814 in the chronological order of their GPS times, were chosen with some unique characteristics of mass, spin and q-values to demonstrate our methodology before executing the pipeline on the entire transient catalogue.

    \end{itemize}

    Any descriptions pertaining to the above dataset will be referred to as \textbf{\textit{"Secondary Data"}} in this paper.

    \subsubsection{Tertiary Data Analysis}
    We can define Tertiary Analysis on LIGO data as the data that is obtained from carrying our mathematical transformations on the secondary data there by giving rise to parameter specific refined measures such as q values, calculation of the evolution of the inward spiral and other measures obtained from theoretical analysis by applying mathematical equations to further evaluate and refine those parameters obtained using secondary analysis.

    \subsection{Objective of Analysis}

    To obtain unique event specific R-K diagrams and to showcase classification of Binary Event Mergers with high Signal to Noise Ratio (SNR) into a particular category of compact objects such as Black Holes, Neutron Stars or Primordial Black Hole Dark Matter.

    Modifications to the R-K pipeline have been done to signify confident binary merger signals and to use them as templates in combination with specific parameters and threshold filters to segregate such signals form false positives and detector noise in future. Furthermore, we have also endeavoured to provide a feasible computational pipeline that would enable the identification and classification of such binary mergers into specific category of compact objects based on secondary data analysis i.e. data obtained after on the various intrinsic physical parameters of such compact objects under consideration. Such secondary analysis is not carried out on Primary (Raw-Strain Data) but on secondary data Bayesian inference to calculate the posterior probability distribution over the parameters (sky location, distance, and/or intrinsic properties of the source) given the observed gravitational-wave signal.

    \subsection{Lens Implementation}

    As established earlier the choice of lens is critical for the basis of ontology and the R-K Model. Hence a binary coalescence or a compact binary merger event was chosen as the primary lens in this case study. Thus all parent nodes of of R-K diagrams on LIGO data would be represented by the binary-merger event label e.g. GW150914 (a typical BH-BH merger famously recorded in September 2015). Moreover, all central Event nodes pertaining to the "merger-event-lens" will also correspond to their respective GPS merger times in phase-space for the classification and analysis of R-K diagrams that are generated from R-K models build upon the choice of this lens. Furthermore, each attribute representing a characteristic physical parameter associated to a parent node or merger event would have a structural ontology with independent physical variable clusters such as all mass measures (M) and all spin parameters (S) clustered together with their dependent attributes such chirp mass, effective spin etc. as explained in section \ref{sec:PhysicalSystems}
    the unique attributes associated with such events would thereby serve as the  foundation for each R-K Diagram in this LIGO data analysis case study.

    \subsection{Preprocessing}

    As the current pipeline uses secondary measures, the preprocessing steps involved in the pipeline was naive. We ingested the secondary data into the pipeline and processed. In future versions of the LIGO R-K Pipeline, the process is intended to be much more elaborate, and eventually involving Bayesian Parameter Estimation over Primary Data as a means of feature extraction.

    \subsection{Modifications to the R-K Pipeline}

    We implemented a standard R-K Pipeline, with 2 stages for optimization. The first stage, involves generating unique topological signatures via a similar method to the store sales data. Training is done to maximize divergence using the same optimization objective function described in the Store Sale's \hyperref[sec:ObjectiveFunction]{Objective Function}.

    After optimization, we encoded the R-K Diagram into a vector and trained an SVM over the vectorized diagrams to produce a binary classifier over the data where X represents each vectorized R-K Diagram and Y are labels of PBH Merger events. Thus the pipeline required two levels of optimization.

    1. Embedding Optimization
    2. Classification Optimization

    \begin{figure}[H]
        \centering
        \includegraphics[width=1.0\linewidth]{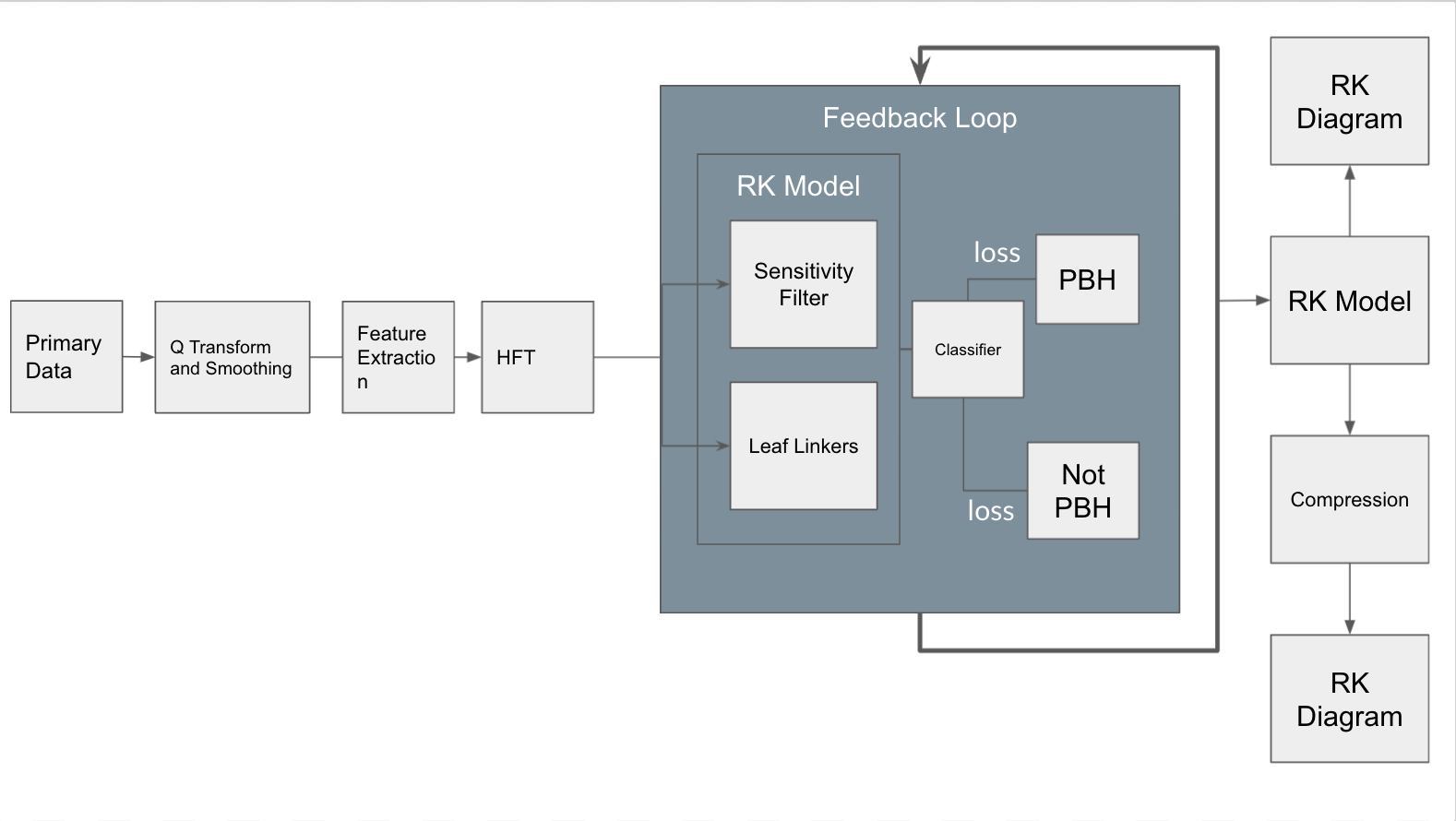}
        \caption{\textit{R-K Pipeline Modifications for LIGO}}
        \label{fig:ligo_pipeline_fig}
    \end{figure}

    These two optimizations, applied sequentially after each-other provide two artifacts:

    1. An R-K Pipeline which generates an R-K Diagram from a single event node
    2. An SVM Classification Boundary to which the R-K Diagrams can be applied against.

    We achieved R-K Diagram vectorization in a similar way to how a Bag of Words vectorization would be applied, where the ``corpus'' is all possible edges for a particular structural graph.

    \begin{figure}[H]
        \centering
        \includegraphics[width=1.0\linewidth]{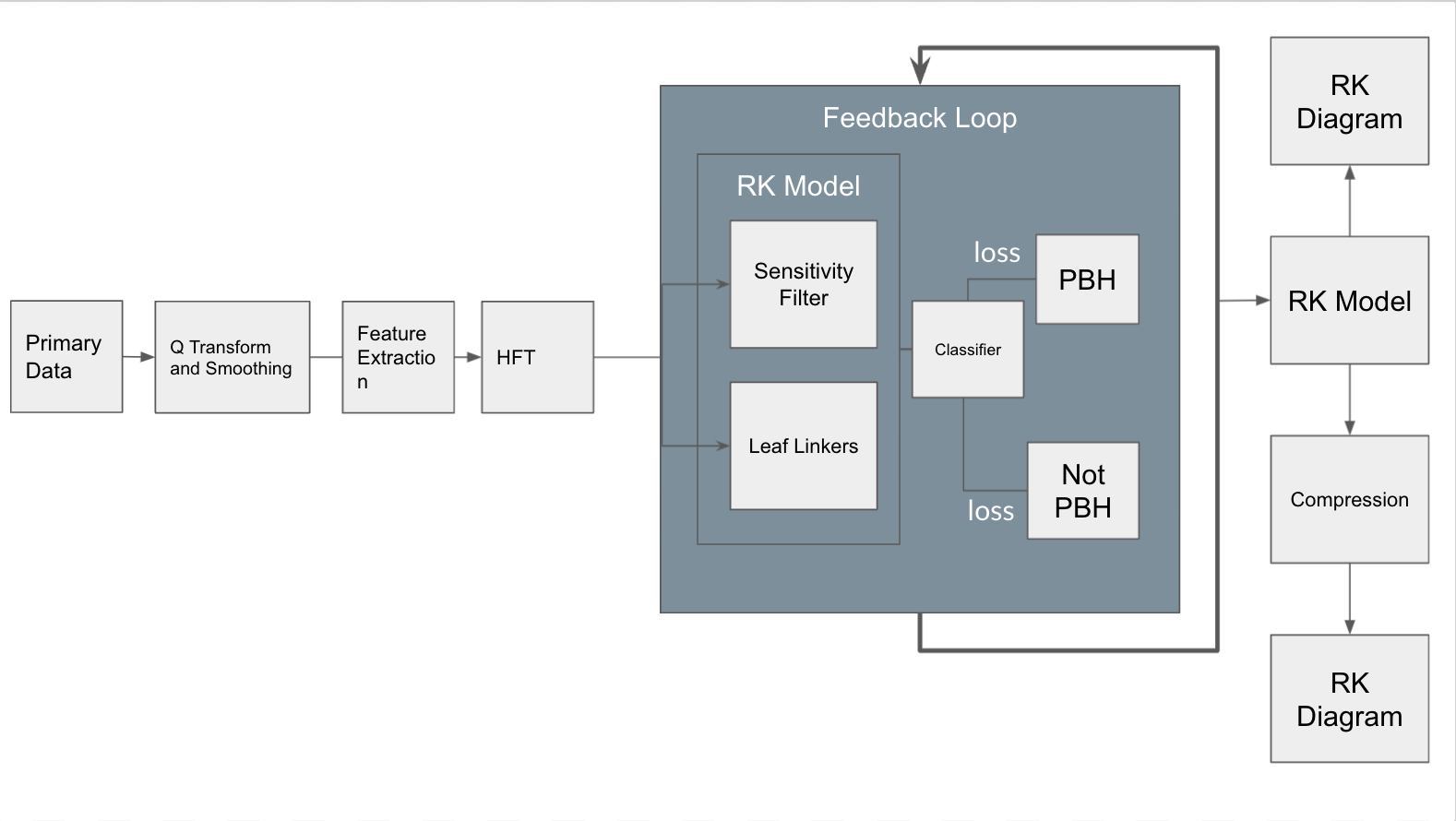}
        \caption{\textit{R-K Pipeline Modifications for LIGO}}
        \label{fig:ligo_pipeline_fig}
    \end{figure}

    \subsubsection{Modification Objectives}
    These modifications to the R-K pipeline were primarily motivated due to the address the following objectives:
    \begin{enumerate}
        \item To serve as a comparative classification frame-work for generating R-K Diagram based templates to classify and categorize compact-binary mergers with an efficient automated pipeline.
        \item To associate unique R-K Diagram classifications having distinct graph and topological properties with their  corresponding amplitude spectral densities (ASD) to allow for unique merger signal classification and segregation based on confidence scores of their corresponding R-K Diagrams in future.
        \item To enable a complete and holistic plug-and play computational framework linked directly to  primary data obtained from the detectors at source followed by effective filtering, noise reduction, Bayesian Parameter Estimation, deriving intrinsic physical parameters to build corresponding R-K models and applying filters and compressors to render the final R-K diagrams.
        \item These R-K diagrams could then serve as feedback templates generating more machine learning templates to optimize loss and refine binary merger signal classifications with further distinction.
    \end{enumerate}

    \subsection{Primary Data Analysis}

    Understanding the noise is crucial to detecting gravitational-wave signals and inferring the properties of the astrophysical sources that generate them. Improper modelling of the noise can result in the significance of an event being incorrectly estimated, and to systematic biases in the parameter estimation. To guard against these unwanted outcomes, detector characterization and noise modelling are significant activities within the LVC.\cite{00.5_GWDetectionNoiseCatalogue} While many textbook treatments of gravitational-wave data analysis describe the idealized case of independent detectors with stationary, Gaussian noise, actual LVC analyses are careful to account for deviations from this ideal.\cite{00.1_2012GWAnalysisFormalism} \cite{00.3_GravitationalWaveResearch}

    The Advanced LIGO and Advanced Virgo detector data have a rich structure in both time and frequency. For a given gravitational-wave source, the noise (as described by its spectral density) governs the measured signal-to-noise ratio (SNR). The spectral frequency content of the LIGO-Livingston detector was averaged over a three-minute period shortly before the first detection of gravitational waves from a binary neutron star merger with the event name GW170817.\cite{00.5_GWDetectionNoiseCatalogue}

    The steep shape at low frequencies is dominated by noise related to ground motion. Above roughly
    $100 Hz,$ the Advanced LIGO detectors are currently quantum noise limited, and their
    noise curves are dominated by shot noise. High amplitude noise features are also present in the data at certain frequencies, including lines due to the AC power grid (harmonics of 60 Hz in the U.S. and 50 Hz in Europe), mechanical resonances of the mirror suspensions, injected calibration lines, and noise entering through the detector control systems. \cite{00.1_2012GWAnalysisFormalism} \cite{00.5_GWDetectionNoiseCatalogue} \cite{00.3_GravitationalWaveResearch}

    \subsubsection{Strain Data}

    The R-K pipeline was retrofitted to run additional precursor steps on strain data as explained in the previous section. This is mainly done as a part of the additional component steps added to the R-K pipeline for reasons explained in section 6.6. The figure below demonstrates plotting of the raw data-strains with initial filters as explained. This process was then extended for all 4 selected events for the purpose of initial analysis and validation, namely:  GW170729, GW170817, GW190521 \& GW190814. The purpose of choosing all 4 events in parallel was done to demonstrate their eventual representations on an expanded topological "Event-scape" that would spread across a plane of GPS times and merger frequencies for all the events plotted together for the sake of future comparison.

    \begin{figure}[H]
        \centering
        \includegraphics[width=1.0\linewidth]{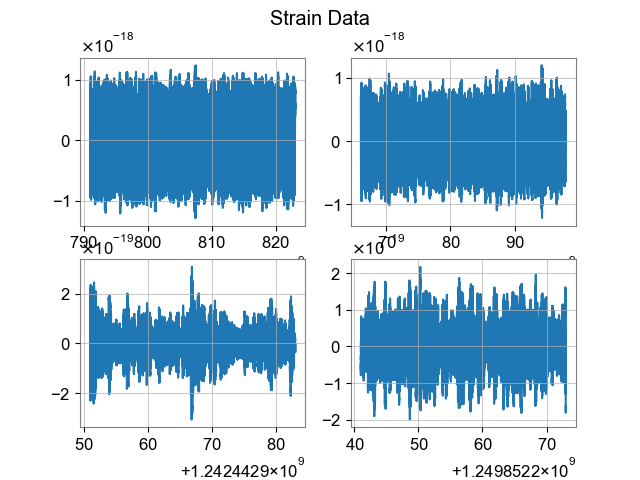}
        \caption{\textit{Raw Data Strain from LIGO}}
        \label{fig:LIGO1_PlaceHolder_fig}
    \end{figure}
    A sequence of processing steps was applied in our modified pipeline to the calibrated strains from the LIGO-Hanford detector such as the one showing $4 s$ of data centered on GPS time 1126259462 \(September 14, 2015 09\:50l\:45 UTC\). This data was obtained form the GW Open Science Center \cite{00_LIGOOpenSciData}. First a Tukey Window with $0.5 s$ is applied, then the data are whitened using an estimate of the noise spectral density. Finally the data are bandpassed filtered to enhance features in the passband (35 Hz; 350 Hz), there by revealing the presence of gravitational-wave signal $GW150914.$ This was done as a first trial run using the modified R-K pipeline in accordance with well established research papers. \cite{00.2_schutz2012GWDataAnalysis} \cite{00.5_GWDetectionNoiseCatalogue}
    The process described in the research papers were then applied to all 4 selected event stains obtained from the GW Open-Science Center and to showcase the capability of our computational framework in terms of processing primary and secondary data with respect to multiple compact binary-merger events in parallel.

    \subsubsection{Spectrograms \& Q-Transforms}

    It was seen that the raw or primary data from the previous stage are dominated by low-frequency noise. Therefore a  Tukey window with 0.5 s transition regions was applied to the raw data. Next, the data-stains were all whitened by dividing the Fourier coefficients by an estimate of the amplitude spectral density of the noise, which ensures that the data in each frequency bin has equal significance by down-weighting frequencies where the noise is loud. The data-strains were were then inverse Fourier transformed to return to the time domain using the following relation:
    \begin{equation}
      d(t) \stackrel{\mathrm{FFT}}{\longrightarrow} \tilde{d}(f) \stackrel{\text { Whiten }}{\longrightarrow} \tilde{d}_{w}(f)
    \end{equation}
    \begin{equation}
      \tilde{d}(f) \stackrel{\text { Whiten }}{\longrightarrow} \tilde{d}_{w}(f) =
      \frac{\tilde{d}(f)}{S_{n}^{1 / 2}(f)} \stackrel{\mathrm{iFFT}}{\longrightarrow}d_{w}(t)
    \end{equation}
The above bandpass technique enhances the visibility of features of interest in this band by removing noise outside of the band - seismic and related noise at low frequencies, and quantum sensing noise at high frequencies. However, it is important to note that Note that such narrow band-passing is only used for visualization purposes and is not employed in the LVC analyses.

    However, as indicated in LIGO research \cite{00.2_schutz2012GWDataAnalysis} \cite{00.3_GravitationalWaveResearch} \cite{00.6_LIGOAnalysisPipeline} \cite{00.5_GWDetectionNoiseCatalogue} Wavelets provide a more
    flexible analysis framework than short-time Fourier transforms. Continuous wavelet transforms are commonly used in LIGO-Virgo data studies to produce spectrograms that provide a visual indication of non-stationary behaviour. Quantitative assessments of non-stationary may also be made by using discrete, orthogonal wavelet transforms. This can be visualised in the best way using normalized q-transforms on the strain data by plotting the noise reduced data strain frequencies against GPS time in a 2-D plane with the normalised energy representing the amplitude spectral densities of each merger-event in the form of the combined plots shown below.

    \begin{figure}[H]
        \centering
        \includegraphics[width=1.0\linewidth]{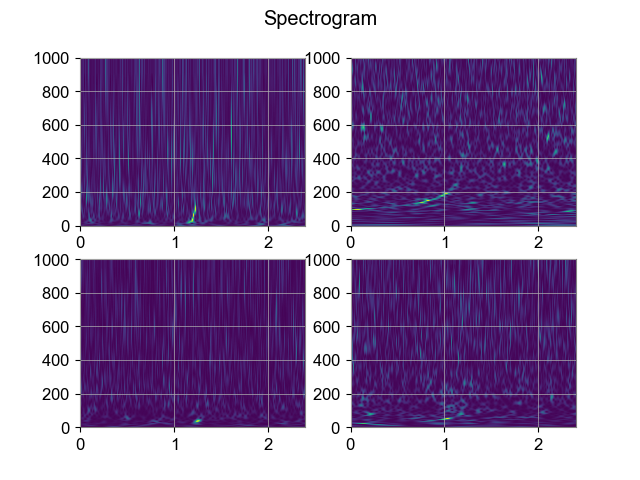}
        \caption{\textit{LIGO Spectrograms with Q Transformed ASD}}
        \label{fig:LIGO2_PlaceHolder_fig}
    \end{figure}

Here the normalised energy  of the Q-transform coefficients can be calculated using the following relation:

\begin{equation}
	E=\frac{|X(\tau, \phi, Q)|^{2}}{\left\langle|X(\tau, \phi, Q)|^{2}\right\rangle}
\end{equation}
Thus, the Q- transform pipeline can be thought of as an optimal matched filter search for minimum uncertainty waveforms of unknown phase in the whitened data streams.

    \subsubsection{Topological Event Scape}

    However, we found that the true power of topological analysis could be utilised in the best way possible by merging all events in 3D for shape rendering and signal filtration using noise reduction techniques. This notion is inspired by topological data visualisation techniques which increase the granularity and bring out the details within high dimensional datasets for further distinction and analysis.

    \begin{figure}[H]
        \centering
        \includegraphics[width=1.0\linewidth]{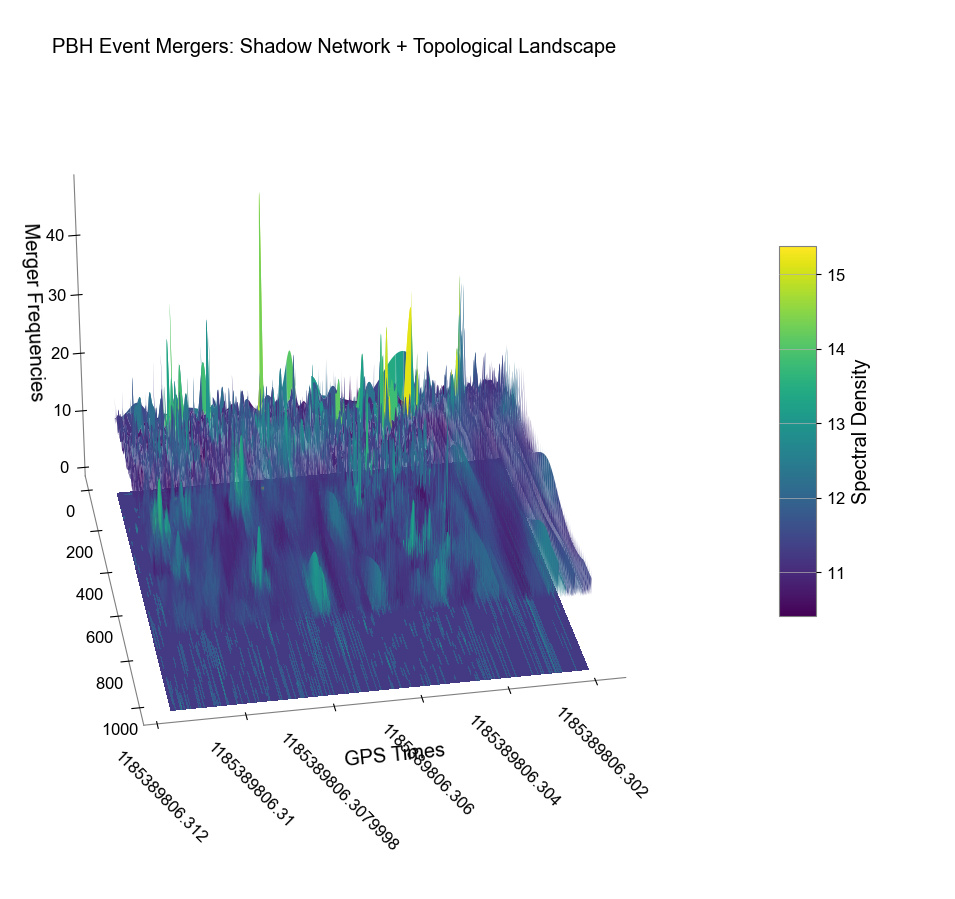}
        \caption{\textit{Topo-Transformed Event-scape with Phase Projected Network of Event Signatures}}
        \label{fig:LIGO3_PlaceHolder_fig}
    \end{figure}

    Thus, the above diagram shows a "Topological Event-Scape" of all 4 merger events with GPS times and merger frequencies representing the 2 axes of the 2D plane while the amplitude of the spectral density representing the normalised energy is plotted in the 3D dimension to bring about more enhanced filtering and noise reduction around the Tukey window with appropriate filters.This is also made possible by the fact that all mergers can be combined together and compared with each other using Universal GPS times and because all mergers tend to have their Fourier transformed frequencies within a comparative scale of values.

    \subsubsection{Filtration \& Noise Reduction}

    In this section we outline how we identify and characterize these noise features so that we can either exclude the bad data or assess the impact of remaining components to search for gravitational-wave signals. This concept can be further enhanced to all data-strains in parallel using the R-K pipeline as shown and can be extended to 'n' number of events in theory. However, we have addressed our 4 selected events  GW170729, GW170817, GW190521 \& GW190814 for the sake of simplicity and clarity.

    In this pipeline, noise reduction filters could be applied to all events simultaneously and we used a combination of Gaussian filters and cosine based Tukey filters such as the one defined below:

    \begin{equation}
        w(x)= \begin{cases}\frac{1}{2}\left\{1+\cos \left(\frac{2 \pi}{r}[x-r / 2]\right)\right\}, \& 0 \leq x<\frac{r}{2} \\ 1, \& \frac{r}{2} \leq x<1-\frac{r}{2} \\ \frac{1}{2}\left\{1+\cos \left(\frac{2 \pi}{r}[x-1+r / 2]\right)\right\}, \& 1-\frac{r}{2} \leq x \leq 1\end{cases}
    \end{equation}

    \begin{figure}[H]
        \centering
        \includegraphics[width=1.0\linewidth]{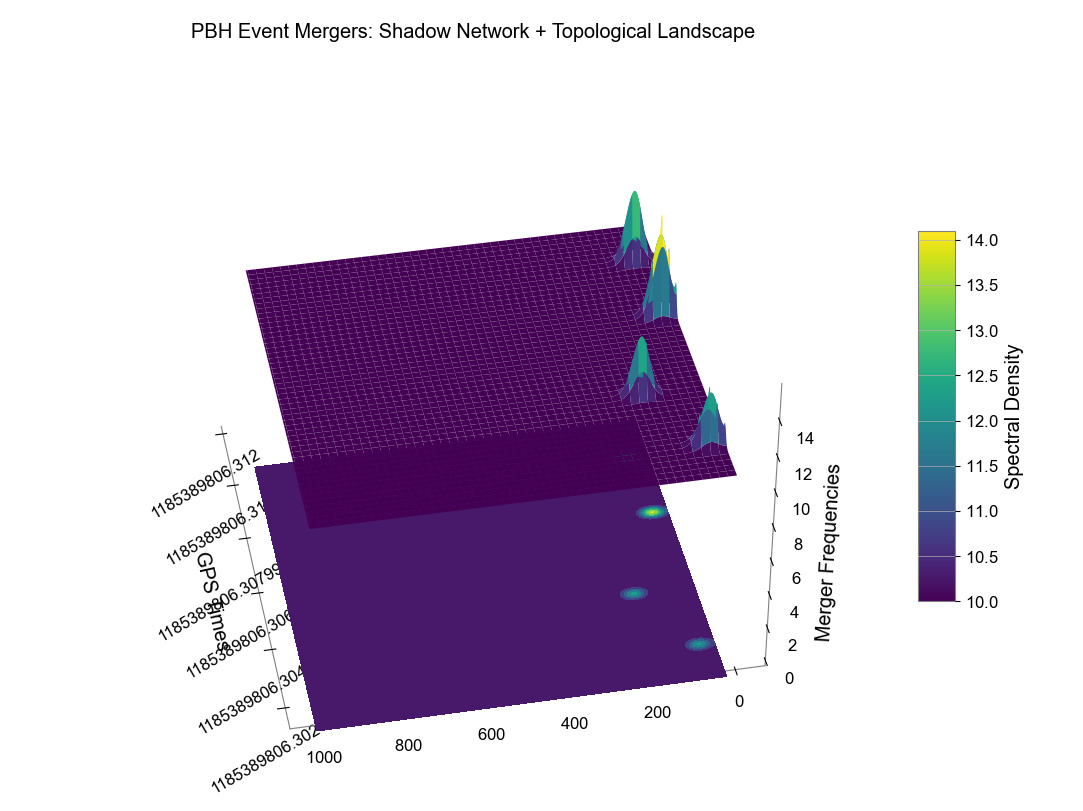}
        \caption{\textit{Topological Noise Reduction on Primary data with phase projections}}
        \label{fig:LIGO4_PlaceHolder_fig}
    \end{figure}

    This allowed us to achieve significant noise elimination while preserving the essential frequencies and spectral energy densities of the merger signals about their corresponding GPS times which could also be reverified using established secondary data from LIGO, thereby establishing the validity of this technique as shown in the above diagram.

    Furthermore, our computational framework allowed for the visualization and mapping of the peak frequency of the 3D wave spectral densities to their corresponding GPS time stamps using phase projection techniques described in the mathematical formalism of this paper.

    \subsubsection{Phase Projections}

    The Phase space projections of the peak frequencies of noise-reduced and filtered gravitational wave merger signals not only correspond to their GPS merger times but is very useful to serve as the choice of lens to build R-K models with  secondary attributes and parameters clustered about these primary event nodes as we shall see in the following sections of this paper.
    \begin{figure}[H]
        \centering
        \includegraphics[width=1.0\linewidth]{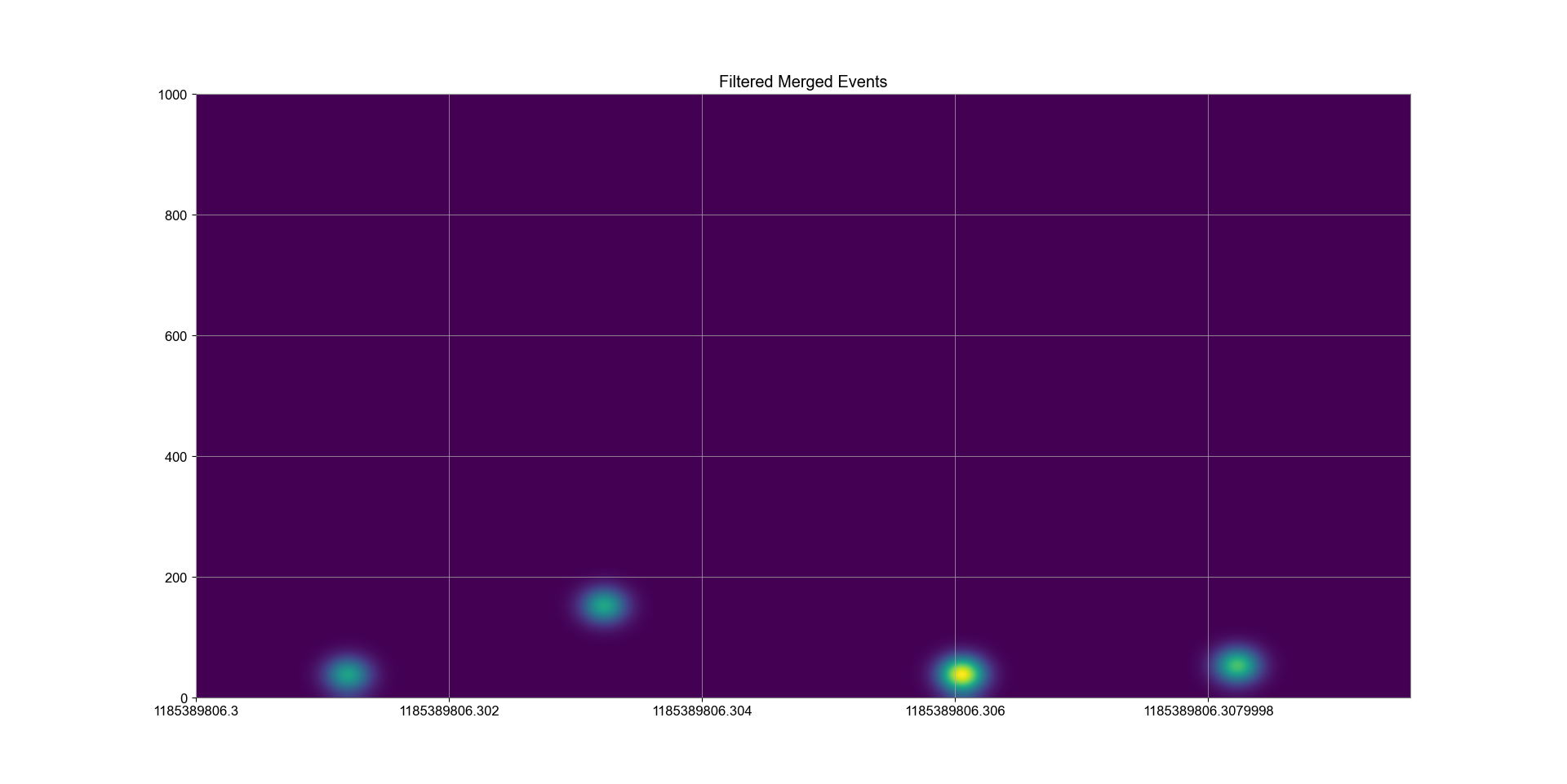}
        \caption{\textit{Phase Projections of Noise Reduced Topological Binary Merger Signals}}
        \label{fig:LIGO5_PlaceHolder_fig}
    \end{figure}

    \subsection{Secondary Analysis}

    We shall now extend the pipeline with the  exploration of secondary measures obtained using posterior probabilities and Bayesian parameter estimates in established research. \cite{24.0_BinaryMergerIdentification} \cite{24.4_CompactBinaryParameterEstimates} \cite{24.5_GWParameterEsitmation} \cite{24.6_LIGOParameterEstimates} It is however, important to note that the current version of the LIGO R-K pipeline does not address the computation of posterior probabilities and Bayesian parameter estimates from Primary Data. This remains to be in the scope of future research which could potentially help automate this entire process thereby generating R-K diagrams directly from detector signal data in near-real time and classifying them into a specific category of compact objects such as Black Holes, Neutron Stars, Primordial Black Hole Candidates etc.

    However, for the sake of the first version of the computational framework and R-K pipeline, we have restricted the scope of this paper in carrying out Secondary Analysis on the estimated physical measures corresponding to each merger event as published. \cite{00_LIGOOpenSciData} \cite{00.7_LIGOBayesianAnalysis} \cite{00.6_LIGOAnalysisPipeline}

    \subsubsection{Omni-view Plots}

    Thus using the flexibility and scalability of the R-K pipeline we have plotted an "Omni-view Plot" of all secondary measures plotted against each other with respect to the entire GWTC currently available.

    \begin{figure}[H]
        \centering
        \includegraphics[width=1.0\linewidth]{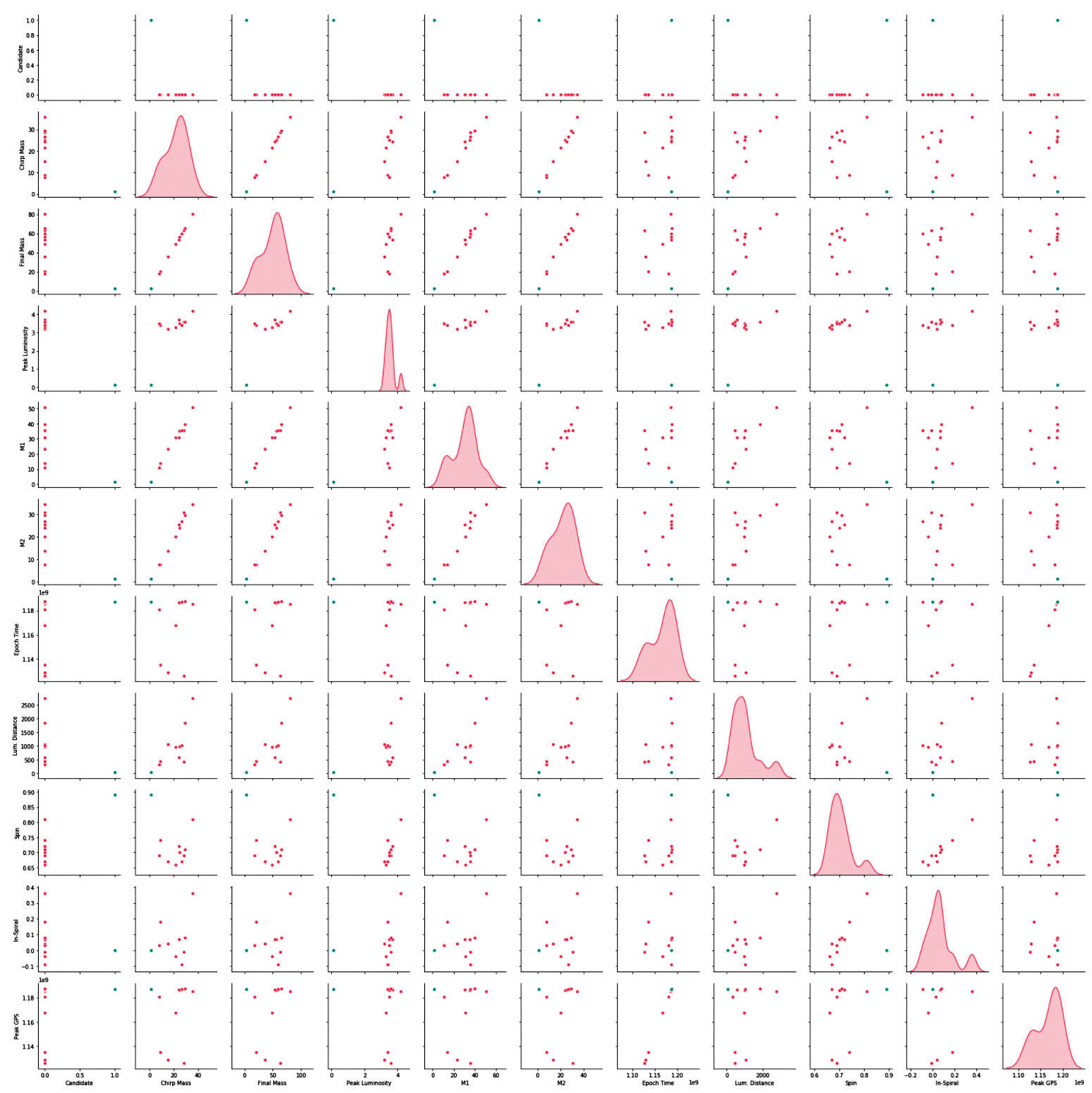}
        \caption{\textit{Omni-view Plot of Secondary-Data-Analysis}}
        \label{fig:LIGO6_PlaceHolder_fig}
    \end{figure}

    This allows for the graphical comparison of any two parameters thereby helping us identify exceptional properties of various merger events and compare them to the standard ones as done in case of the 4 merger events ( GW170729, GW170817, GW190521 \& GW190814) selected using this exercise. Now by removing the leading diagonal and all repeated  in graphical plots he omni-plot shown in the above diagram, one can easily cluster dependent and independent variable clusters for building R-K models. A more detailed and usable version of this omni-plot is available on \url{https://dev.topobot.ai/}

    \subsubsection{Event Dendograms}

    The we have built an ontology of from the LIGO data based on dependent and independent variables pertaining to the various intrinsic physical parameters of the compact binary mergers which describes the hierarchic relationships between all physical measures to be taken into consideration for building the respective R-K models. The various selected attributes obtained using posterior probabilities and Bayesian parameter estimates are grouped as follows:
   
   \begin{itemize}
   	\item Primary Mass $(M_1)$, Secondary Mass $(M_2)$,Total Mass $(M_\odot)$, Chirp Mass $(M_\odot)$, Detector Frame Chirp Mass $(M_\odot)$ and Final Mass $(M_\odot)$ are all grouped together under the mass cluster which is dimensionally independent of all other clusters in the Dendogram. 
   	\item Similarly, all spin measures pertaining to $\chi_eff $ , In-spiral \& Final Spin get grouped together in the dimensionally independent spin cluster
   	\item We also consider 2 separate dimensionless ratios representing measures of Red-shift \& Q-values (ratio of Primary Mass $(M_1)$ \& Secondary Mass $(M_2)$,) in 2 separate independent clusters for comparing characteristic differences between Black Holes, Neutron Stars \& candidate Primordial Black Holes.
   \end{itemize}  
   
   it is important to note that it is not essential to pre-select a limited set of parameters from the source data (LIGO or otherwise) for the hierarchical feature extraction and building Dendograms using the R-K pipeline. We have chosen a subset of these specific parameters for the sake of simplicity and clarity in the first version of our implementation with reference to the latest developments in the filed of Gravitational Wave Astronomy.
   
   \begin{figure}[H]
   	\centering
   	\includegraphics[width=1.0\linewidth]{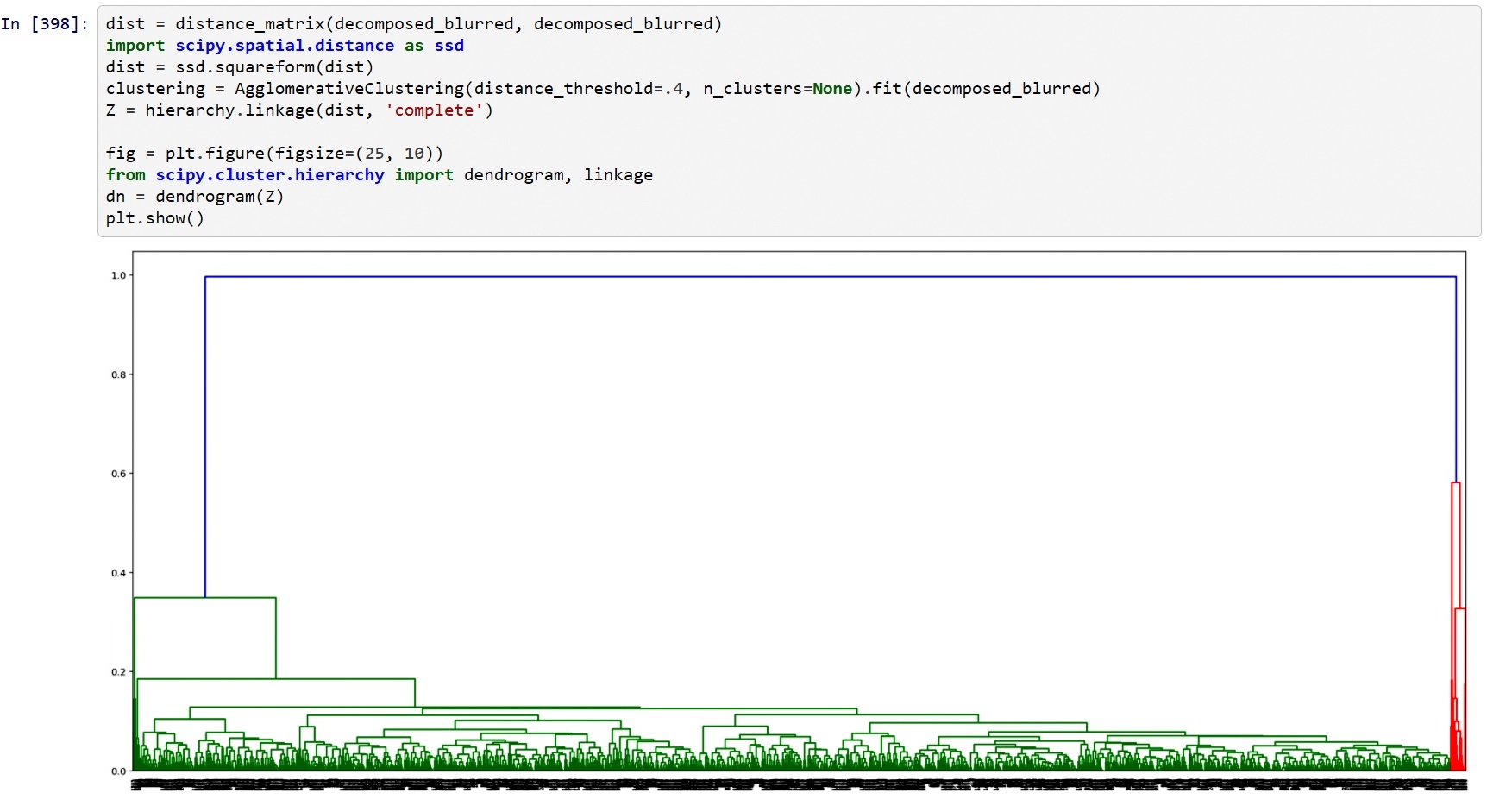}
   	\caption{\textit{Dendograms for Event-Based Topological Clustering }}
   	\label{fig:LIGO7_PlaceHolder_fig}
   \end{figure}
   
   The above diagram shows the primary event node marked in 'red'. This serves as the event node to which the entire mass cluster has been linked via Hierarchical Feature Extraction techniques clustering all the  various mass attributes, i.e. Primary Mass $(M_1)$, Secondary Mass $(M_2)$,Total Mass $(M_\odot)$, Chirp Mass $(M_\odot)$, Detector Frame Chirp Mass $(M_\odot)$ and Final Mass $(M_\odot)$ grouped together in terms of their interdependencies as shown in the diagram.
   
    \subsubsection{Clustering of Bayesian Parameters}
    
In this section, we will address the hierarchy defined within each cluster with the example of the spin cluster as shown in the diagrams below. We shall also consider a single event GW 170729 for the purpose of demonstrating each step in this process. The R-K pipeline allows for the flexibility to scale and run clustering on all merger events in parallel resulting in the fundamental structural graph for Gravitational Wave Analysis based on LIGO data.

    \begin{figure}[H]
        \centering
        \includegraphics[width=1.0\linewidth]{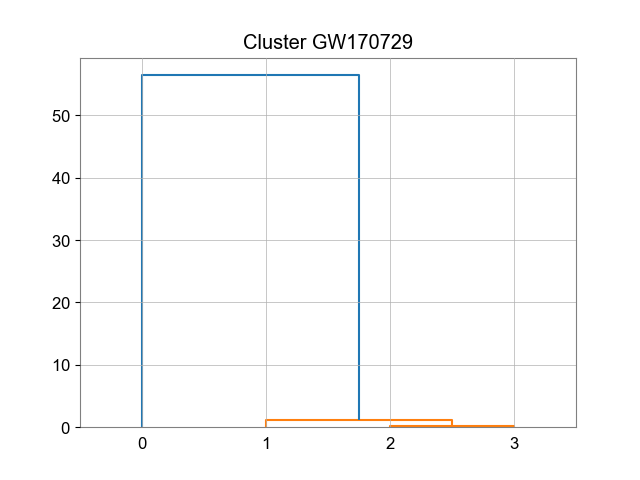}
        \caption{\textit{Single Event Dendogram GW170729}}
        \label{fig:LIGO8_PlaceHolder_fig}
    \end{figure}

The diagram above addresses the very first step of associated with a single event Dendogram. In this case the parent node or the 'event node' of the binary-merger event GW170729 first gets linked to one cluster at a time. In this case the diagram shows the linking of the spin cluster. All steps in this process take place in the Topological Phase Space and are independent of any choice of coordinates thereby ensuring scalability and flexibility of this framework.

    \begin{figure}[H]
        \centering
        \includegraphics[width=1.0\linewidth]{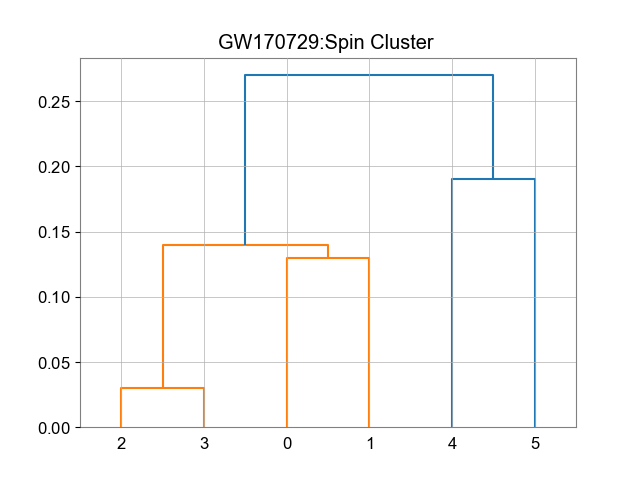}
        \caption{\textit{Spin Clustering of GW170729}}
        \label{fig:LIGO9_PlaceHolder_fig}
    \end{figure}

Now, we address the hierarchical grouping and ontological relationships within each cluster. The 3 measures of effective spin, in-spiral and final spin are ontologically linked and clustered together in an optimal way as shown in the figure above.

    \begin{figure}[H]
        \centering
        \includegraphics[width=1.0\linewidth]{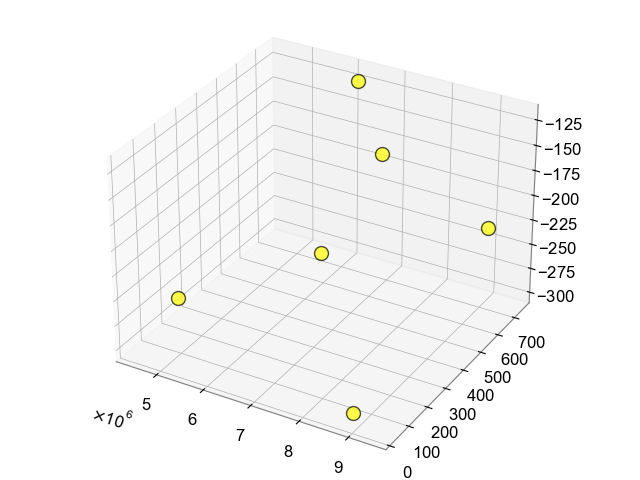}
        \caption{\textit{Spin Nodes of 170729}}
        \label{fig:LIGO10_PlaceHolder_fig}
    \end{figure}
The next stage addresses all the nodes of each cluster separately in Phase Space. In this case it takes into account all values of each spin measure within the entire range of estimation error as indicated in the GWTC. for example in case of GW170729 the $\chi_eff = 0.37 $ with an error range of +0.21 to -0.25. Hence all min-max error range values are encoded into nodes along with the most probable  $\chi_eff = 0.37 $. 
    \begin{figure}[H]
        \centering
        \includegraphics[width=1.0\linewidth]{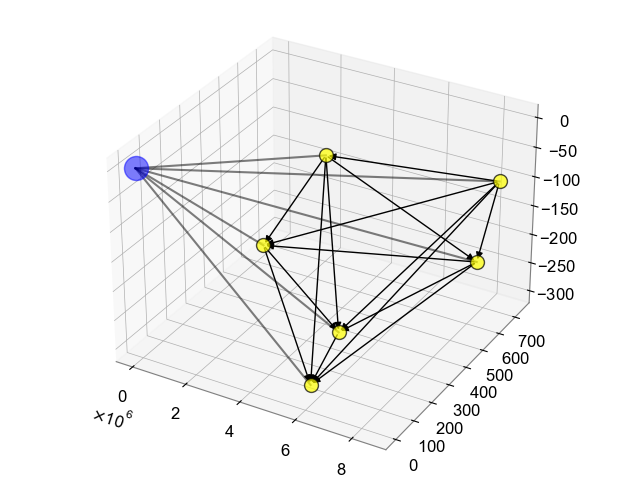}
        \caption{\textit{Spin Simplex of 170729 linking with Central Binary Merger Event node marked in blue}}
        \label{fig:LIGO11_PlaceHolder_fig}
    \end{figure}

The final step links all the nodes in the correct manner directed to the highest or lowest value according to the choice of the user. However, in this case we have used the Leaf Linker to connect all the node values with edges within the cluster pointing towards the highest value of each measure. The edges also represent the entire range of possible values with any amount of granularity between the maximum and minimum range of each probable measure. This also gives rise to DAGs and simplicial complexes within each independent cluster of an R-K model. As a final step all nodes in each cluster are then linked to the parent node or the "event node" represented in blue in the above diagram. We have also carried out isometric-compression of such structural graphs based on the euclidean distance using the Leaf Linker algorithm. This single linkage clustering was also done via the Leaf Linker available as a part of the R-K toolkit.

\subsubsection{Unfiltered Structural Graphs}

The steps described in the above section were carried out in parallel on each of the selected binary-merger events. This gave rise to 4 independent structural graphs for the binary merger events:  GW170729, GW170817, GW190521 \& GW190814, as shown in the diagram below.

    \begin{figure}[H]
        \centering
        \includegraphics[width=1.0\linewidth]{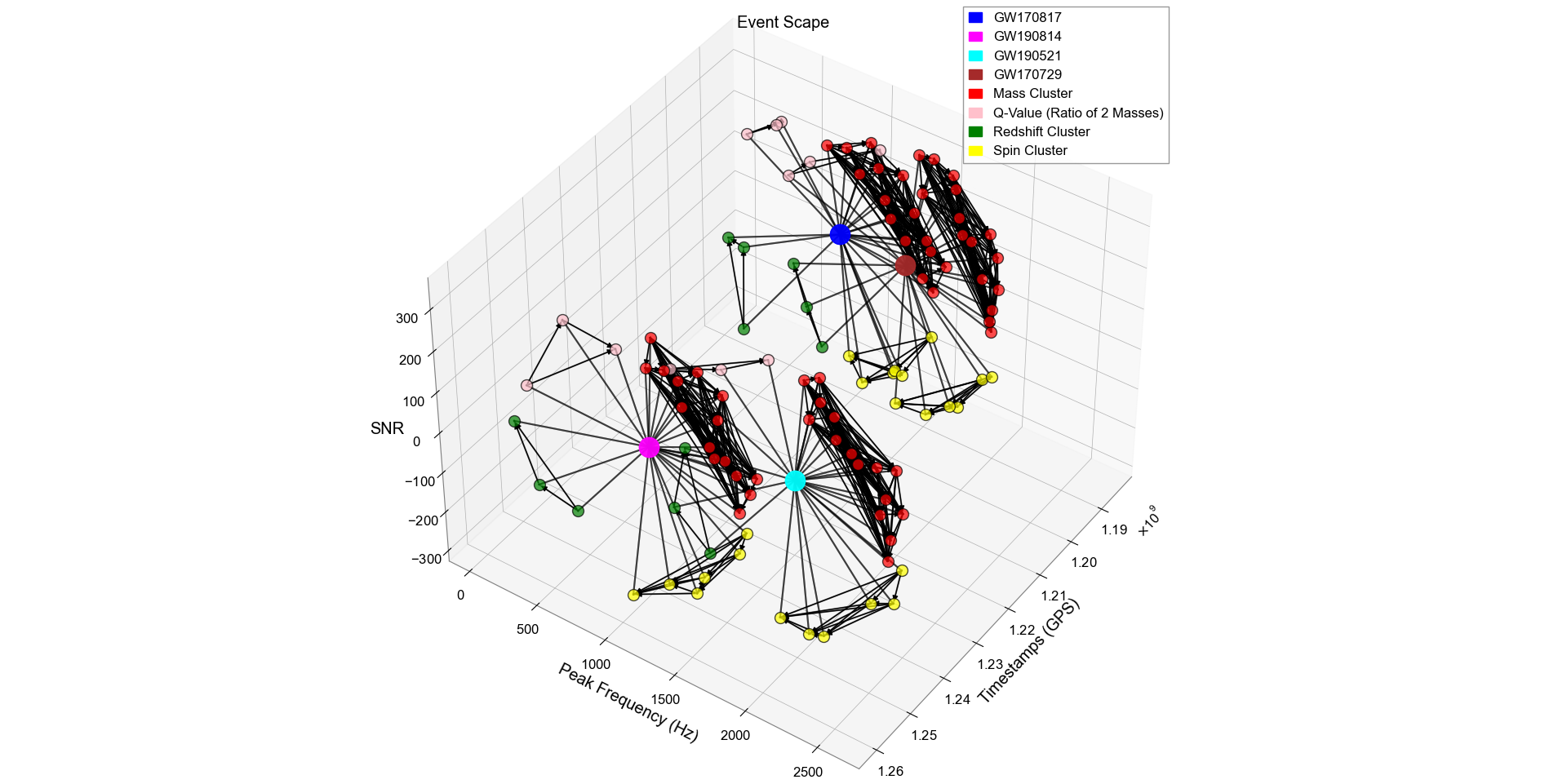}
        \caption{\textit{Event-scape of 4  Unfiltered Structural Topologies with linked multi-graphs as generated by the R-K Pipeline}}
        \label{fig:LIGO12_PlaceHolder_fig}
    \end{figure}
However, as detailed in the previous case-study with store-sales data, these structural graphs rise purely out of all data-points available in the columns of the high-dimensional dataset. Thus node masks and filter functions need to be applied to obtain unique event-driven R-K Diagrams as discussed in section \ref{sec:Filters} .

\subsubsection{Data Filters \& Sensitivity Thresholds}

The modified R-K pipeline for LIGO comes with a built-in visualizer and a GUI to track each step of the process from loading primary/raw strain data all the way up to plotting R-K diagrams. This is given to users to enable the smooth application of node masks and range filters to all the selected clusters of unfiltered R-K models for all binary merger events. One can use the GUI to run through and validate each step of the process described in this case study. 

    \begin{figure}[H]
        \centering
        \includegraphics[width=1.0\linewidth]{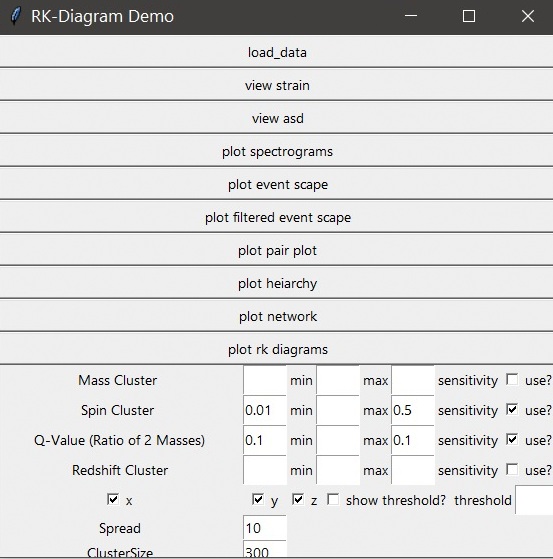}
        \caption{\textit{Sensitivity Thresholds and domain specific filters applied to the R-K Pipeline }}
        \label{fig:LIGO13_PlaceHolder_fig}
    \end{figure}

Hence the GUI in the above diagram allows parameter based topological filtering of the base structural graph or the R-K model in a way such that node masks and range filters either suppress or express components of each independent cluster based on any chosen measure and its corresponding filter parameters. In this case our objective was to find R-K diagrams with all spin measures close to 0 and low q-value ratios of around 0.1 in order to segregate potential neutron-star mergers from stellar black holes and primordial black holes. The mass cluster values add further distinction in terms of determining whether the compact binaries are in the mass gap. The choice of these parameters were driven by well established research in this field. \cite{00.7_LIGOBayesianAnalysis} \cite{00.6_LIGOAnalysisPipeline} \cite{24.5_GWParameterEsitmation} \cite{24.7_qvalueestimation} \cite{24.8_PBHdetectionparameters} \cite{24.9_EffectiveSpin}

    \subsubsection{
      Final R-K Diagrams}
    
Thus Parameter Based Topological Filtering was carried out on selected clusters such as Mass, Spin, Q-Ratio \& Red-shift with the Range Filter algorithms applied within pre-set ranges of $\chi_eff=0.01$ \&  $q=0.1$  \cite{24.7_qvalueestimation} \cite{24.9_EffectiveSpin} \cite{00.6_LIGOAnalysisPipeline}considering all error bars to obtain the corresponding R-K diagrams of the 4 selected events as described blow in the order of their GPS merger times.

    \begin{figure}[H]
        \centering
        \includegraphics[width=1.0\linewidth]{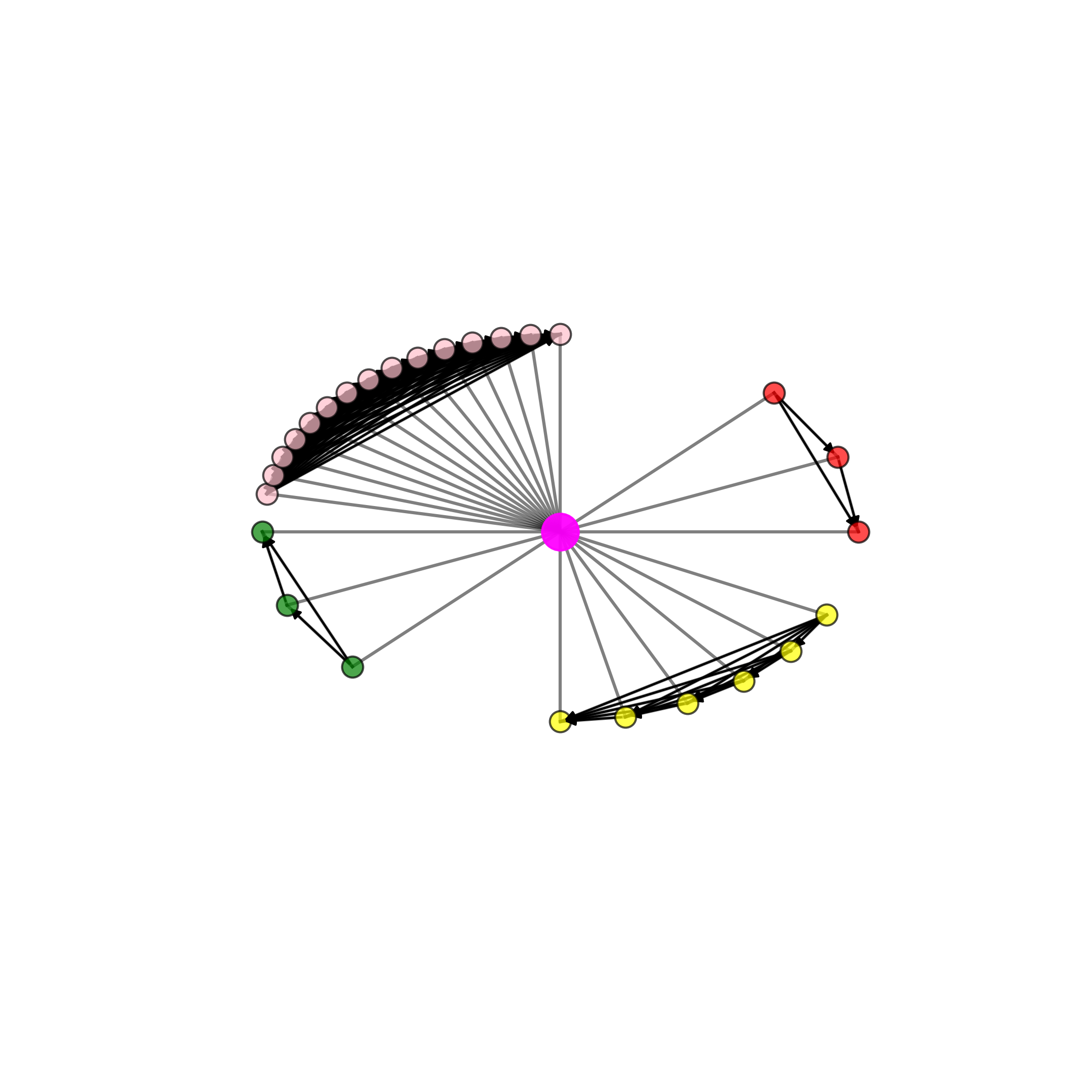}
        \caption{\textit{Final R-K Diagram of GW170729}}
        \label{fig:LIGO14_PlaceHolder1_fig}
    \end{figure}

The first R-K Diagram under consideration is that of the event GW170729 as shown above. In this case we find $\chi_eff \le 0.058$ \&  $q\ge 1.09$. The mass values don't fall in the mass gap range and hence represent all characteristics of a typical stellar black hole.

    \begin{figure}[H]
        \centering
        \includegraphics[width=1.0\linewidth]{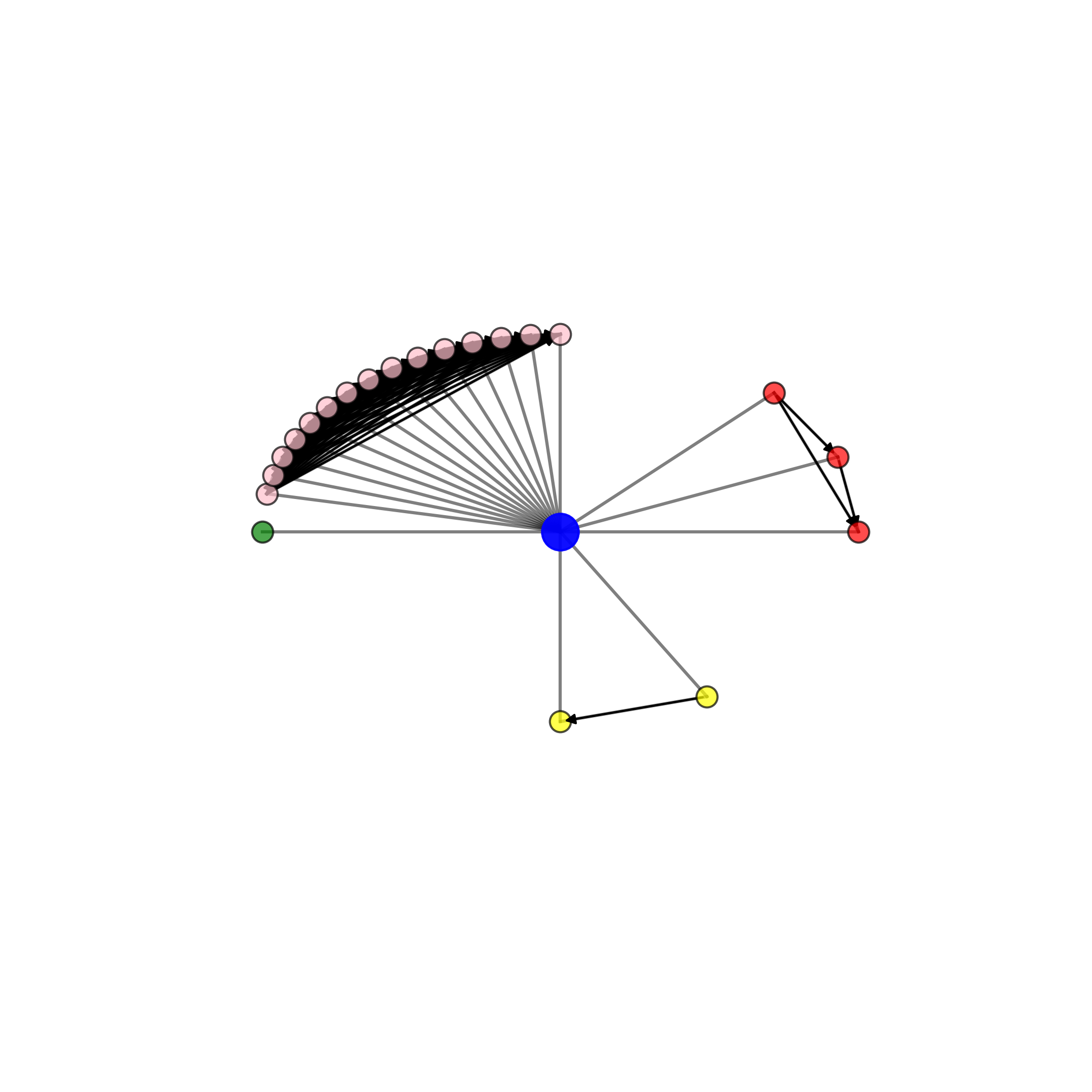}
        \caption{\textit{Final R-K Diagram of GW170817}}
        \label{fig:LIGO14_PlaceHolder2_fig}
    \end{figure}

The second R-K Diagram under consideration is that of the event GW170817 as shown above. In this case we find $\chi_eff = 0.01$ \&  $q\le 0.89$. The mass values don't fall in the mass gap range but the q value differs significantly from that of a candidate primordial black hole. This can be further verified by adding an electromagnetic counter-part from multi-messenger data which has been left out of scope for the purpose of this paper but could be easily added to the R-K models by extending the R-K pipeline in future. Thus the detection of an electromagnetic counterpart, with low $\chi_eff$ close to 0 and q value close to 1 establishes this event as a compact binary NS-NS (Neutron Star) merger.

    \begin{figure}[H]
        \centering
        \includegraphics[width=1.0\linewidth]{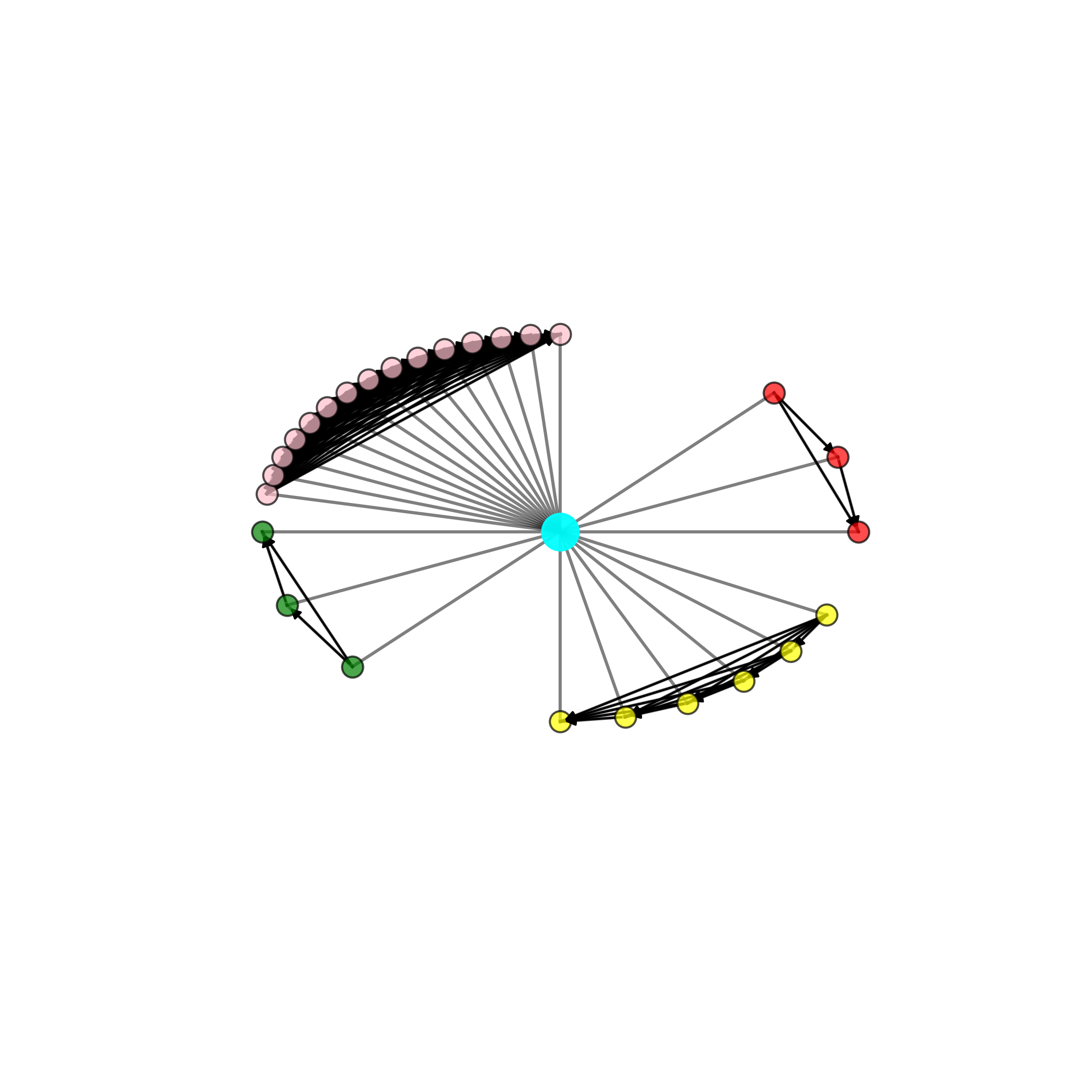}
        \caption{\textit{Final R-K Diagram of GW190521}}
        \label{fig:LIGO14_PlaceHolder3_fig}
    \end{figure}

The third R-K Diagram under consideration is that of the event GW190521 as shown above. In this case we find $\chi_eff \le 0.9$ \&  $q\ge 4$. The mass values don't fall in the mass gap range and hence this binary merger event also represents all characteristics of a typical stellar black hole much like GW170729 and both have topologically similar R-K diagrams.

    \begin{figure}[H]
        \centering
        \includegraphics[width=1.0\linewidth]{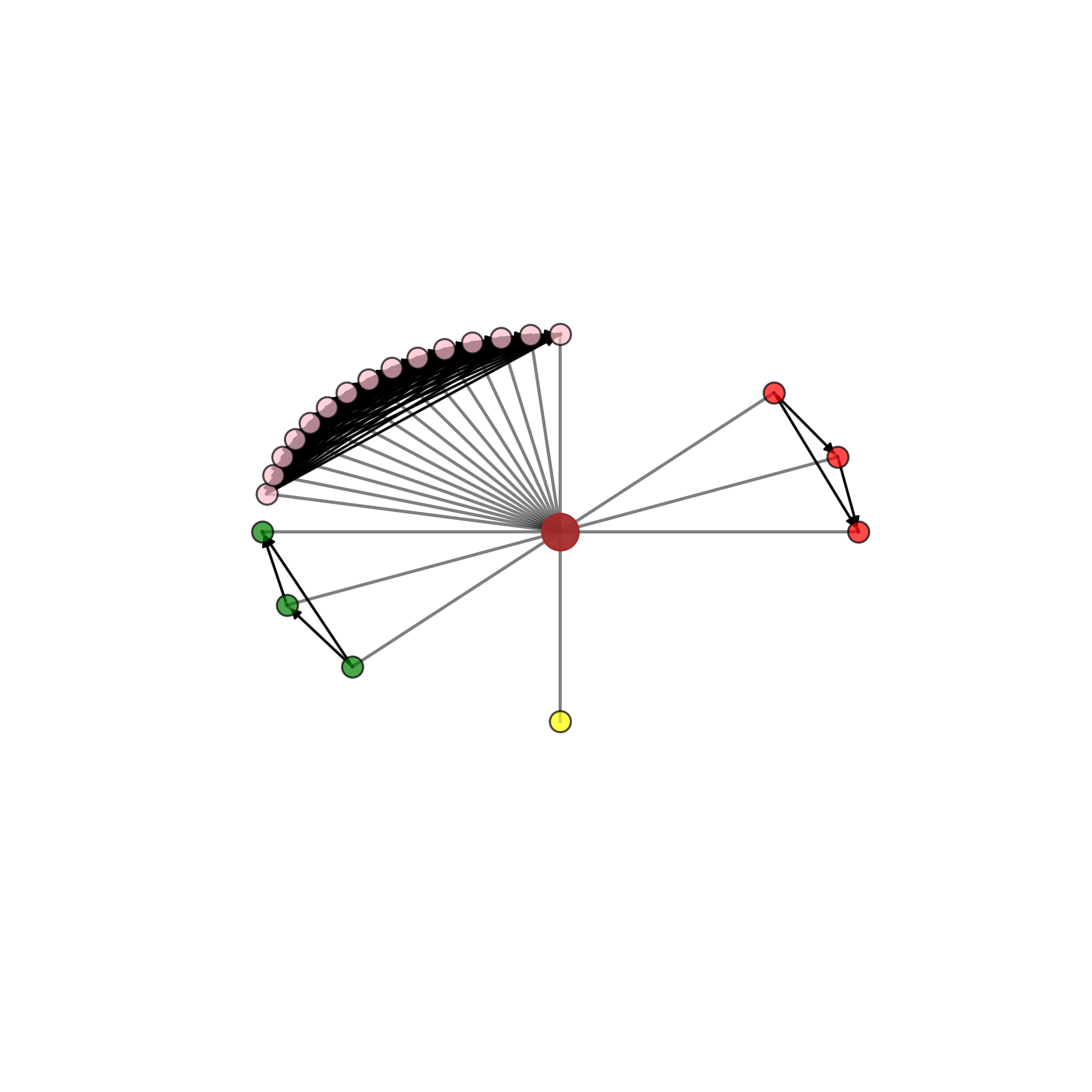}
        \caption{\textit{Final R-K Diagram of GW190814}}
        \label{fig:LIGO14_PlaceHolder_fig}
    \end{figure}

The fourth and final R-K Diagram under consideration is that of the event GW190814 as shown above. This is also the most interesting one under consideration as in this case we find $0.04 \le \chi_eff \le 0.05$ \&  $q = 0.1$. The mass values  definitely fall in the mass gap with no electromagnetic counterpart. Hence it has a lot of prominent features that are indicative of a candidate Primordial Black Hole (Dark Matter). However, further increase in detector sensitivity combined with added parameters such as more accurate detection of high Red-shifts which would be a distinguishing characteristic of candidate Primordial Black Holes. However, LIGO and VIRGO are currently not sensitive enough to detect such high red-shifts and therefore we require more sensitive probes like the Einstein Telescope for further parameter based distinction in future. But as discussed earlier, the R-K pipeline is designed in a highly scalable manner to include Multi-messenger data-streams in future to provide further distinction between R-K diagrams of different compact binary mergers.  

\subsubsection{Untuned R-K Diagram Event-Scape}

Thus all R-K diagrams of any given set of compact binary merges can be represented in a transformed Event-scape in 3D space with the x-axis representing the GPS time, y-axis representing the peak frequency values and the z axis representing the amount of SNR or signal to noise ratio as shown in the diagram below. The higher the SNR value the more confident we can be with respect to the detection and classification of distinct compact binaries.

    \begin{figure}[H]
        \centering
        \includegraphics[width=1.0\linewidth]{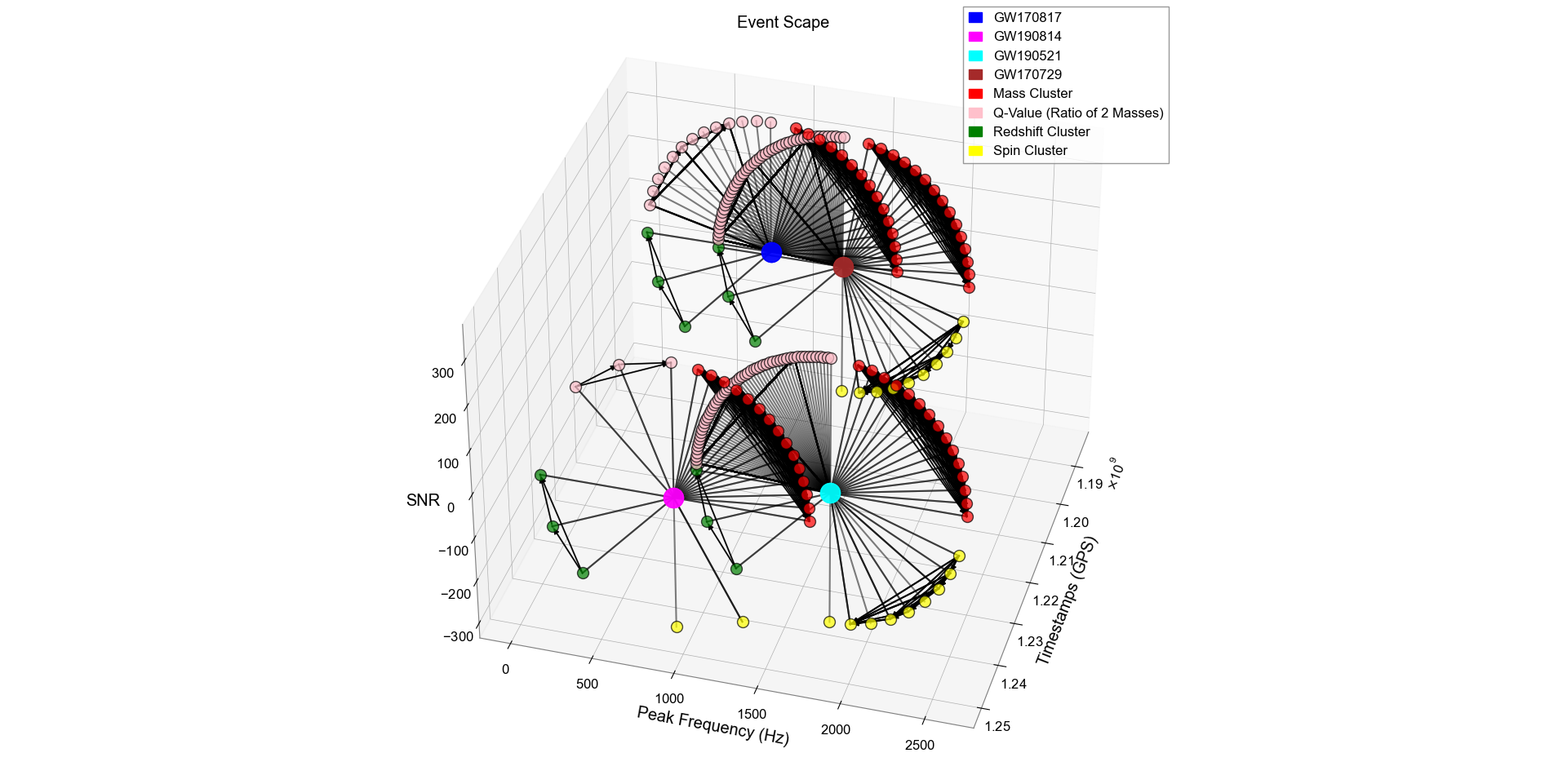}
        \caption{\textit{All 4 Event Based R-K Diagrams in 3D}}
        \label{fig:LIGO15_PlaceHolder_fig}
    \end{figure}

The above diagram shows multiple merger events plotted in the event-scape for which the topological distance or divergence and similarity measures were computed using templates from matched filters with respect to the 4 selected events. In order to test the accuracy and effectiveness of classification we set template labels on the 4 classified events. In order to show a proof of concept we assigned absolute labels to all 4 merger events such that GW170729 and GW190521 were assigned the label of BH-BH merger, GW170817 was labelled as a NS-NS merger and GW190814 was labelled as a PBH-PBH merger.

This resulted in an initial average intra-class topological distance separation of \textbf{0.17} between BH-BH mergers upon applying the parameter specific Range Filters and Node Masks discussed in the previous sections. This would imply an intra-class similarity of \textbf{0.83} between BH-BH mergers. It was also interesting to note an inter-class topological distance separation of \textbf{0.32} between BH-BH and confident NS-NS mergers which would imply a lower order inter-class similarity of \textbf{0.68} and verifies the effectiveness of R-K diagrams and the R-K pipeline to a significant extent. Comparative studies based on similarity measures of the candidate PBH was left for the next section as no clear or conclusive results were obtained using the untuned  models.

\subsubsection{Tuned Results for Template based Classifications}

In order to maximize divergence across R-K Diagrams, we trained our models using Facebook's Nevergrad \cite{a2020_nevergrad} optimizers. Thus, as demonstrated in the below sections, by providing an objective function and iterating over the various R-K models in a stochastic batch, we were able to reduce total loss of the set across different labelled classes over 1000 iterations.

After optimization, we encoded the R-K Diagram into a vector and trained an SVM over the vectorized diagrams to produce a binary classifier over the data where X represents each vectorized R-K Diagram and Y are labels of PBH Merger events. Thus the pipeline required two levels of optimization. The SVM generated a plane of separation between R-K diagrams as shown in the form of a grid in the diagram below.

    \begin{figure}[H]
        \centering
 	\includegraphics[width=1.0\linewidth]{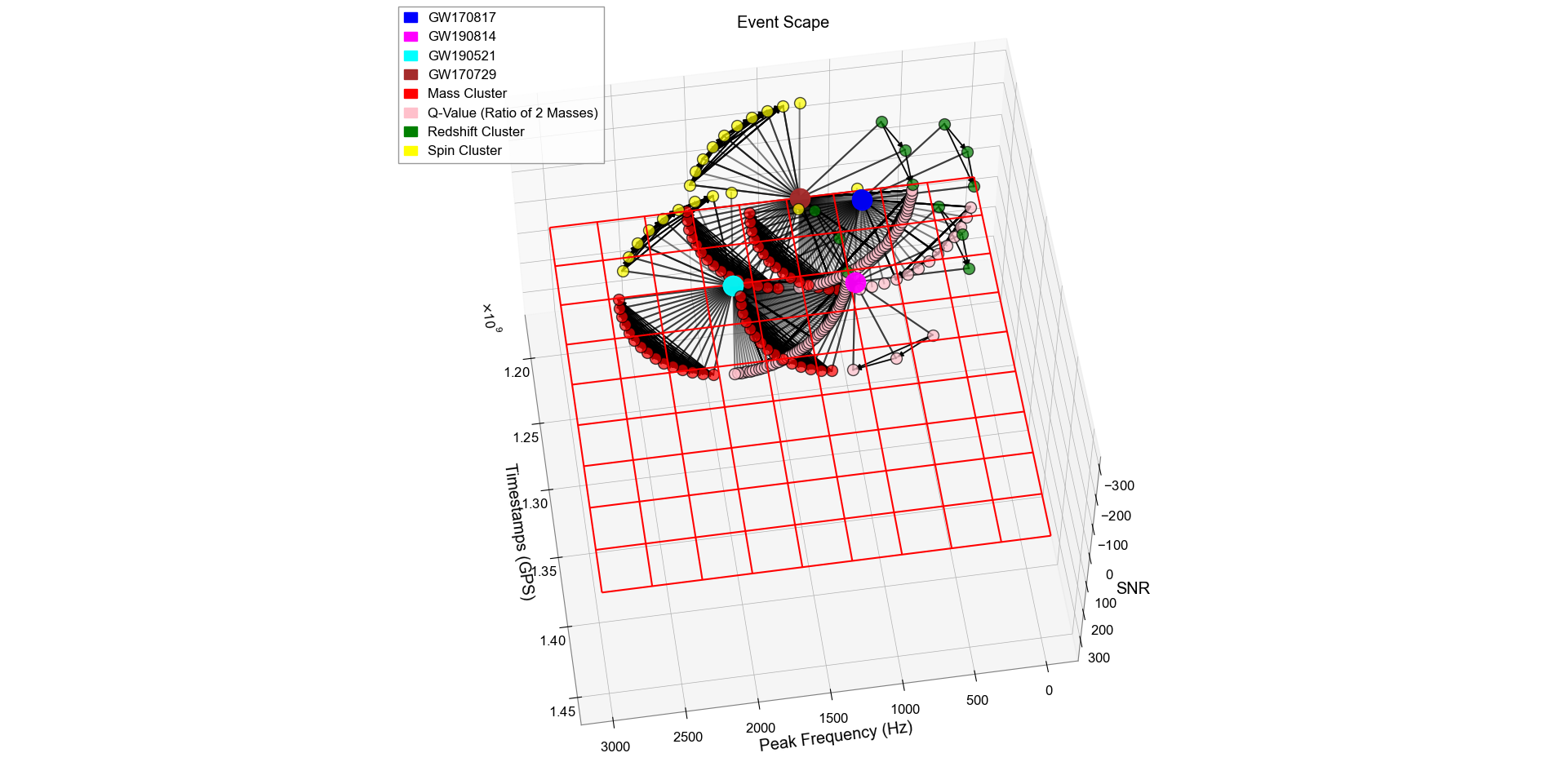}
 	\caption{\textit{All 4 Event Based R-K Diagrams in 3D with SVM threshold grid}}
 	\label{fig:LIGO16_PlaceHolder_fig}
 \end{figure}

This gave us closer intra-class topological separation between BH-BH mergers which were recorded with an average similarity measure between the range of $\lbrack0.9, 0.96\rbrack$ and NS-NS Mergers between a range   $\lbrack0.59, 0.65\rbrack$. The candidate PBH measured an interclass similarity of 0.71 with w.r.t. to NS-NS mergers and 0.62 w.r.t. BH-BH mergers with distinctly different R-K Diagrams. However, a more detailed study of the unique topological signatures and R-K diagrams of Primordial Black Holes is definitely promising but beyond the scope of this paper and should be left for future research due to the lack of conclusive data and sufficient observational parameters at present.

 \subsubsection{R-K Event-Scape vs Primary Analysis}
 
 The fine-tuned results of the R-K Event-Scape provides distinct categorical R-K diagrams representing a specific class of compact binary mergers as shown in the previous section. However, it is important to note that these fine-tuned R-K diagrams can then be remapped to the Q-transformed spectral densities from Primary Analysis.  
 
 \begin{figure}[H]
 	\centering
 	\includegraphics[width=1.0\linewidth]{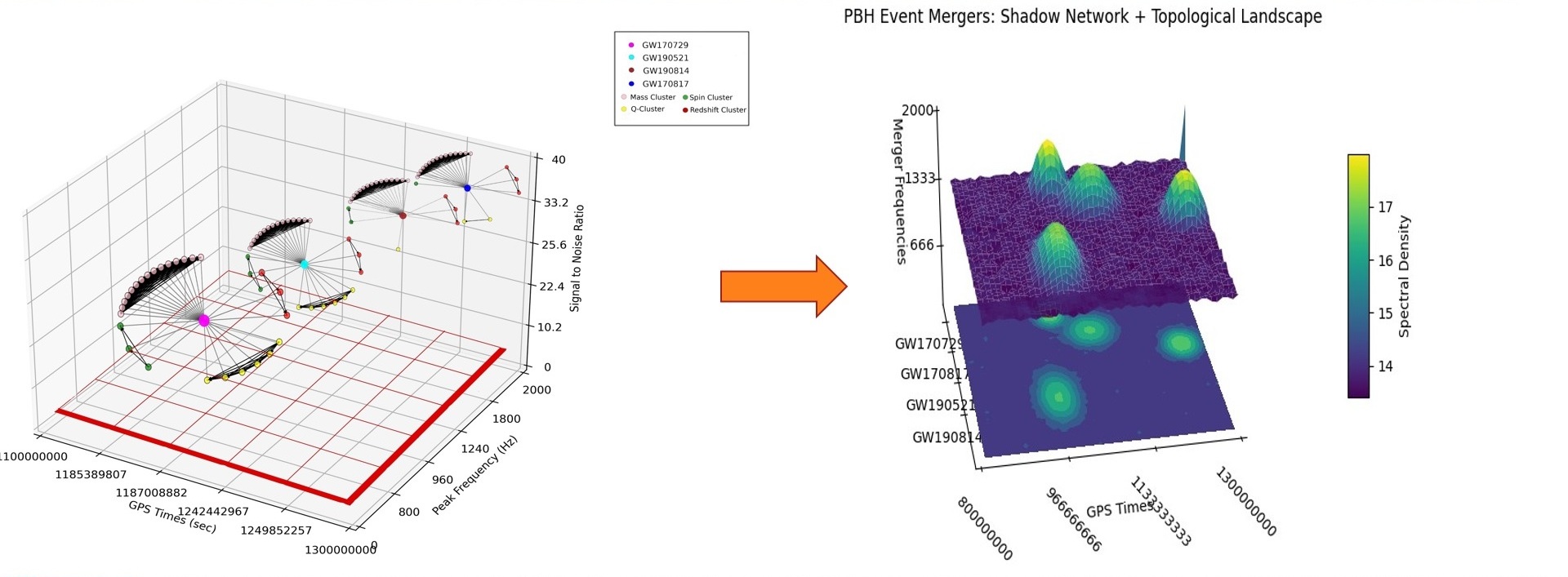}
	\caption{\textit{Mapping R-K Event-Scape to Primary Data}}
 	\label{fig:LIGO17_PlaceHolder_fig}
 \end{figure}

A study could then be conducted using the automated R-K pipeline to find any meaningful correlations between the the topological characteristics of the merger signals obtained form the detectors at source and their corresponding labelled counterparts obtained in the form of R-K Diagrams from the R-K pipeline. Since the final R-K Diagrams have been vectorized for the purpose of classification using SVMs, therefore they could serve as templates for the segregation and classification of future gravitational wave merger signals  obtained from compact binary mergers after the established steps of noise-filtering are carried out on the same. This would allow for near-real time segregation and classification of confident binary mergers which would keep improving over time with the addition of more SVM parameters from Multi-messenger data-streams and future advancements in LIGO. 

 \subsection{Scope of Future Work}

In this section we have showcased an example of a schematic representation of how the R-K pipeline could complement and augment the capabilities of  the current LIGO Analysis Pipeline which has been published by Abbott et.al. \cite{00.6_LIGOAnalysisPipeline} as shown in the diagram below.

\begin{figure}[H]
	\centering
	\includegraphics[width=1.0\linewidth]{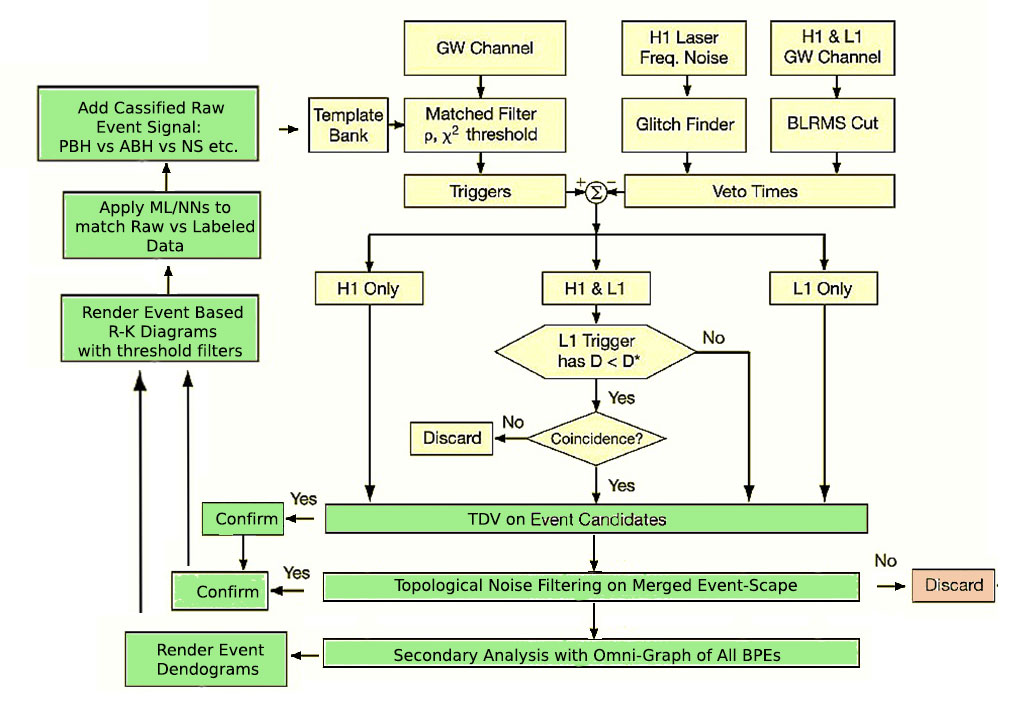}
	\caption{\textit{Modifying and augmenting a current version of the LIGO Analysis Pipeline with
			Topology, Graph Theory \& Machine Learning driven components of the R-K pipeline in future as shown in the diagram}}
	\label{fig:LIGO2_PlaceHolder_fig}
\end{figure}

In the diagram above, the components marked in yellow represent the existing components of the LIGO Analysis pipeline  as established by Abbott et.al. However, the green components of the diagram represent components of the R-K pipeline that could further augment and enhance its existing capabilities in terms of gravitational wave signal identification and the automated segregation and classification of compact-binary mergers. This can be achieved through effective computational frameworks involving Topology and Graph Theory where conventional machine learning techniques and neural networks have limited scope of application as proposed in this paper. Furthermore, R-K Diagrams can not only be used to explore  interesting ways of providing compact binary merger classifications, but they can also enrich the template banks for future identification, segregation and classification of gravitational wave merger signals detected at source.

An example of such a template bank has been conceptualised in the diagram below. In this case, any tertiary analysis  parameters obtained from future research papers could be easily added to this TDA pipeline  to better classify compact binary merges. Even though each binary merger could render R-K diagrams that vary in geometry, they would have strong categorical correlation with respect to topological similarity measures demonstrated in this paper.

 \begin{figure}[H]
	\centering
	\includegraphics[width=1.0\linewidth]{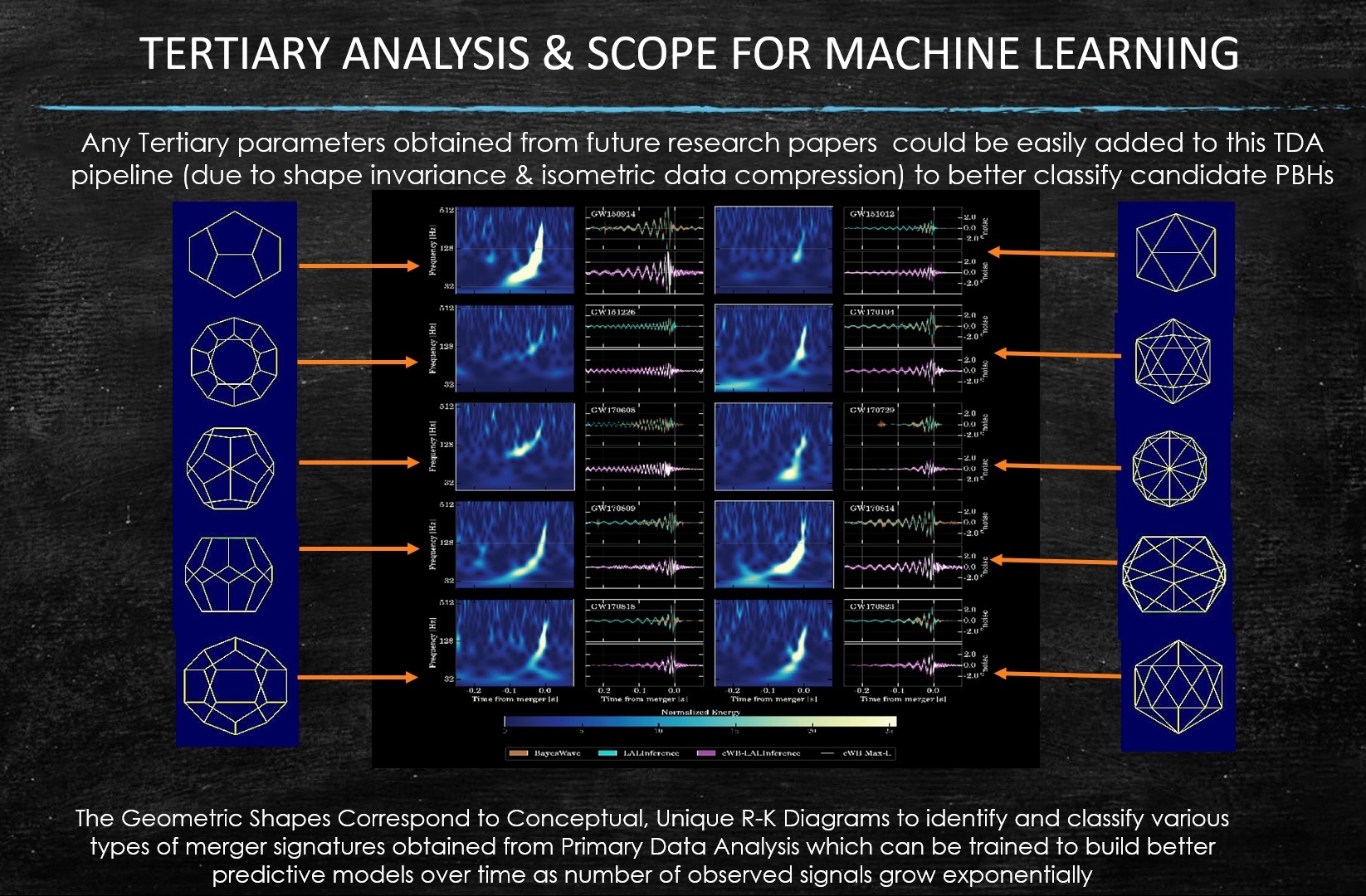}
	\caption{\textit{Binary Merger Event Classifications based on Tertiary Analysis and ML Predictions based on training of Similarity \& Divergence Measures between Topological Signatures of Merger Events }}
	\label{fig:LIGO18_PlaceHolder_fig}
\end{figure}

In the above diagram, the geometric shapes correspond to conceptual, unique R-K Diagrams to identify and classify various types of merger signatures obtained from Primary Data Analysis which can be trained to build better predictive models over time as number of observed signals grow exponentially along with the addition of new parameters to R-K models using Multi-messenger data-streams.

\subsection{Results \& Discussions}

Thus, the LIGO case study has clearly demonstrated the validity of this novel approach to Topological Graph Theory using R-K diagrams by providing a robust, flexible and scalable computational framework for the purpose of gravitational wave detector signal identification and compact binary merger classifications. It addresses some of the existing challenges with respect to the analysis of low mass-low spin compact binary mergers pertaining to Neutron Stars and candidate Primordial Black Holes by addressing the topological similarity and divergence across classes with effective ways of addressing computational complexity. It also provides an automated classification framework where conventional methods using machine learning and neural networks fall short due to lack of training data and limitations in terms of optimizing loss on multiple high dimensional data parameters all at once. One of the most unique advantages of this novel approach using R-K diagrams lies in ability to extract ontology driven effective summaries from high dimensional datasets containing significant amount of noise. it is also able to reduce analytical complexities significantly by providing a coordinate independent framework in phase space where discreet similarity or divergence measures can be obtained by varying changes on multiple data parameters all at once. Thereby it reduces complex classification problems to topological similarity and divergence measures as a measure of overall distance between any 2 R-K diagrams or their corresponding clusters in topological phase space. It can also enrich template banks for future predictive modelling on raw/primary gravitational wave data obtained from detectors at source and could effectively scale to include all multi-messenger data-streams to provide meaningful data summaries using R-K models and filtered, compressible R-K diagrams. Thus, novel techniques such as this could be eventually used to gain unique perspectives on all of Dark Matter using Multi-Messenger data sources in the near future.

 \section{Summary and Conclusions}

We introduced a novel approach toward \textbf{Topological Graph Theory Analysis}, by combining TDA and Graph Theories using a foundational ontology. We encoded vectorized associations between data points for smooth transitions between Graph and Topological Data on two disparate datasets: (1) A generalised implementation on a standard store sales dataset\cite{TableauSuperStore} (2) A specialised scientific implementation on LIGO Open Science Gravitational Wave data. This resulted in in filter specific, Homotopic self-expressive, event-driven unique topological signatures which we have referred as \textbf{``R-K Diagrams''}. With respect to store sales data, unique topological structures were derived between distinct purchase events with a loss ( measured in similarity between a set of R-K Diagrams ) in the range of $\lbrack0.78, 0.88\rbrack$. In the case of the classification case study with LIGO data analysis, we recorded a high accuracy of classification of classification with respect to known merger signals and a average similarity measure of BH-BH Mergers falling between the range of  $\lbrack0.9, 0.96\rbrack$ and NS-NS Mergers between a range   $\lbrack0.59, 0.65\rbrack$. The candidate PBH measured an inter-class similarity of 0.71 with w.r.t. to NS-NS mergers and 0.62 w.r.t. BH-BH mergers with distinctly different R-K Diagrams. This definitely presents an interesting case for further classification studies in future with more sensitive and varied data-streams. Therefore, we believe the findings of our work will lay the foundation for many future scientific and engineering applications of stable, high-dimensional data analysis with the combined effectiveness of \textbf{Topological Graph Theory} transformations.

The results obtained verify the basis of this novel approach in the following ways: \textbf{(1)} Distinct topological structures were created from the data and we have done so both on a store sales data and LIGO data. \textbf{(2)} These topological structures have emergent properties that when evaluated and compared to, have the capacity to provide meaningful insights into the data that standard data analysis techniques would not identify. For example, topological differences were exposed in the analysis that could not be exposed over metric based analysis such as euclidean distance, 'Mahalanobis Distance', and other standard metric based distance measures. \textbf{(3)} The resulting structures provide an extensible representation, which can be applied different methods of analysis, such as classification, identification, segmentation, etc.

We acknowledge that this novel approach is still very young and can mature significantly over time. We identify the following key areas for improvement:

\begin{enumerate}
    \item{Better filters and linkage functions.}
    \item{Better encoding and standadization formats for hierarchical feature transformation.}
    \item{Improved methods for optimization and training of R-K Pipelines.}
    \item{Better distance functions against R-K Diagrams.}
    \item{Improved visualizations.}
    \item{More complicated pipelines.}
    \item{Better encoding methods for graph pipelines.}
    \item{Additional applications and use cases.}
    \item{Improved software and toolkit maturity.}
\end{enumerate}

We believe that this computational framework and its varied applications will mature and expand over time, combining the advantages of topology and graph theory analysis with an underlying ontology can provide a novel and powerful method of analysing data, in which hidden properties and underlying patterns undiscovered by other data analysis techniques will emerge with unique event-driven R-K Diagrams.

\end{multicols}


\clearpage
\phantomsection
\addcontentsline{toc}{chapter}{Bibliography}

\printbibliography


\end{document}